\documentclass[12pt]{article}
\usepackage{a4,epsf}
\renewcommand{\Re}{{\rm Re}}
\renewcommand{\Im}{{\rm Im}}

\newcommand{\BLDgamma}{\gamma\hspace{-1.24ex}\gamma
                            \hspace{-1.24ex}\gamma}
\newcommand{\bfp}{{\bf p}}
\newcommand{\rmp}{{\rm p}}
\newcommand{\bfk}{{\bf k}}
\newcommand{\rmk}{{\rm k}}

\newcommand{\bfx}{{\bf x}}
\newcommand{\Z}{{\rm Z}}
\newcommand{\calc}{{\cal C}}
\newcommand{\slsh}[1]{\mbox{$#1 \hspace{-.45em} /$}}
\def\bld#1{\mbox{\boldmath$#1$}}
\def\bldsml#1{\mbox{\boldmath\scriptsize$#1$}}

\begin{document}
\begin{titlepage}
\title{Top Quark Polarization in Polarized\\
$e^+e^-$ Annihilation near Threshold\footnote{
    Work supported in part
    by the Polish State Committee for Scientific Research
    (KBN) grants 2P30225206 and 2P30207607,
    by Graduiertenkolleg ``Elementarteilchenphysik an Beschleunigern''
    and BMBF under contract 05-6KA93P(6)
    and by EEC contract ERBCIPDCT 940016.}
}
\author{ \\
R. Harlander$^a$,
M. Je\.zabek$^{b}$,
J.H. K\"uhn$^a$ and M. Peter$^a$\\
\\
{\normalsize \it $^a$ Institut f\"ur Theoretische Teilchenphysik,
D-76128 Karlsruhe, Germany}\\
{\normalsize \it $^{b}$ Institute of Nuclear Physics,
Kawiory 26a, PL-30055 Cracow, Poland}
}
\date{}
\maketitle
\thispagestyle{empty}
\vspace{-4.0truein}
\begin{flushright}
{\bf TTP 95-48\footnote{
    The complete paper, including figures, is
    also available via anonymous ftp at
    ttpux2.physik.uni-karlsruhe.de (129.13.102.139) as
    /ttp95-48/ttp95-48.ps,
    or via www at
    http://ttpux2.physik.uni-karlsruhe.de/cgi-bin/preprints}
}\\
{\bf March 1996}\\
{\bf hep-ph/9604328}
\end{flushright}
\vspace{3.0truein}
\begin{abstract}
\noindent
{\small
Top quark polarization in $e^+e^-$ annihilation into $t\bar t$
is calculated for linearly polarized beams.
The Green function formalism is applied
to this reaction near threshold.
The Lippmann--Schwinger equations for the $S$-wave and
$P$-wave Green functions are solved numerically for the
QCD chromostatic potential given by the two-loop formula
for large momentum transfer and Richardson's ansatz
for intermediate and small momenta.
$S$-$P$--wave interference contributes to all
components of the top quark polarization vector. Rescattering of
the decay products is considered.
The mean values $\langle n \ell \rangle$
of the charged lepton four-momentum projections on appropriately chosen 
directions $n$ in semileptonic top decays are proposed as experimentally
observable quantities sensitive
to top quark polarization.
The results for $\langle n \ell \rangle$ are obtained
including $S$-$P$--wave interference and rescattering of the decay
products.
It is demonstrated that for the longitudinally polarized
electron beam a highly polarized sample of top quarks can
be produced.\\
\vskip1cm
\begin{center}
\normalsize\it
Submitted for publication to Zeitschrift f\"ur Physik C
\end{center}
}
\end{abstract}
\end{titlepage}
\section{Introduction}
The top quark is the heaviest known elementary object. 
Precision studies of its interactions will lead to profound
progress in particle physics and may open an exciting new window towards the
very high mass scale.  Its large mass allows to probe deeply into the
QCD potential which governs the dynamics of the nonrelativistic $t\bar
t$ system. Such a system will provide a unique opportunity for a
variety of novel QCD studies.  It is likely that precise studies of
top quark production and decays will reveal new important information
on the mechanism of electroweak symmetry breaking.  The analysis of
polarized top quarks and their decays has recently attracted
considerable attention \cite{Kuehn3,teupitz,kzfest}.  For
nonrelativistic top quarks the polarization studies are free from
hadronization ambiguities. This is due to the short lifetime of the
top quark which competes favorably with the formation time of top mesons and
toponium resonances.  Therefore top decays interrupt the process of
hadronization at an early stage and practically eliminate associated
non-perturbative effects.

Threshold production of top quarks at a future
electron-positron
collider will allow to study their properties with
extremely
high precision. The dynamics of the top quark
is strongly influenced by its large width
$\Gamma_t\approx 1.5$ GeV.  Individual quarkonium resonances
can no longer be resolved, hadronization effects
are irrelevant and an effective cutoff of the
large distance (small momentum) part of the hadronic interaction
is introduced \cite{K,BDKKZ,FadKhoz1}.
This in turn allows to measure the short distance part
of the potential, leading to a precise determination
of the strong coupling constant \cite{strassler}.
The analysis of the total cross section combined with
the top quark momentum distribution
will determine
its mass $m_t$ with an accuracy of at least 300 MeV and its
width $\Gamma_t$ to
about 10\%. For a Higgs boson mass of order 100 GeV
even the $t\bar t H$ Yukawa coupling could be indirectly
deduced from its contribution to the vertex
correction \cite{work}.
Additional constraints on these parameters can be derived
from the forward-backward asymmetry of top quarks and from
measurements of the top quark spin.
Close to threshold, for $E=\sqrt{s}-2m_t\ll m_t$,
the total cross section and similarly the momentum
distribution of the quarks are essentially governed by
the $S$-wave amplitude, with $P$-waves suppressed
$\sim \beta^2 \sim
\sqrt{E^2+\Gamma_t^2}/m_t \approx 10^{-2}$.
The forward-backward asymmetry and, likewise,
the transverse component of the top quark spin originate
from the interference between $S$- and $P$-wave
amplitudes and are, therefore, of order $\beta \approx 10^{-1}$
even close to threshold.  Note that the expectation value
of the momentum is always different
from zero as a consequence of the large top width
and the uncertainty principle, even for $E=0$.
\par\noindent
It has been demonstrated \cite{FadKhoz1,strassler}
that the Green function technique is particularly
suited to calculate the total cross section in the threshold region.
The method has been extended
to predict the top quark momentum distribution.
Independent approaches have been developed for solving the Schr\"odinger
equation in position space \cite{MurSum1} and the Lippmann--Schwinger
equation in momentum space \cite{JKT,JT}.
The results of these two methods agree very well.
One of the most important future applications will be the
determination of $m_t$ and $\alpha_{\rm s}$ \cite{work}.
A further generalization then leads to the inclusion of
$P$-waves and, as a consequence, allows to predict
the forward-backward asymmetry \cite{MurSum2}.
It has been shown \cite{hjkt}  that the same function
$\varphi_{\rm _R}(\bfp,E)$ which results from the $S$-$P$--wave
interference governs the dynamical behaviour of the forward-backward
asymmetry as well as the angular dependence of the transverse
part of the top quark polarization.  The close
relation between this result and the tree level prediction,
expanded up to linear terms in $\beta$, has been emphasized.
The relative importance of $Z$ versus $\gamma$ and
of axial versus vector couplings depend on the
electron (and/or positron) beam polarization.
All predictions can, therefore, be further tested by exploiting
their dependence on beam polarization.
In fact the reaction $e^+e^-\to t\bar t$ with longitudinally
polarized beams is the most efficient and flexible source
of polarized top quarks. At the same time the longitudinal
polarization of the electron beam is an obvious option for
a future linear collider.
\par
In the present article the polarization dependent angular and momentum
distributions of top quarks are studied (neglecting effects of $CP$
violation, the corresponding distributions for $\bar t$ antiquark can
be obtained through a $CP$ transformation) and the results of
\cite{hjkt} are expanded in three directions:
\begin{itemize}
\item[1.] 
{\it Normal polarization.} The calculation of the polarization normal
to the production plane is a straightforward extension of the
previous work \cite{hjkt}. It is based on the same nonrelativistic
Green function as before, involving, however, the imaginary part of
the interference term $\varphi_{\rm _I}(\bfp,E)$.  A component of the
top quark polarization normal to the production plane may also be
induced by time reversal odd components of the $\gamma t \bar t$- or
$\Z t \bar t$-coupling with an electric dipole moment as most
prominent example. Such an effect would be a clear signal for
physics beyond the standard model.  The relative sign of particle
versus antiparticle polarizations is opposite for the QCD-induced
and the $T$-odd terms respectively which allows to discriminate between the
two effects. Nevertheless it is clear that a complete understanding
of the QCD-induced component is mandatory for a convincing analysis
of the $T$-odd contribution. 

\item[2.]  
{\it Rescattering of decay products.} Both $t$ quark and $\bar t$
antiquark are unstable and decay into $W^+b$ and $W^-\bar b$, respectively.
Neither $b$ nor $\bar b$ can be considered as freely propagating particles.
Rescattering in the $t\bar b$ and $b\bar t$ systems affects not only the
momenta of the decay products but also the polarization of the top quark.
Moreover, in the latter case, when the top quark decays first and its colored
decay product $b$ is rescattered in a Coulomb-like chromostatic potential of
the spectator $\bar t$, the top polarization is not a well defined
quantity. Instead one can consider other observables, like the angular momentum
of the $Wb$ subsystem, which coincides with the spin of top quark if
rescattering is absent.  Rescattering corrections are suppressed by
$\alpha_{\rm s}$.  The resulting modifications of the momentum distribution
are, therefore, minor and as far as the total cross section is
concerned can even be shown to vanish \cite{MY94,SThes}. In contrast the
forward-backward asymmetry as well as the transverse and normal parts of the
top quark spin are suppressed by a factor $\sim \beta$.  Hence they are
relatively more sensitive towards rescattering corrections.  The treatment of
rescattering follows to some extent the formalism of \cite{MurSum1,MurSum2}.
However, instead of the Coulomb potential we employ the full QCD potential.

The lifetime of the $t\bar t$ system is about $(2\Gamma_t)^{-1}$. 
At the time of the
decay of the first constituent the typical potential energy of $t\bar b$ 
(or $b\bar t$) is about 
$\alpha_{\rm s}^2\, m_t\sim\alpha_{\rm s}\sqrt{m_t\Gamma_t}$ 
where the approximate identity between the two quantities
is a numerical coincidence valid for $m_t$ of about 180 GeV. At the time of the
second (spectator) decay the mean separation of $b$ and $\bar b$ has increased
to $\Gamma_t$ and the corresponding potential energy is reduced to 
$\alpha_{\rm s}\,\Gamma_t$.  
Thus, rescattering in the $b\bar b$ system is less important
and will be neglected.
\item[3.]
{\it Moments of the lepton angular distribution.} 
The direction of the charged lepton in
semileptonic decays is optimally suited to analyze the polarization of the top
quark \cite{teupitz}. 
The reason is \cite{JK89b} that in the top quark rest frame
the double differential energy-angular distribution of the charged
lepton is a product of the energy and the angular dependent factors.
The angular dependence is of the form $(1 + P\cos\theta)$, where $P$
denotes the top quark polarization and $\theta$ is the angle between
the polarization three-vector and the direction of the charged
lepton. Gluon radiation and virtual corrections in the top quark
decay practically do not affect these welcome
properties \cite{CJK91}. It is therefore quite natural to perform
polarization studies by measuring the inclusive distributions of say
$\mu^+$ in the process $e^+e^-\to t(\mu^+\nu_\mu b)\bar t(jets)$.
This can be also convenient from the experimental point of view
because there is no missing energy-momentum for the $\bar t$
subsystem. From the theoretical point of view the direction of the
charged lepton is advantageous since it is equivalent to the top
quark polarization when rescattering is absent and remains defined
even after $b\bar t$ rescattering is included.  However, the
semi-analytic calculation of this effect is difficult because
production and decay mechanisms are coupled.  In this article
moments of experimentally measurable distributions are proposed
which are less difficult to calculate but still retain the important
information on top quark polarization.  These moments, the mean
values $\langle n \ell \rangle$ of the charged lepton
four-momentum projections in semileptonic top decays are calculated
including $S$-$P$--wave interference and ${\cal O}(\alpha_{\rm s})$
rescattering corrections.
\end{itemize}

The outline of this paper is as follows: In Sect.~\ref{technic_sec}
the relations between the momentum, angular and spin distributions of
top quarks and $S$- and $P$-wave Green functions are established.  The
formalism is developed in sufficient generality such that the
application to other reactions like $\gamma\gamma$-fusion into $t\bar
t$ is obvious. Simple rules for the translation from tree level
approximation to the case including the QCD potential are presented.
This section expands and details our previous results, as far as the
polarization in the production plane is concerned, and presents new
results for the normal polarization.  In Sect.~\ref{rescatt} a
formalism for the rescattering of the top quark decay products is
presented using the spin projection operators for the top quark.  The
influence of the rescattering on the various differential
(momentum, angular, polarization)
distributions is studied and limitations of the formalism in the case
of $b\bar t$ rescattering are discussed.  In Sect.~\ref{moments} the
moments $\langle n \ell \rangle$ are defined and calculated.  In
Sect.~\ref{results_sec} the numerical solutions of the
Lippmann--Schwinger integral equations are presented. Predictions are
given for the longitudinal, perpendicular and normal polarization
including predictions for the corresponding moments $\langle
n \ell \rangle$.  The dependence on the various input parameters
and the effects of rescattering are studied.  Various technical
aspects are relegated into the appendices.

\section{Green functions, angular distributions
        and quark polarization\label{technic_sec}}
\subsection{The nonrelativistic limit\label{nonrel_subs}}
Top quark production in the threshold region is conveniently
described by the Green function method which allows
to introduce in a natural way the effects of the large top
decay rate $\Gamma_t$ and avoids the summation of many
overlapping resonances. The total cross section can be
obtained from the imaginary part of the Green function
$G(\bfx=0,\bfx'=0,E)$ via the optical theorem. To predict
the differential momentum distribution, however, the
complete $\bfx$-dependence of $G(\bfx,\bfx'=0,E)$ (or, more precisely,
its Fourier transform) is required. In a calculation with
non-interacting quarks close to threshold the
forward-backward asymmetry, the angular dependent term
($\sim \cos\vartheta$) of the longitudinal part
and the transverse part of the top quark
polarization are all proportional to the quark velocity
$\beta$ and originate from the interference of an $S$-wave
with a $P$-wave amplitude.
These distributions are described
by $\nabla' \cdot G(\bfx,\bfx',E)|_{\bfx'=0}$ or,
equivalently, by the component of the Green
function with angular momentum one. The connection between
the relativistic treatment and the nonrelativistic
Lippmann--Schwinger equation has been discussed in
the literature \cite{strassler,MurSum2}. The subsequent discussion
follows these lines.  It includes, however, also the spin degrees
of freedom and is, furthermore, formulated sufficiently general
such that it is immediately applicable to other reactions of interest.
The main ingredient in the derivation of the nonrelativistic limit
is the ladder approximation for the vertex function $\Gamma_\calc$.
This vertex function is the solution of the following integral equation
(Fig.~\ref{lipp.ps}):
\begin{figure}
\begin{center}
  \leavevmode
  \epsfxsize=12.cm
  \epsffile[50 220 540 350]{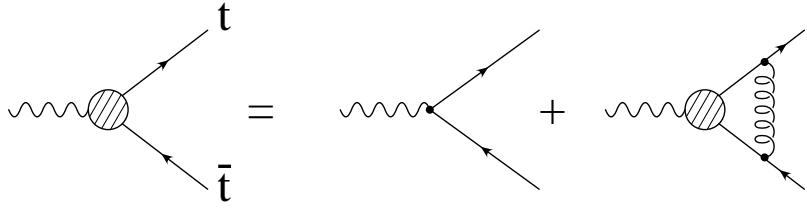}\\
  \caption[]{
        \label{lipp.ps}\sloppy Lippmann--Schwinger equation (\ref{int_eq})
        in diagrammatical form.}
\end{center}
\end{figure}
\begin{figure}[htb]
  \begin{flushleft}
    \leavevmode
    \epsfxsize=5.cm
    \epsffile[165 200 400 370]{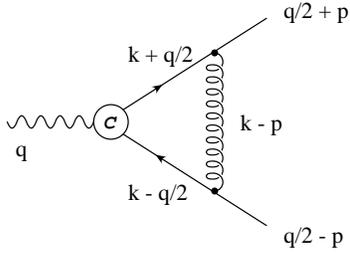}\\[-1.cm]
    \hfill
    \parbox{8.cm}{
      \caption[]{\label{1loopc.ps}\sloppy Definition of the four-momenta.}}
  \end{flushleft}
\end{figure}
\begin{equation}\label{int_eq}
  \Gamma_\calc = \calc + \int {d^4k \over (2 \pi)^4}
  \left( -{4 \over 3} 4 \pi \alpha_{\rm s} \right)
    D_{\mu \nu}(p-k) \gamma^\mu S_{\rm F}(k+{q \over 2})
    \Gamma_\calc(k,q) S_{\rm F}(k-{q\over2}) \gamma^\nu ,
\end{equation}
with $\calc=\gamma_\mu$ or $\gamma_\mu\gamma_5$
in the cases of interest for $e^+e^-$-annihilation.
The conventions for the flow of momenta are illustrated in
Fig.~\ref{1loopc.ps}.
The four-momenta are related to the nonrelativistic
variables by
\begin{eqnarray}
q &=& (2 m_t + E,{\bf 0}) \nonumber\\
p &=& (0, \bfp) \label{impulse}\\
k &=& (k_0,\bfk) .\nonumber
\end{eqnarray}
In perturbation theory the ladder approximation is motivated
by the observation that for each additional rung the energy
denominator after loop integration compensates the coupling
constant attached to the gluon propagator. This is demonstrated most
easily in Coulomb gauge. Contributions from transverse gluons
as well as those from other diagrams are suppressed by higher powers of
$\beta\sim\alpha_{\rm s}$.
The gluon propagator is thus replaced by the
instantaneous nonrelativistic potential:
\begin{equation}\label{gluon_prop}
- {4\over 3}\, 4\pi \alpha_{\rm s} D_{\mu\nu}(p) \rightarrow
         i\, V(\bfp)\, \delta_{\mu 0} \delta_{\nu 0} .
\end{equation}
The dominant contribution to the integral originates from the region
where $|\bfk|\ll m_t$. Including terms linear in $\bfk$,
quark and antiquark
propagators are approximated by\\
\parbox{32.em}{
\begin{eqnarray*}
\hspace{10.ex}
S_{\rm F}(k+{q\over2}) &=& i {\Lambda_+ -
     {\bfk\cdot\bldsml\gamma \over 2 m_t} \over
     {E \over 2} + k_0 - {\bfk^2 \over 2 m_t} + i {\Gamma_t \over 2}} \\
S_{\rm F}(k-{q\over2}) &=& i {\Lambda_- -
      {\bfk\cdot \bldsml\gamma \over 2 m_t} \over
      {E \over 2} - k_0 - {\bfk^2 \over 2 m_t} + i {\Gamma_t \over 2}} \, ,\\
\Lambda_\pm &=& {1\pm\gamma^0\over 2} \, .
\end{eqnarray*}}
\hfill\parbox{5.ex}{
        \begin{eqnarray} \label{fermion_props} \end{eqnarray}}
The ``elementary'' vertex $\calc$ is independent of $k_0$.  (Within the
present approximations this is even true if $\calc$ does depend on $\bfk$ as is
the case in the analogous treatment of $\gamma\gamma\to t\bar t$
discussed below.)
Up to and including order $\beta$ terms
a self-consistent solution of the integral equation (\ref{int_eq})
can be obtained if $\Gamma_\calc$ is taken
independent of $k_0$ and the nonrelativistic spins of $t$ and $\bar t$.
The $k_0$ integration is then easily performed and the
integral equation simplified to
\begin{equation}\label{int_eq_2}
\Gamma_\calc = \calc - \int {d^3 k \over (2 \pi)^3}
        V(\bfp - \bfk) \gamma^0
        (\Lambda_+ - {\bfk\cdot \BLDgamma \over 2 m_t})
        {\Gamma_\calc(\bfk,E) \over E - {\bfk^2 \over m_t} + i \Gamma_t}
        (\Lambda_- - {\bfk\cdot \BLDgamma \over 2 m_t})
        \gamma^0  .
\end{equation}
In the calculation of the cross section for the production of
$t$ plus $\bar t$
with momenta $q/2\pm p$ and spins $s_\pm$ respectively traces ${\cal H}$ of the
following structure will arise:
\begin{eqnarray}
{\cal H} &=& {\rm Tr} \{ {\cal P}_+({q \over 2} + p,s_+)
  \Gamma_\calc {\cal P}_-({q \over 2} - p,s_-) \bar{\Gamma}_{\calc'} \} , \\
  {\cal P}_\pm(p,s) &=& {\slsh p \pm m_t \over 2 m_t}\,
        {1+\gamma_5 \slsh s \over 2} ,
\end{eqnarray}
where we allowed for mixed terms with $\calc$ different from $\calc'$
arising e.g.~from vector-axialvector interference.
Expanding again up to terms linear in $\bfk$,
this trace can be transformed into
\begin{eqnarray}
{\cal H} &=&
        {\rm Tr} \{ {\cal S}_+ \widetilde\Gamma_\calc {\cal S}_-
        \bar{\widetilde\Gamma}_{\calc'} \} , \\
{\cal S}_\pm &=& {1 \pm {\bf s}_\pm \cdot\bld{\Sigma} \over 2}  ,\\
\bld{\Sigma} &=& \BLDgamma \gamma_5 \gamma^0 =
        \left(\begin{array}{cc} \bld{\sigma} & 0 \\
                                0            & \bld{\sigma}
                \end{array}
        \right) ,
\end{eqnarray}
with the nonrelativistic reduction defined through
\begin{equation}\label{tilde}
\widetilde{\Gamma}_\calc(\bfp,E)
        = \Lambda_+ \, (1-{\bfp\cdot \BLDgamma \over 2 m_t}) \,
        \Gamma_\calc(\bfp,E)\,(1-{\bfp\cdot \BLDgamma \over 2 m_t}) \,
        \Lambda_-  .
\end{equation}
It is thus sufficient to calculate the ``reduced'' vertex
function $\widetilde \Gamma_\calc$. Dropping again terms
of order $\bfk^2$, the corresponding integral
equation is cast into a particularly simple form
\begin{equation}\label{int_eq_3}
\widetilde{\Gamma}_\calc(\bfp,E) = \widetilde{\calc} (\bfp) +
        \int {d^3 k \over (2 \pi)^3} V(\bfp - \bfk)
        {\widetilde{\Gamma}_\calc(\bfk,E) \over E -
        {\bfk^2 \over m_t} + i \Gamma_t}  ,
\end{equation}
where $\widetilde{\calc}(\bfp)$ is defined in analogy to
$\widetilde{\Gamma}_\calc(\bfp,E)$ in (\ref{tilde}).
Consistent with the nonrelativistic approximation only the
constant and the linear term in the Taylor expansion of the
elementary vertex will be considered\footnote{
In the notation of Ref.\cite{kks} one gets:
  $\widetilde\calc(0)=\Lambda_+{\cal O}_0\Lambda_-$ and
  ${\bf D} = \Lambda_+\left[-{1\over 2m_t}\{{\cal O}_0,\BLDgamma\}_+ +
        \hat{\bld{\cal O}}\right]\Lambda_-$.}
\begin{equation}
\widetilde{\calc}(\bfp) = \widetilde{\calc}(0) + {\bf D}\cdot \bfp .
\end{equation}
The matrices  $\widetilde\calc(0)$ and ${\bf D}$ may in general
depend on external momenta, polarization vectors or Lorentz
indices. A self-consistent solution for the vertex
$\widetilde\Gamma_\calc$ is then given by
\begin{equation}\label{ansatz}
\widetilde\Gamma_\calc(\bfp,E) = \widetilde\calc (0)
{\cal K}_{\rm S}(\rmp,E) +
        {\bf D}\cdot \bfp \, {\cal K}_{\rm P}(\rmp,E) .
\end{equation}
The scalar vertex functions ${\cal K}_{\rm S,P}$ depend on the modulus of the
three momentum
$$  \rmp = |\bfp| $$
(which should be distinguished from the four-momentum $p$) and 
the nonrelativistic energy $E$ only.
They are solutions of the nonrelativistic integral equations
\begin{eqnarray}
{\cal K}_{\rm S} (\rmp,E) &=& 1 + \int {d^3 k \over (2 \pi)^3}
        V(\bfp - \bfk) {{\cal K}_{\rm S} (\rmk,E) \over E -
        {\bfk^2 \over m_t} + i \Gamma_t}  \\
{\cal K}_{\rm P} (\rmp,E) &=& 1 + \int {d^3 k \over (2 \pi)^3}
        {\bfp\cdot \bfk \over \bfp^2}
        V(\bfp - \bfk) {{\cal K}_{\rm P} (\rmk,E) \over E -
        {\bfk^2 \over m_t} + i \Gamma_t}
\end{eqnarray}
and are closely related to the Green function
${\cal G}(\bfp,\bfx,E)$ which, in turn, is a solution of the
Lippmann--Schwinger equation
\begin{equation}\label{LS}
\bigg[E - {\bfp^2 \over m_t} + i \Gamma_t \bigg]
        {\cal G} (\bfp,\bfx,E) = e^{i \bfp \cdot \bfx} +
        \int {d^3 k \over (2 \pi)^3} V(\bfp - \bfk)
        {\cal G} (\bfk,\bfx,E) .
\end{equation}
Let us denote the first two terms of the Taylor series with respect
to $\bfx$ by $G$ and $F$ respectively:
\[
{\cal G}(\bfp,\bfx,E) =
G(\rmp,E) + \bfx \cdot i \bfp \, F(\rmp,E) + \ldots
\]
They are solutions of the integral equations
\begin{eqnarray}
G(\rmp,E) &=& G_0(\rmp,E) + G_0(\rmp,E) \int {d^3 k\over (2 \pi)^3}
        V(\bfp - \bfk) G(\rmk,E)                   \label{lipps} \\
F(\rmp,E) &=& G_0(\rmp,E) + G_0(\rmp,E) \int {d^3 k\over (2 \pi)^3}
        {\bfp \cdot \bfk \over \bfp^2}
        V(\bfp - \bfk) F(\rmk,E)  ,                \label{lippp}
\end{eqnarray}
with
\begin{equation}\label{free}
G_0(\rmp,E) = {1\over E - {\bfp^2 \over m_t} + i \Gamma_t} ,
\end{equation}
and the relation between Green function and vertex function
\begin{equation}
  G(\rmp,E) = G_0(\rmp,E) {\cal K}_{\rm S} (\rmp,E) , \qquad
  F(\rmp,E) = G_0(\rmp,E) {\cal K}_{\rm P} (\rmp,E)
\end{equation}
is  evident.
In the case of $e^+e^-$-annihilation top production proceeds
through the space components of the vector and axial vector
current.  The relevant elementary
vertex $\widetilde\calc(\bfp)$ is given by
\begin{eqnarray}
\widetilde{\gamma_j}(\bfp) &=& \Lambda_+ \gamma_j \Lambda_- \\
\widetilde{\gamma_j\gamma_5}(\bfp) &=& \Lambda_+({i\over m_t})
        (\bld{\gamma}\times\bfp)_j \Lambda_-
\end{eqnarray}
for vector and axial current respectively. Production of $t\bar t$ in
$\gamma\gamma$-fusion would lead to an elementary vertex of the form
\begin{eqnarray}
  \widetilde{\calc}(0) &\propto& i (\bld{\epsilon}_1
  \times \bld{\epsilon}_2)\cdot
  {\bf n}_{e^-} \Lambda_+ \gamma_5 \Lambda_-\\
  {\bf D} &\propto& {1\over m_t} \Lambda_+ \left[
  (\bld{\epsilon}_1 \cdot \bld{\epsilon}_2)
  ({\bf n}_{e^-}\cdot \BLDgamma) {\bf n}_{e^-} +
  (\bld{\epsilon}_2\cdot \BLDgamma) \, \bld{\epsilon}_1 +
  (\bld{\epsilon}_1\cdot \BLDgamma) \, \bld{\epsilon}_2 \right]
  \Lambda_-,
\end{eqnarray}
with $\bld{\epsilon}_1$, $\bld{\epsilon}_2$
the polarization vectors of the
photons,
and the present formalism applies equally well. This case has
been studied in \cite{FKotsky}.
\par\noindent

\subsection{Cross sections and distributions\label{cross_sec}}
 
In the nonrelativistic limit (and defining $E_t= p^0_t-m_t$)
the two particle phase space for the production of a stable
$t$-  and $\bar t$-quark is conveniently written in the form
\begin{eqnarray}
\lefteqn{\int d{\rm PS} = \int{d^4 p_t\over (2\pi)^4} 
        \int{d^4 p_{\bar t}\over (2\pi)^4}
        (2\pi)^4 \delta^{(4)}(p_t+p_{\bar t}-q) \, \cdot} \nonumber\\
& &     \mbox{\hspace{3em}}\cdot \,{1\over 2 p_t^0} 
        2\pi \delta(p_t^0-\sqrt{m_t^2+\bfp_t^2}) \,
     {1\over 2 p_{\bar t}^0} 2\pi 
    \delta(p_{\bar t}^0-\sqrt{m_t^2+\bfp_{\bar t}^2}) =\\
  &=& {1\over 4 m_t^2}\int {dE_t\over 2\pi} {dE_{\bar t}\over 2\pi} 
        {d^3 p\over (2\pi)^3} 2\pi \delta (E_t-{\bfp^2\over 2m_t})
     2\pi \delta (E_{\bar t}-{\bfp^2\over 2m_t}) 2\pi\delta(E-E_t-E_{\bar t}) .
        \nonumber
\end{eqnarray}

The first two delta functions represent the on-shell
conditions for top and antitop quarks. In the case of
unstable particles they are replaced by Breit--Wigner
functions
\begin{equation}\label{br-wig}
2\pi\delta(E-{\bfp^2\over 2m_t}) \rightarrow
        {\Gamma_t\over (E-{\bfp^2\over 2m_t})^2 + (\Gamma_t/2)^2}\quad,
\end{equation}
see also discussion in Sect.~\ref{rescatt}.1~.
The $dE_{\bar t}$ integration is trivial. The $dE_t$
integral can be performed explicitly with the help of the
residue theorem. For this second step it is essential that
the matrix element is independent of $E_t$ for fixed $E$, an
assumption evidently fulfilled by the amplitude derived
above. The phase space integration is thus cast into the
following form
\begin{equation}
\int d{\rm PS} = {1\over 4m_t^2} \int {d^3p\over (2\pi)^3} 
        {2\Gamma_t\over (E-{\bfp^2\over m_t})^2 + \Gamma_t^2} .
\end{equation}

Therefore only the combinations
\begin{eqnarray} 
|G_0 {\cal K}_{\rm S}|^2 &=& |G|^2 \\
{\cal K}_{\rm S} G_0 \, {\cal K}_{\rm P}^* G_0^* &=& G F^*
\end{eqnarray} 
appear in the calculation of the $S$-wave dominated
differential cross sections and the $S$-$P$--wave
interference terms.

\subsection{Top production in electron positron annihilation
\label{eplus_subs}}
With these ingredients it is straightforward to calculate the
differential momentum distribution and the polarization of
top quarks produced in electron positron annihilation.
Let us introduce the following conventions for the fermion
couplings
\begin{equation}
v_f = 2 I^3_f - 4 q_f \sin^2\theta_{\rm W} , \qquad  a_f = 2 I^3_f .
\end{equation}
$P_\pm$ denotes the longitudinal electron/positron
polarization and
\begin{equation}\label{chi}
\chi={P_+-P_-\over1-P_+P_-}
\end{equation}
can be interpreted as effective longitudinal polarization of
the virtual intermediate photon or $Z$ boson.
The following abbreviations will be useful below:
\begin{eqnarray}
a_1 &=& q_e^2 q_t^2 + (v_e^2 + a_e^2) v_t^2 d^2 +
        2 q_e q_t v_e v_t d \nonumber \\
a_2 &=& 2 v_e a_e v_t^2 d^2 + 2 q_e q_t a_e v_t d \nonumber \\
a_3 &=& 4 v_e a_e v_t a_t d^2 + 2 q_e q_t a_e a_t d \label{coupl}\\
a_4 &=& 2 (v_e^2 + a_e^2) v_t a_t d^2 + 2 q_e q_t v_e a_t d \nonumber\\
d &=& {1\over 16 \sin^2\theta_{\rm W}\cos^2\theta_{\rm W}}\,{s\over s - M_Z^2}.
    \nonumber
\end{eqnarray}
The differential cross section, summed over polarizations of quarks
and including $S$-wave and $S$-$P$--interference contributions,
is thus given by
\begin{eqnarray}
{d^3\sigma \over dp^3} &=&
{3 \alpha^2 \Gamma_t \over 4 \pi m_t^4} (1-P_+P_-)
\left[ { (a_1 + \chi a_2)
        \left(1-{16 \alpha_{\rm s} \over 3 \pi} \right)
        \left|G(\rmp,E)\right|^2 + }\right. \nonumber\\
& & 
\left. {+(a_3+\chi a_4)
\left( 1-{12\alpha_{\rm s} \over 3 \pi} \right)
{\rmp \over m_t} \Re \left(\,G(\rmp,E) F^*(\rmp,E)\,\right)\,
\cos\vartheta} \right]  \label{dsig_d3p} .
\end{eqnarray}
The vertex corrections from hard gluon exchange for $S$-wave \cite{barbieri}
and $P$-wave \cite{KZ} amplitudes are included in this
formula. It leads to the following forward-backward asymmetry
\begin{equation}\label{afb}
{\cal A}_{\rm FB}(\rmp,E) = C_{\rm FB}(\chi)\, \varphi_{\rm _R}(\rmp,E),
\end{equation}
with
\begin{equation}
   C_{\rm FB}(\chi) = {1 \over 2}\, {a_3
    + \chi a_4 \over a_1 + \chi a_2} ,
\end{equation}
$\varphi_{\rm _R} = \Re\,\varphi$,  and
\begin{equation}\label{phi}
\varphi(\rmp,E) =
{(1-{4 \alpha_{\rm s}/3 \pi})\over (1-{8 \alpha_{\rm s}/3 \pi})}\,
        {\rmp \over m_t}\,
        {F^* \!(\rmp,E) \over G^* \!(\rmp,E)}  .
\end{equation}
This result is still differential in the top quark momentum.
Replacing $\varphi(\rmp,E)$ by
\begin{equation}\label{cap_phi}
\Phi(E) =
{(1-{4 \alpha_{\rm s}/3 \pi})\over (1-{8 \alpha_{\rm s}/3 \pi})}\,
{\int_0^{\rmp_m} d\rmp\,
{\rmp^3 \over m_t}\, F^*(\rmp,E)G(\rmp,E) \over
\int_0^{\rmp_m} d\rmp \, \rmp^2 \left|G(\rmp,E)\right|^2}  .
\end{equation}
one obtains the integrated forward-backward
asymmetry\footnote{For the case without beam polarization
this coincides with the earlier result \cite{MurSum2}, as far as the
Green function is concerned. It differs, however, in the
correction originating from hard gluon exchange.}.
The cutoff $\rmp_m$ must be introduced to eliminate the
logarithmic divergence of the integral. For free particles
(or sufficiently far above threshold) one finds for example
\begin{equation}\label{cut_dyn}
\Phi(E) = \sqrt{\frac{E}{m_t}} +
\frac{2 \Gamma_t}{\sqrt{m_t E}\pi}\ln {\rmp_m\over m_t} \, .
\end{equation}
This logarithmic divergence is a consequence of the fact
that the nonrelativistic approximation is used outside its
range of validity. One may either choose a cutoff of order
$m_t$ or replace the nonrelativistic phase space element
$\rmp\,d\rmp/m_t$ by $\rmp\,d\rmp/\sqrt{m_t^2 + \rmp^2}$. In practical
applications a cutoff will be introduced by the experimental
procedure used to define $t\bar t$-events.

\subsection{Polarization\label{pol_subs}}
\begin{figure}[ht]
\begin{flushleft}
\leavevmode
\epsfxsize=8cm
\epsffile[100 370 500 520]{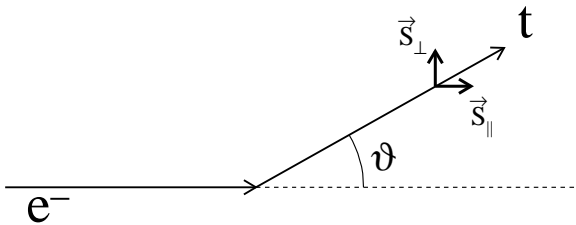}\\[-1.5cm]
\hfill
\parbox{6.cm}{\small
\caption[]{\label{dyn.ps}\sloppy Definition of the spin directions.
        The normal component ${\bf s}_{\rm N}$ points out of the plane.}}
\end{flushleft}
\end{figure}
To describe top quark production in the threshold region it
is convenient to align the reference system with the beam
direction (Fig.~\ref{dyn.ps}) and to define
\begin{eqnarray}
{\bf s}_{\|} &=& {\bf n}_{e^-} \nonumber\\
{\bf s}_{\rm N} &=& {{\bf n}_{e^-} \times {\bf n}_t \over
        |{\bf n}_{e^-} \times {\bf n}_t|}               \label{basis}\\
{\bf s}_\bot &=& {\bf s}_{\rm N} \times {\bf s}_{\|} .  \nonumber
\end{eqnarray}
In the limit of small $\beta$ the quark spin is essentially
aligned with the beam direction apart from small
corrections
proportional to $\beta$, which depend on the production
angle. A system of reference with ${\bf s}_\|$ defined with respect
to the top quark momentum \cite{krz}
is convenient in the high energy limit but evidently becomes
less convenient close to threshold.
\par\noindent
 The differential cross section for production of a top quark
of three-momentum $\bfp$ and spin projection $+1/2$ on direction ${\bf s}_+$ 
reads\footnote{In the non-relativistic approximation ${\bf s}_+$
is the same in the top quark rest frame and in the $t\bar t$ center-of-mass
frame up to corrections of order $\beta^2$ which are neglected.}
\begin{equation}
{d^3\sigma(\bfp,{\bf s}_+) \over dp^3} \;= \;
{1\over 2} \, \left( 1 +\mbox{\boldmath$\cal P$}\cdot{\bf s}_+ \right)\,
{d^3\sigma \over dp^3}\;  .
\end{equation}
The three-vector {\boldmath$\cal P$} characterizes the polarization of the
top quark.  Its components ${\cal P}^i$ can be obtained from the
formula
\begin{equation}
{\cal P}^i \; =\;
\left[\, {d^3\sigma(\bfp,{\bf s}_+^{(i)})\over dp^3} 
- {d^3\sigma(\bfp,{\bf s}_+^{(i)})\over dp^3}\, \right] 
\; \left/ \;
\left[\, {d^3\sigma(\bfp,{\bf s}_+^{(i)})\over dp^3} 
+ {d^3\sigma(\bfp,{\bf s}_+^{(i)})\over dp^3}\, \right]  \right.  
\end{equation}
where ${\bf s}_+^{(i)}\; =\; {\bf s}_{_{\|}},\; {\bf s}_\bot$ and
${\bf s}_{\rm N}$.
Including the QCD potential one obtains for the three components
of the polarization
\begin{eqnarray}
{\cal P}_\|(\bfp,E,\chi) &=& C_\|^0(\chi)
+ C_\|^1(\chi)\, \varphi_{\rm _R}(\rmp,E)\,\cos\vartheta\,
 \label{thr_long}\\
{\cal P}_\bot(\bfp,E,\chi) &=& C_\bot(\chi)\,
\varphi_{\rm _R}(\rmp,E)\,
\sin\vartheta\,
\label{thr_perp}\\
{\cal P}_{\rm N}(\bfp,E,\chi) &=& C_{\rm N}(\chi)
\varphi_{\rm _I}(\rmp,E)
\sin\vartheta\,
        \label{thr_norm} ,
\end{eqnarray}\\
\parbox{75.ex}{
\begin{eqnarray*}
& &\hspace{5.ex}C_\|^0 (\chi) =
-{a_2 + \chi a_1 \over a_1 + \chi a_2} ,\hspace{6.9ex}
  C_\|^1 (\chi) = \left( 1-\chi^2 \right) {a_2 a_3 - a_1 a_4   \over
        \left(a_1 + \chi a_2 \right)^2} ,\\
& &\hspace{5.ex}C_\bot(\chi)  = -{1\over 2} \,
{a_4 + \chi a_3 \over a_1 + \chi a_2} ,
    \qquad C_{\rm N}(\chi) =-{1 \over 2}\, {a_3
    + \chi a_4 \over a_1 + \chi a_2}\, =\, - C_{\rm FB}(\chi) ,
\end{eqnarray*}}
\hfill
\parbox{5.ex}{
\begin{eqnarray} \label{coefs} \end{eqnarray} }
with $\varphi_{\rm _I} = \Im\,\varphi$, and $\varphi(\rmp,E)$ as defined
in (\ref{phi}).  The momentum integrated quantities are obtained by
the replacement $\varphi(\rmp,E) \to \Phi(E)$. The case of
non-interacting stable quarks is recovered by the replacement
$\Phi\to\beta$, an obvious consequence of (\ref{cap_phi}).
\par
\begin{figure}[ht]
 \begin{center}
  \leavevmode
  \epsfxsize=14cm
  \epsffile[40 290 535 525]{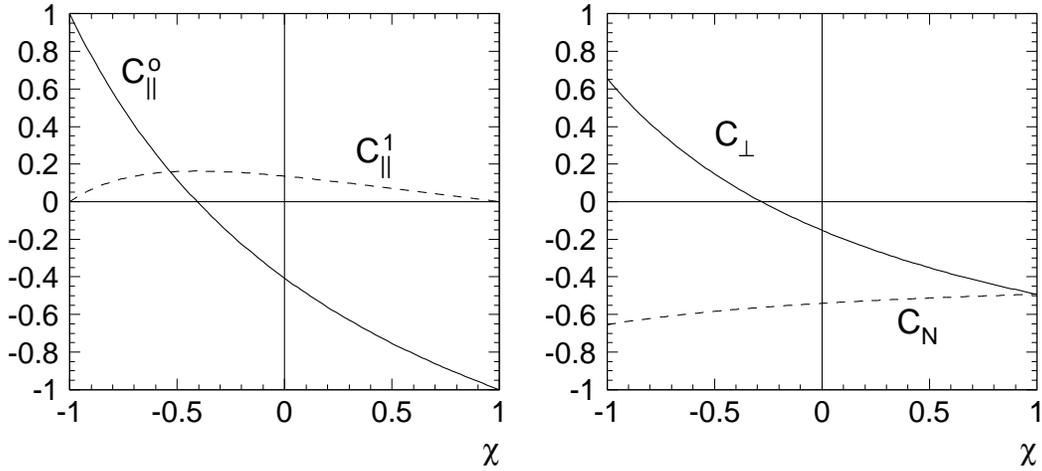}
  \caption[]{\label{pol_coefs.ps}\sloppy
        The coefficients (\ref{coefs}) for $\sqrt{s}/2 = m_t=180$ GeV.}
 \end{center}
\end{figure}

\noindent
Let us emphasize the main qualitative features of the result:
\begin{itemize}
\item Top quarks in the threshold region are highly polarized.  Even
  for unpolarized beams the longitudinal polarization amounts to about
  $-0.41$ and reaches $\pm1$ for fully polarized electron beams. This
  later feature is of purely kinematical origin and independent of the
  structure of top quark couplings.  Precision studies of polarized
  top decays are therefore feasible.
\item Corrections to this idealized picture arise from the small
  admixture of $P$-waves. The transverse and the normal components of
  the polarization are of order 10\%. The angular dependent part of
  the parallel polarization is even more suppressed.  Moreover, as a
  consequence of the angular dependence its contribution vanishes upon
  angular integration.
\item The QCD dynamics is solely contained in the functions $\varphi$
  or $\Phi$ which is the same for the angular distribution and the
  various components of the polarization. (However, this
  ``universality'' is affected by the rescattering corrections as
  discussed in Sect.~\ref{rescatt}.)  These functions which evidently
  depend on QCD dynamics can thus be studied in a variety of ways.
\item The relative importance of $P$-waves increases with energy,
  $\Phi\sim\sqrt{E/m_t}$.  This is expected from the close analogy
  between $\Phi_{\rm R}=\Re\,\Phi$ and $\beta$. 
  In fact, the order of magnitude of
  the various components of the polarization above, but close to
  threshold, can be estimated by replacing $\Phi_{\rm R}\to \rmp/m_t$.
\end{itemize}
The $C_i$ are displayed in Fig.~\ref{pol_coefs.ps} as functions of the
variable $\chi$.  For the weak mixing angle a value
$\sin^2\!\theta_{\rm W}= 0.2317$ is adopted, for the top mass $m_t=180$ GeV.
As discussed before, $C_\|^0$ assumes its maximal value $\pm 1$ for
$\chi=\mp 1$ and the coefficient $C_\|^1$ is small throughout. The
coefficient $C_\bot$ varies between $+0.7$ and $-0.5$ whereas $C_{\rm
  N}$ is typically around $-0.5$.  The dynamical factors $\Phi$ are
around $0.1$ or larger, such that the $P$-wave induced effects should
be observable experimentally.
\par\noindent
The normal component of the polarization which is proportional to
$\varphi_{\rm _I}$ has been predicted for stable quarks in the framework of
perturbative QCD \cite{dev,krz}. In the threshold region the phase can be
traced to the $t\bar t$ rescattering by the QCD potential. For 
stable quarks, assuming a pure Coulomb potential $V=-4\alpha_{\rm s}/3r$, 
the nonrelativistic problem can be solved analytically \cite{FKotsky} and
one finds
\begin{eqnarray}
\lim_{\Gamma_t \rightarrow 0 \atop E\rightarrow \bfp^2/m_t} 
        \left(E - {\bfp^2\over m_t} + i\Gamma_t \right) G(\rmp,E) &=&
        \exp\left(\pi {\bar k} \over 2 \rmp\right)\,
        \Gamma(1+{i{\bar k}/\rmp}) \label{lim1_gt0}\\
\lim_{\Gamma_t \rightarrow 0 \atop E\rightarrow \bfp^2/m_t} 
        \left(E - {\bfp^2\over m_t} + i\Gamma_t \right) F(\rmp,E) &=&
        \left(1 - i{{\bar k} \over \rmp}\right) 
        \exp\left(\pi {\bar k} \over 2 \rmp\right)\,
        \Gamma(1+{i{\bar k}/\rmp}) \label{lim2_gt0},\ \ \ \ \
\end{eqnarray}
with ${\bar k} = 2m_t\alpha_s/3$
and hence
\begin{eqnarray}
\varphi_{\rm _I}(\rmp,E) &\rightarrow&
{2\over 3}\alpha_{\rm s}{1-4\alpha_{\rm s}/3\pi\over
        1-8\alpha_{\rm s}/3\pi} \label{phi_gt0}\\
\Phi_{\rm I}(\rmp,E) &\rightarrow&
{2\over 3}\alpha_{\rm s}{1-4\alpha_{\rm s}/3\pi\over
        1-8\alpha_{\rm s}/3\pi} . \label{cap_phi_gt0}
\end{eqnarray}
The component of the polarization normal to
the production plane is thus
approximately independent of $E$ and essentially measures the strong
coupling constant. In fact one can argue that this is a unique way to
get a handle on the scattering of heavy quarks through the QCD
potential.

\section{\label{rescatt}Rescattering corrections to the differential
         cross section}

\subsection{Introductory remarks}

Close to threshold, the velocity $\beta$ of the top quark is of the
same order of magnitude as the strong coupling. To include only the
${\cal O}(\beta)$ contributions to the differential cross section and
polarization and discard rescattering corrections is thus
inconsistent. To obtain reliable predictions for the observables it is
mandatory to calculate the ${\cal O}(\alpha_{\rm s})$ corrections.
Some of these can be implemented without much effort as they are
known since several years, and in fact they have already been taken
into account in the preceding formul\ae:
\begin{itemize}
\item 
  the modification of the $t\bar tV$ and $t\bar tA$ vertices due
  to transverse gluon exchange for small $\beta$ simply reduces to a
  multiplication by constant form factors \cite{JLZ82}:
  \begin{eqnarray} 
    G(\rmp,E) & \longrightarrow & \Big(1-\frac{8\alpha_{\rm s}}{3\pi}\Big)
               G(\rmp,E) \\
    F(\rmp,E) & \longrightarrow & \Big(1-\frac{4\alpha_{\rm s}}{3\pi}\Big)
               F(\rmp,E).
  \end{eqnarray}
  This explains the corresponding factors in (\ref{dsig_d3p}) and in
  the definition of $\varphi({\bf p},E)$ and $\Phi(E)$ (\ref{phi}).

\item 
  Corrections to $t$ or $\bar t$ decay can be taken into account
  by using the full 1-loop result for $\Gamma_t$ \cite{JK89a}.
\end{itemize}

\begin{figure}[h]\begin{center}
\begin{tabular}{cc}
  \epsfxsize 63mm \mbox{\epsffile[25 570 200 650]{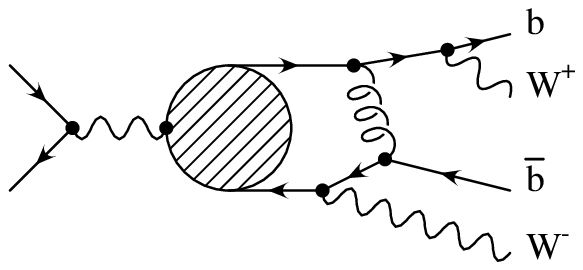}} &
  \epsfxsize 63mm \mbox{\epsffile[25 570 200 650]{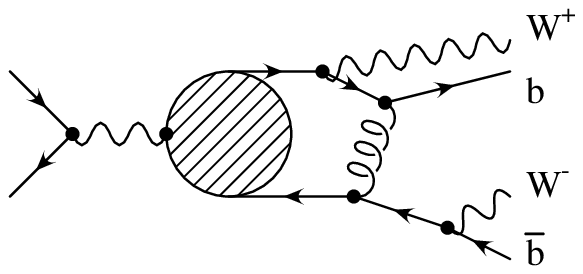}} \\
  a) & b)
\end{tabular}\end{center}
\caption{\label{tbbar}Lowest order rescattering diagrams.}
\end{figure}

If the total cross section were the only quantity of interest, all
next-to-leading contributions would be included this way, because it is well
known since quite some time that the excitation curve is free of (other)
${\cal O}(\alpha_{\rm s})$ corrections \cite{Ube60}. This result has also
been re-established recently \cite{MY94,SThes}.
But this statement does not hold
for the differential cross section or polarization, since these are
affected through the diagrams shown in Figs. \ref{tbbar}a and \ref{tbbar}b
(and a third diagram with gluon exchange between the two $b$-quarks, which
can, however, be neglected), i.e.\ corrections caused by interactions
between the decay products of one of the top quarks with the second top.
They will also be called rescattering corrections.

Their calculation obviously requires a modification of the approach as
at least the decay of one of the two top (anti)quarks has to be
included. In this section a formalism will be pursued which attempts to
retain the notion of quark polarization in the presence of
rescattering of decay products. In Sect.~\ref{moments} a slightly
different strategy will be adopted which aims directly at predictions
for experimentally observable lepton distributions, viz.\ 
appropriately chosen moments. Let us start with the first approach and
write the differential cross section in the following
form\footnote{with the 3-axis identical to the electron beam axis} :
\begin{eqnarray}\label{dsig}
  \frac{d^3\sigma(\bf p,s_+)}{dp^3} &=&
        \frac{3\alpha^2}{64m_t^6}\frac{
        1-P_{e^+}P_{e^-}}{4\pi}\sum_{i,j=1,2}(a_1+a_2\chi)\Big(L_s^{ij}
        +C_\parallel^0L_a^{ij}\Big)H^{VV}_{ij}\nonumber \\ &&
        +{\cal O}(\beta)\mbox{ terms}.
\end{eqnarray}
The three-vector ${\bf s_+}$ describes the spin of top quark, which in this subsection
is considered as a well defined quantity. The leptonic
tensors $L^{ij}$ are defined through
\begin{equation}\label{eq:Lij}
 L_s^{ij}=8m_t^2\delta^{ij}~,~L_a^{ij}=-8m_t^2i\epsilon^{ij}~,~i,j=1..2~.
\end{equation}
As already discussed in the preceding section the terms of order $\beta$
arise from $S$-$P$--wave interference, i.e.\ from interference between the
amplitudes corresponding to vector and axialvector couplings $t\bar t\gamma$
and $t\bar t Z^0$. In the approximation of the present article contributions
suppressed by $\alpha_s\beta$ are neglected. Hence it suffices to consider
order $\alpha_s$ corrections to contributions involving only vector couplings
and the corresponding vertex functions ${\cal K}_{\rm S}(\rmp,E)$.
$H^{VV}_{ij}$ represents the leading (vector-vector) part of the
hadronic tensor
\begin{equation}\label{hij}
   H^{VV}_{ij}=\int\!\frac{dp_0}{2\pi}\int\!d\mbox{PS}_2(t;bW^+)\int\!d
        \mbox{PS}_2(\bar t;\bar bW^-)\,
        \sum_{\epsilon_\pm}\,{\cal M}_i{\cal M}_j^\dagger~+~\mbox{c.c.}
\end{equation}
where ${\cal M}_i$ denotes the amplitude for $V \to t\bar t\to b\bar
bW^+W^-$. The momenta of top and antitop are given by $t=b+W^+=q/2+p$,
$\bar t=\bar b +W^-=q/2-p$ and ${\epsilon_\pm}$ denote the polarization
vectors of $W^\pm$.

This approach can be adopted in a straightforward way as long as only
factorizable diagrams are considered
where the top quark decay products do not rescatter.  Diagram
Fig.~\ref{tbbar}b, however, as a truly non-factorizable contribution
leads to difficulties when discussing the top quark polarization as will become
clear below.
Nevertheless this formalism will be presented for pedagogical reasons before
a more powerful approach will be introduced in Sect.~\ref{moments}.

To lowest order in the nonrelativistic approximation
the amplitude ${\cal M}_i$ is given by
\begin{eqnarray} 
  {\cal M}^{(0)}_i({\bf s}_+)
        &=&-\Big(\frac{ig_w}{\sqrt8}\Big)^2\bar u(b)\not\!\epsilon_+
        (1-\gamma^5)\frac{1+\gamma_5\!\not\!s_+}{2}\Lambda_+\gamma_i\Lambda_-
        \not\!\epsilon_-(1-\gamma^5)v(\bar b) \nonumber \\ &&
        \cdot G(\rmp,E)\Big(G_t(p,E)+G_{\bar t}(p,E)\Big)
\label{eq:Mi0}
\end{eqnarray}
with the nonrelativistic propagators introduced in (4)
(see also App.~\ref{RApp}),
$$ S_{\rm F}(p\pm\frac{q}{2}) = 
    i\Big(\Lambda_\pm-\frac{\bf p\cdot\bld\gamma}{2m}\Big)
   \cdot G_{t/\bar t}(p,E), $$
and where
\begin{eqnarray}
 -S_{\rm F}(p+q/2)\widetilde\Gamma^i_V S_{\rm F}(p-q/2) &\approx& 
    \Lambda_+\gamma^i\Lambda_-
     {\cal K}_{\rm S}(\rmp,E)G_t(p,E)G_{\bar t}(p,E) \nonumber \\
 &=& \Lambda_+\gamma^i\Lambda_- G(\rmp,E)(G_t(p,E)+G_{\bar t}(p,E))
     \label{vred}
\end{eqnarray}
has been used.  The amplitude ${\cal M}^{(0)}_i(s_+)$ depends on $s_+$
which for a stable top quark would be interpreted as its polarization
four-vector. For unstable top quark this interpretation is still valid
if the decay products do not interact. In such a case the total
angular momentum of the $W^+b$ subsystem is conserved.  Thus, in
principle at least, the angular momentum state $|{1\over 2},m\rangle$
of $W^+b$ can be inferred from the measurement of the spin-dependent
angular distributions of $W$ and $b$.  The quantum number $m$ denotes
an eigenvalue of the angular momentum projection on direction $\bf
s_+$. It is evident that one can identify $m$ with the projection of
top quark spin if there is no interaction changing momenta and
polarizations of $W$ and $b$ after the decay.

Inserting the expression for ${\cal M}$ into (\ref{dsig}) and (\ref{hij}),
summing over polarizations $\epsilon_+$ and $\epsilon_-$,
and performing the $(bW^+)$ and $(\bar bW^-)$ phase space integrations with the
help of
\begin{equation}\label{psp}
  \frac{g_w^2}{4}\int\!d\mbox{PS}_2(t;bW^+)\sum_{\epsilon_+}\not\!\epsilon_+
   \not\!b\not\!\epsilon_+(1-\gamma^5)=\frac{\Gamma_t}{m_t}\not\!t(1-\gamma^5)
\end{equation}
and the corresponding identity for the $(\bar bW^-)$ system, one re-derives the
results given in the preceding section. The inclusion of the axialvector
couplings and calculation of the $S$-$P$--wave interference effects can be done
in an analogous way.  

\subsection{Rescattering in the $t\bar b$ system}

The rescattering in the $t\bar b$ system is given by
diagram \ref{tbbar}a and leads to an additional amplitude of the
form
\begin{eqnarray}
  {\cal M}^{a}_i &=& \Big(\frac{ig_w}{\sqrt8}\Big)^2\int\!\frac{d^4k}{
     (2\pi)^4}\bar u(b)\not\!\epsilon_+(1\!-\gamma^5)i\Lambda_+\gamma^\mu
     \Lambda_+\gamma_i\Lambda_-\!\not\!\epsilon_-(1-\gamma^5)
     S_{\rm F}(k\!-\!p\!-\!\bar b) \nonumber \\ && \gamma^\nu v(\bar b)\cdot 
     G_t(p,E)G(\rmk,E)\Big(G_t(k,E)+G_{\bar t}(k,E)\Big)D_{\mu\nu}(k-p)
     g_s^2C_F \quad .
\end{eqnarray}
In this amplitude all
the top quark propagators 
have  been substituted
by their nonrelativistic limits
\[ S_{\rm F}(t) \longrightarrow i\Lambda_+ G_t(p,E). \]

As a consequence the top quark couples to gluons
via the projection $\Lambda_+\gamma^\mu\Lambda_+$, which implies
that only instantenous Coulomb-like gluon exchange contributes to 
leading order. In the calculation of the ${\cal O}(\alpha_{\rm s})$ 
corrections to the differential cross sections
$$ \sim\; \int\!\frac{dp_0}{2\pi}\int\!
   d\mbox{PS}_2(t;bW^+)\int\!d\mbox{PS}_2(\bar t;\bar b W^-)\,
   \sum_{\epsilon_+,\epsilon_-} \,
   \Big( {\cal M}^{(0)}_i(s_+){\cal M}^{(a)\dagger}_j+
   {\cal M}^{(a)}_i{\cal M}^{(0)\dagger}_j(s_+)\Big)
$$
only the interference of ${\cal M}^{a}_i$
with the lowest order amplitude ${\cal M}^{(0)}_i(s_+)$
contributes. A more detailed description of the calculation 
of this contribution is presented in App.~\ref{RApp}.
At this point it should suffice to mention that it essentially involves
choosing the appropriate integration contours and using the residue
theorem. The result can be written in the following compact form:
\begin{equation}
  \frac{d\sigma^{(a)}}{d\rmp d\Omega}=
  \frac{1}{4\pi}\frac{d\sigma^{(0)}}{d\rmp}
    \frac{1}{4}\Big(\psi_1({\rmp},E)+C^0_\parallel\,\kappa\,
    \psi_{\rm _R}(\rmp,E)
    \cos\vartheta+{\bf P\cdot s}_+\Big),
\end{equation}
where 
$d\sigma^{(0)}/d\rmp$ denotes the top quark momentum distribution
in the lowest order approximation,
\begin{eqnarray} 
  P_\parallel & = &C_\parallel^0\,\psi_1(\rmp,E)
      + \kappa\, \psi_{\rm _R}(\rmp,E)\cos\vartheta~, \\
    P_\perp & = & P_{\rm N}=0~, \\
  \kappa & = & \frac{1-2y}{1+2y} \qquad \mbox{and} \qquad y = \frac{m_W^2}{m_t^2}.
  \label{eq:kappay}
\end{eqnarray}
The two functions $\psi_1$ and $\psi_{\rm _R}$ are defined as the following
convolutions:
\begin{eqnarray}
   \psi_1(\rmp,E) &=& 2\,\Im\int\!\frac{d^3k}{(2\pi)^3}V(|{\bf k}-{\bf p}|)
        \frac{G(\rmk,E)}{G(\rmp,E)}\frac{\arctan{\frac{|{\bf k}-{\bf p}|}
        {\Gamma_t}}}{|{\bf k}-{\bf p}|} \\
   \psi_2(\rmp,E) &=& 2\int\!\frac{d^3k}{(2\pi)^3}V(|{\bf
        k}-{\bf p}|)\frac{{\bf p}\cdot({\bf k-p})}{|{\bf p}||{\bf k- 
        p}|^2}\frac{G(\rmk,E)}{G(\rmp,E)}\nonumber \\ && \times
        \Big(1-\frac{\Gamma_t}{|{\bf k}-{\bf p}|}\arctan{\frac{|{\bf k}-{\bf
        p}|}{\Gamma_t}}\Big)\\
   \psi_{\rm _R}(\rmp,E) &=& \Re\;\psi_2(\rmp,E).
\end{eqnarray}

It is noteworthy that rescattering in the $t\bar b$ system does not
affect the transverse and normal components of the top quark polarization.
The top quark width $\Gamma_t$ acts as an infrared
cutoff: in the limit $\Gamma_t\to0$ the real part of $\psi_2$ becomes
infrared divergent.

\subsection{Rescattering in the $(Wb)\bar t$ system}

The corrections due to rescattering of the top quark decay products on the
spectator $\bar t$ can be calculated in an analogous way. They result from
the interference of diagram \ref{tbbar}b and the lowest order amplitude
${\cal M}^{(0)}_i(s_+)$. However, for diagram \ref{tbbar}b 
it is not possible to identify the total angular momentum $j$ of the $W^+b$ system
and the spin of the top quark. In fact the latter quantity is not well defined
and $j$ can be different from $1/2$. However, only for 
$j=1/2$ and $m=\pm 1/2$ there are non-zero contributions to order $\alpha_s$
corrections 
$$
   \sim\; \int\!\frac{dp_0}{2\pi}\int\!
   d\mbox{PS}_2(t;bW^+)\int\!d\mbox{PS}_2(\bar t;\bar b W^-)\,
   \sum_{\epsilon_+,\epsilon_-} \,
   \Big( {\cal M}^{(0)}_i(s_+)
   {\cal M}^{(b)\dagger}_j+
   {\cal M}^{(b)}_i{\cal M}^{(0)\dagger}_j(s_+)\Big) .
$$
Summing over $W^\pm$ polarizations and calculating the phase space integrals one obtains
\begin{equation}
  \frac{d\sigma^{(b)}}{d\rmp d\Omega}=
  \frac{1}{4\pi}\frac{d\sigma^{(0)}}{d\rmp}
    \frac{1}{4}\Big(\psi_1(\rmp,E)+C^0_\parallel\,\kappa\,
    \psi_{\rm _R}(\rmp,E)\cos\vartheta+{\bf P\cdot s}_+\Big)
\end{equation}
with
\begin{eqnarray}
  P_\parallel & = & C_\parallel^0\,\psi_1(\rmp,E)
        +\kappa\,\psi_{\rm _R}(\rmp,E)\cos\vartheta \\
  P_\perp & = & \kappa\,\psi_{\rm _R}(\rmp,E)\sin\vartheta \\
  P_{\rm N} & = & C^0_\parallel\,\kappa\,\psi_{\rm _I}(\rmp,E)\sin
   \vartheta
\end{eqnarray}
where the additional function 
\begin{equation}
\psi_{\rm _I}(\rmp,E)=\Im\,\psi_2(\rmp,E)
\end{equation}
appears.

At first glance this result may be surprising.  One might have thought
that any interaction after top decay cannot influence top
polarization.  There is however a problem with such a simple picture,
because the spin of an intermediate (virtual) particle is not a well
defined quantity and thus no observable. What has been calculated by
using the spin projection operator should be interpreted as the $(bW)$
total angular momentum, which coincides with top spin on Born level
only. Clearly the angular momentum of $Wb$ may well be influenced by
final state interactions. From a practical point of view it will be
quite difficult if not impossible to measure this quantity.

It definitely makes more sense to consider the angular distribution of
the charged leptons from the decay $t\to W^+b\to l^+\nu_lb$ instead.
Rescattering in the $(Wb)\bar t$ system produces a contribution which is 
non-factorizable, i.e.\ it is
impossible to write the final expression in the form
\begin{equation}\label{dsdpdedo}
   \frac{d^6\sigma(e^+e^-\to l^+\nu_lbW^-\bar b)}{dp^3dE_ld\Omega_l} =
   \frac{d^3\sigma(e^+e^-\to t\bar t)}{dp^3}\frac{1}{\Gamma_t}
   \frac{d\Gamma_{t\to bl\nu}({\bf s}_+)}{dE_ld\Omega_l}
\end{equation}
which is possible for all the other diagrams.

\subsection{Momentum distribution and polarizations}

Adding the contributions discussed in the preceding subsections
one can calculate the order $\alpha_s$ correction to  
the differential cross section summed over spins.
The momentum distribution is affected by $\psi_1$ only:
\begin{equation}\label{dsig_lept}
  \frac{d\sigma^{(0)}}{d\rmp} \to \frac{d\sigma^{(1)}}{d\rm p}=
       \frac{d\sigma^{(0)}}{d\rm p}\cdot\Big(1+
       \psi_1(\rmp,E) \Big).
\end{equation}
It follows that at order $\alpha_s$
the total cross section for $t\bar t$ production is unaffected 
by rescattering corrections\cite{MY94,SThes}. Indeed
\begin{equation}
\int\!\frac{d^3p}{(2\pi)^3}|G(\rmp,E)|^2\psi_1(\rmp,E)=0
\end{equation}
because for any real function $f(|{\bf k}-{\bf p}|)$ the integral
\[
  J = \int\!\frac{d^3p}{(2\pi)^3}\int\!\frac{d^3k}{(2\pi)^3}G({\rmp},E)
      G^*({\rmk},E)f(|{\bf k}-{\bf p}|) = J^*
\]
is real.

The function
$\psi_2({\rmp},E)$ gives the order $\alpha_s$ corrections
to the forward-backward
asymmetry ${\cal A}_{\rm FB}$ and the polarization vector {\boldmath$\cal P$}
of the $bW$ system. It is interesting to 
note similarities between these formul\ae\ and the ${\cal O}(\beta)$
expressions: 
\begin{eqnarray}
  {\cal A}_{\rm FB}^{^{\rm resc}}(\rmp,E) & = & \frac{C_\parallel^0}{2}
        \kappa\,\psi_{\rm _R}(\rmp,E)\\
  {\cal P}_\parallel^{^{\rm resc}}(\rmp,E) & = & \Big(1-
        (C_\parallel^0)^2\Big)\kappa\,\psi_{\rm _R}(\rmp,E)\cos\vartheta \\
  {\cal P}_\perp^{^{\rm resc}}(\rmp,E)     & = & \frac{1}{2}
        \kappa\,\psi_{\rm _R}(\rmp,E)\sin\vartheta \\
  {\cal P}_{\rm N}^{^{\rm resc}}(\rmp,E)    & = & \frac{C_\parallel^0}{2}
        \kappa\,\psi_{\rm _I}(\rmp,E)\sin\vartheta.
\end{eqnarray}

Let us close this section with two remarks.
The results for the unpolarized case have been first derived
in \cite{SThes} and are in accordance with ours. 
Following \cite{SThes,FMY94} one can consider also
the diagram with gluon exchange between $b$ and $\bar b$. 
However, it has already been argued in the Introduction that this 
contribution is suppressed. Our conclusion is supported 
by the results of \cite{SThes,FMY94} for unpolarized beams.

\section{\label{moments}Moments of the lepton spectrum}

It has already been stated that the optimal way to experimentally
determine top quark polarization is based on the analysis
of the charged lepton distribution in the process
\[ e^+e^- \to t\bar t\to bl\nu\bar bW^-. \]
A particularly convenient distribution is the charged lepton angular 
distribution, which for free top quark decay  in its rest frame is 
given by the formula
\begin{equation}
  \frac{d\Gamma}{d\Omega_l} = \frac{\Gamma}{4\pi}\Big(1+|{\mbox{\boldmath$\cal
   P$}}|\cos\theta\Big)
\end{equation}
where $\theta$ denotes the angle between the lepton momentum and top
polarization {\boldmath$\cal P$}.  It is quite straightforward to
generalize this formula to the case of factorizable contributions to
$t\bar t$ production, c.f.\ (\ref{dsdpdedo}). In
fact this problem can be solved by using the Lorentz transformation
from the top quark rest frame to the center-of-mass frame for the
$t\bar t$ system. However the diagram of Fig.\ref{tbbar}b leads to a
non-factorizable contribution with production and decay mechanisms
coupled in an intricate way. The corresponding expression for the
angular distribution of the charged lepton involves a complicated
integral over the phase space and the three-momentum transfer between
$b$ and $\bar t$. In particular integration over the phase space of
the top quark is difficult because the direction of the charged lepton
in $t\to bW \to bl\nu$ is fixed and one cannot use Lorentz symmetry to
perform integrations. We were unable to simplify this expression. It
is more convenient to consider moments of the charged lepton
distribution which are given by manifestly Lorentz invariant integrals
over the three-body Lorentz invariant phase space ${\rm
  PS}_3(t;bl\nu)$. Moreover, these moments are measurable and contain
important information on polarization.  From the calculational point
of view Lorentz symmetry simplifies the problem considerably and
closed expressions can be derived even for non-factorizable
contributions.

Let $l$ be the four-momentum of the charged lepton and $n$ denote one
of the following four-vectors, which in the $t\bar t$ rest frame read,
c.f. (\ref{basis}):
\begin{eqnarray}
n_{\|} &=& \left( 0, {\bf s}_{\|}\,\right)  \\
n_\bot &=& \left( 0, {\bf s}_\bot\,\right)  \\
n_{\rm N} &=& \left( 0, {\bf s}_{\rm N}\right)  \\
n_{(0)} &=&  (1,0,0,0) 
\end{eqnarray}
The moments are defined as the average values of the scalar products $(nl)$.
One can consider the moments $\langle nl\rangle$ for a fixed three-momentum
of the top quark,
\begin{equation}
  \langle nl\rangle \equiv
       \Big(\frac{d^3\sigma}{d\rmp d\Omega_p}\Big)^{-1}\int dE_ld\Omega_l
       \frac{d\sigma(e^+e^-\to bl\nu\bar bW^-)}{d\rmp d\Omega_pd\Omega_ldE_l}
       (nl)
\end{equation}
or, integrating over the direction of the top quark, the moments
\begin{equation}
  \langle\langle nl\rangle\rangle \equiv 
       \Big(\frac{d\sigma}{d\rmp}\Big)^{-1}\int dE_ld\Omega_ld\Omega_p
       \frac{d\sigma(e^+e^-\to bl\nu\bar bW^-)}{d\rmp d\Omega_pd\Omega_ldE_l}
       (nl)
\end{equation}
or, finally the moments $\langle\langle\langle nl\rangle\rangle\rangle$
integrated also over $\rmp$. 
The calculation of these quantities is straightforward because the integral
over the three-body phase space $\mbox{PS}_3(t;bl\nu)$
can be calculated analytically.
The hadronic tensor for the differential cross section is given by
\begin{equation}
   H_{ij}=\int\!\frac{dp_0}{2\pi}\int\frac{dW^2}{2\pi}\int d\mbox{PS}_3(t;bl\nu)        
      \int d\mbox{PS}_2(\bar t;\bar b W^-) \, \sum_{\epsilon_-}
      {\cal M}_i{\cal M}_j^\dagger
\end{equation}
with the appropriate amplitudes including 
semileptonic $W^+$ decays. The corresponding
tensor needed for the determination of the moments will be called $H_{ij;n}$
and simply contains the additional factor
$(nl)$ under the integrals
\begin{equation}
   H_{ij;n}=\int\!\frac{dp_0}{2\pi}\int\frac{dW^2}{2\pi}\int d\mbox{PS}_3(t;bl\nu)        
      \int d\mbox{PS}_2(\bar t;\bar b W^-) \,\sum_{\epsilon_-}
      {\cal M}_i{\cal M}_j^\dagger\, (nl)
\end{equation}
Using the following decomposition of the three-body phase space
\begin{equation} 
\int d\mbox{PS}_3(t;bl\nu)=\frac{1}{2\pi}\int dW^2\int d\mbox{PS}_2
(t;bW)\int d\mbox{PS}_2(W;l\nu)
\end{equation} 
one can reduce the calculation to a sequence of Lorentz invariant integrations
over two-body phase spaces. 
A more detailed description  of the calculation
can be found in App.~\ref{app:moments}.

Let us, for illustrative purpose, discuss the Born- and $S$-$P$--wave
interference part. In this case one arrives at
\begin{equation}
 \langle nl\rangle = \mbox{BR}(t\to bl\nu) \frac{1+2y+3y^2}{4(1+2y)}
    \Big[ (tn)+\frac{m_t}{3}(n{\cal P}) \Big]
\end{equation}
with $y$ (and $\kappa$) as defined in (\ref{eq:kappay}).
In the $t\bar t$ center-of-mass frame  the space components of
${\cal P}$ are given by (\ref{thr_long})--(\ref{thr_norm}) and
\begin{eqnarray}
  t^\mu & = & (m_t, {\bf p}) \\
  {\cal P}_0   &=& \frac{1}{m_t}{\mbox{\boldmath$\cal P$}\cdot\bf p}.
\end{eqnarray}
This clearly demonstrates that for the factorizable contributions
the moments contain information equivalent to those obtained by
using the approach of Sect.~2 and Sect.~3.
Integrating over the emission angle of the top quark one derives the
following expression for the angular independent part of the longitudinal
polarization
\begin{equation}
  {\cal P}_\parallel^{(0)} \equiv C_\parallel^0 = -3 {\langle\langle n_{\|}l
     \rangle\rangle}\left/ {\langle\langle n_{(0)}l \rangle\rangle} \right. . 
\label{slong}
\end{equation}
Similarly the coefficient characterising normal polarization can be
obtained from (cf.(\ref{thr_norm}))
\begin{equation}
  C_{\rm N}\, \Im\;\varphi(\rmp,E) =  -\frac{12}{\pi}
     \langle\langle n_{\rm N}l\rangle\rangle \left/
        \langle\langle n_{(0)}l \rangle\rangle \right.  . 
\label{snorm}
\end{equation}

Let us now return to the case with rescattering included. It requires
the introduction of a third convolution function that only enters
through diagram \ref{tbbar}b:
\begin{eqnarray}
  \psi_3(\rmp,E) &=& \Im\int\!\!\frac{d^3k}{(2\pi)^3}V(|{\bf k}-{\bf p}|)
   \frac{G(\rmk,E)}{G(\rmp,E)}\frac{\arctan\frac{|{\bf k-p}|}{\Gamma_t}}
    {|{\bf k}-{\bf p}|}\Bigg[3\Big(\frac{{\bf p\cdot(k-p)}}{|{\bf p}||{\bf k}-
    {\bf p}|}\Big)^2-1\Bigg].
\end{eqnarray}
The complete result for the moments, correct up to (and including)
${\cal O}(\beta,\alpha_{\rm s})$ reads
\begin{eqnarray}
 \langle nl\rangle &=& \mbox{BR}(t\to bl\nu) \frac{1+2y+3y^2}{4(1+2y)}\Bigg[
      \nonumber \\
      && (tn)\Big( 1+ \frac{y(1-y)}{(1+2y)(1+2y+3y^2)}C_\parallel^0
         \psi_{\rm _R}(\rmp,E)\cos\vartheta\Big) \nonumber \\
      && +\frac{m_t}{3} n{\cal P}+\frac{m_t}{3}n\delta{\cal P} \Bigg]
        \label{full_mom}
\end{eqnarray}
where the complete differential cross section as given by 
\begin{eqnarray}
  \frac{d\sigma}{d\rmp d\Omega_p} &=& \frac{d\sigma^{(0)}}{d\rmp}\cdot\Big(
       1+\psi_1(\rmp,E)\Big)\frac{1+2{\cal A}_{\rm FB}(\rmp,E)\cos\vartheta}{4\pi}
       \qquad\mbox{with} \\
  {\cal A}_{\rm FB}(\rmp,E) & = & C_{\rm FB}\,\varphi_{\rm _R}(\rmp,E)+
       \frac{C_\parallel^0}{2}\kappa\,\psi_{\rm _R}(\rmp,E)
\end{eqnarray}
should be used as normalization. The change in ``polarization'' caused
by rescattering looks as follows:
\begin{eqnarray}
 \delta{\cal P}_\parallel &=& \Bigg[\frac{2+3y-5y^2-12y^3}{(1+2y)(1+2y+3y^2)}-
    \kappa(C_\parallel^0)^2\Bigg]\psi_{\rm _R}(\rmp,E)\cos\vartheta \nonumber \\
    && +\frac{1-4y+3y^2}{4(1+2y+3y^2)}(1-3\cos^2\vartheta)C_\parallel^0
    \psi_3(\rmp,E) \label{p2long}\\
 \delta{\cal P}_\perp &=& \frac{3(1-3y^2)}{2(1+2y+3y^2)}
    \psi_{\rm _R}(\rmp,E)\sin\vartheta \nonumber \\ 
    && -\frac{3(1-4y+3y^2)}{8(1+2y+3y^2)}C_\parallel^0
    \psi_3(\rmp,E)\sin2\vartheta  \label{p2perp}\\
 \delta{\cal P}_{\rm N} &=& 0,\qquad \delta{\cal P}_0 = 0.
\end{eqnarray}

An important consequence of these expressions -- which obviously differ from
those given in Sect.~\ref{rescatt} -- is that the normal
component of the polarization vector is unaffected by rescattering
corrections, and that (\ref{slong}) and (\ref{snorm}) remain valid. The
additional term multiplying $tn$ in (\ref{full_mom}) and the two
$\psi_3$ dependent terms on the other hand seem to spoil the beauty of and the
similarity with the Born result. But it had to be expected that the
rescattering corrections would destroy such a simple form.
Numerically however, because of the large mass of the top quark, it
will turn out below that these three terms can be neglected.

\section{Numerical results\label{results_sec}}

The QCD potential used in the numerical analysis has been described in
\cite{JT}. It depends on the strength of the coupling constant. If not
stated otherwise $\alpha_{ \overline{\rm MS}}(M_Z^2)=0.125$
is adopted. For the top mass a value of $180$ GeV is used and, correspondingly,
$\Gamma_t=$1.55 GeV \cite{JK89a}. The effects of a momentum dependent width on
cross section and momentum distributions have been studied in \cite{JT} and 
shown to be well under control and small.

Step by step the predictions will be presented for the momentum distribution,
the forward-backward asymmetry and the top quark polarization and thus for
observables of increasing degree of complication. They will be shown to
involve increasingly complicated rescattering corrections. For each of them
the prediction based on $S$-$P$--wave interference will be given which allows
to understand the qualitative features of the result. Subsequently the
modifications due to rescattering will be introduced.

\subsubsection{Momentum distribution}
\begin{figure}
 \begin{center}
  \leavevmode
   \epsfxsize=135mm
   \epsffile[70 130 510 735]{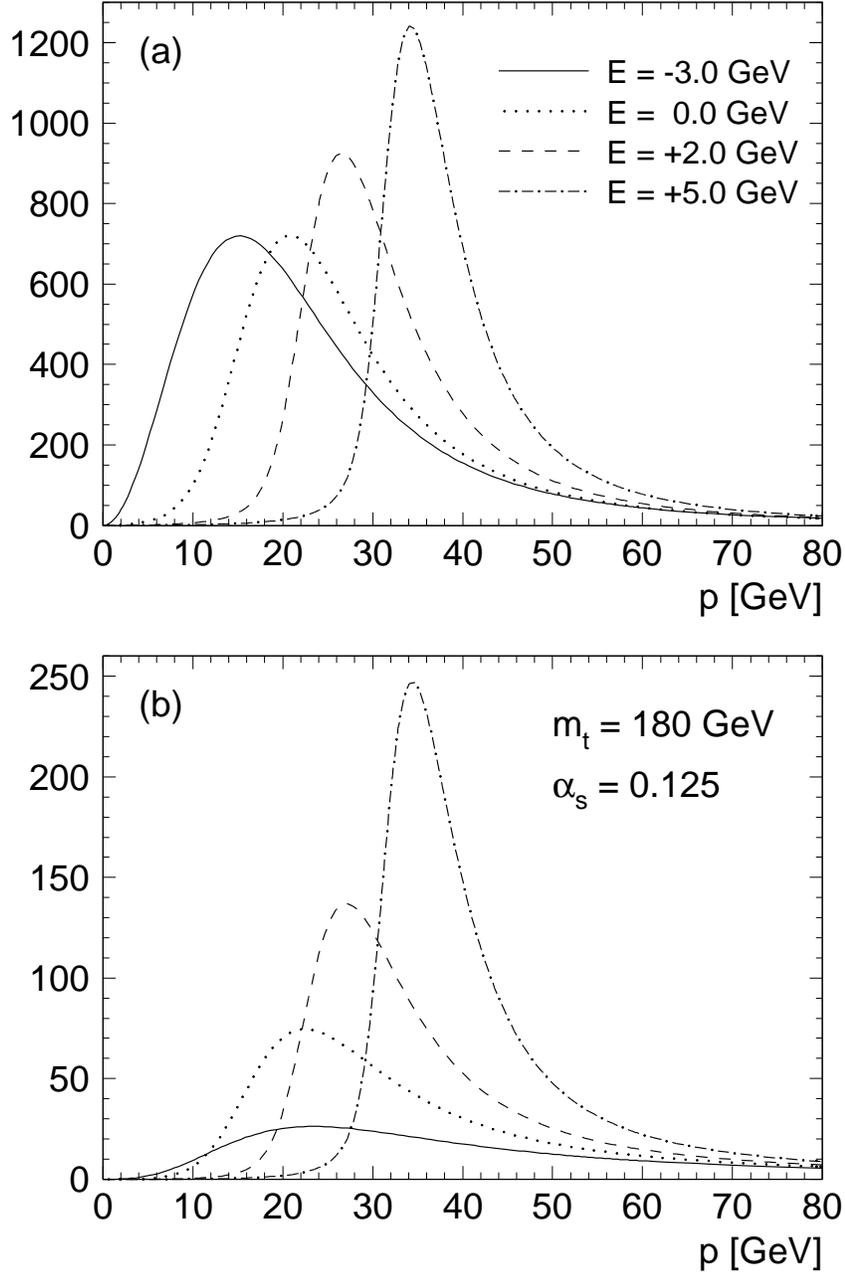}
  \caption[]{\label{ggfg.ps}\sloppy (a) $|\bfp \, G(\bfp,E)|^2$ and 
   (b) the $S$-$P$--interference term $|\bfp|^3/m_t\cdot 
        \Re(G(\bfp,E)\, F^*(\bfp,E))$ for $m_t=180$ GeV, 
        $\alpha_{\rm s} = 0.125$
        and three different energies, where the solid line corresponds to the
        energy of the $1S$-Peak.}
 \end{center}
\end{figure}

In Fig.~\ref{ggfg.ps} the squared $S$-wave Green function multiplied by
$\rmp^2$ and the $S$-$P$--interference term are displayed for four different
energies. The former characterizes the shape of the momentum distribution
$d\sigma/d$p, the latter is responsible for the forward-backward asymmetry and
angular dependent spin effects, neglecting of course rescattering.  The shift
towards larger momenta and the narrowing of the distribution with increasing
energy are clearly visible, as is the strong growth of the $P$-wave
contribution.

\begin{figure}
 \begin{center}
  \leavevmode
  \epsfxsize=135mm
  \epsffile[40 185 535 630]{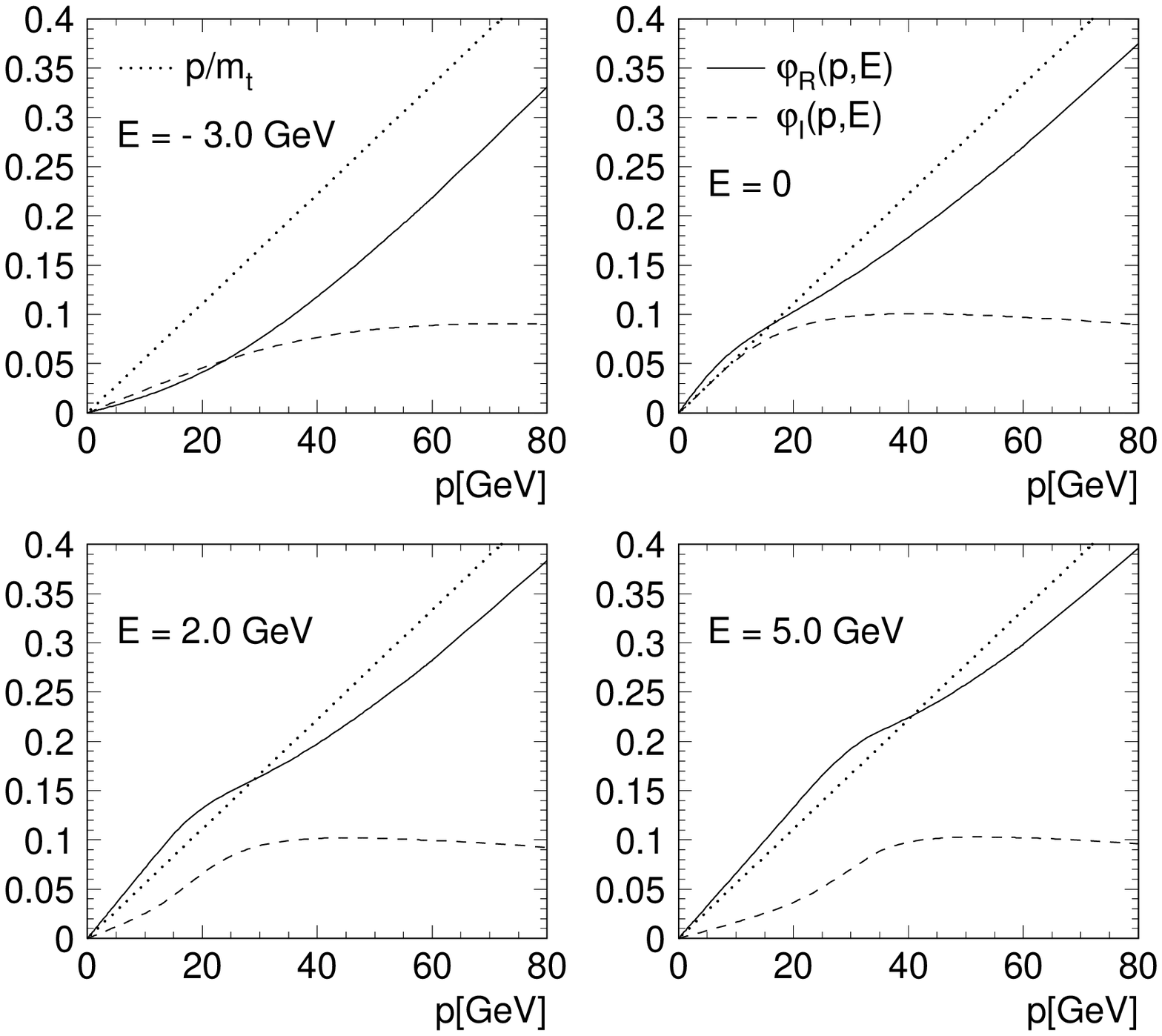}
  \caption[]{\label{phi.ps}\sloppy 
        Real (solid) and imaginary (dashed) part of the function
        $\varphi(\rmp,E)$ for $m_t = 180$ GeV, $\alpha_{\rm s}=0.125$ 
        and four
        different energies. The dotted line shows the free particle case
        $\Re\,\varphi = \mbox{p}/m_t$.}
 \end{center}
\end{figure}

In Fig.~\ref{phi.ps} we display real and imaginary parts of the function 
$\varphi$, which is essentially given by the ration between 
$S$-$P$--interference and the squared $S$-wave. 
In the limit of non-interacting quarks this corresponds to $\rmp/m_t$. 
The function $\varphi_{\rm _R}$ will become relevant in
the discussion of angular distributions, the polarization is affected by
$\varphi_{\rm _R}$ and $\varphi_{\rm _I}$.

\begin{figure}
\begin{tabular}{cc}
  \epsfxsize 70mm \mbox{\epsffile[0 0 567 454]{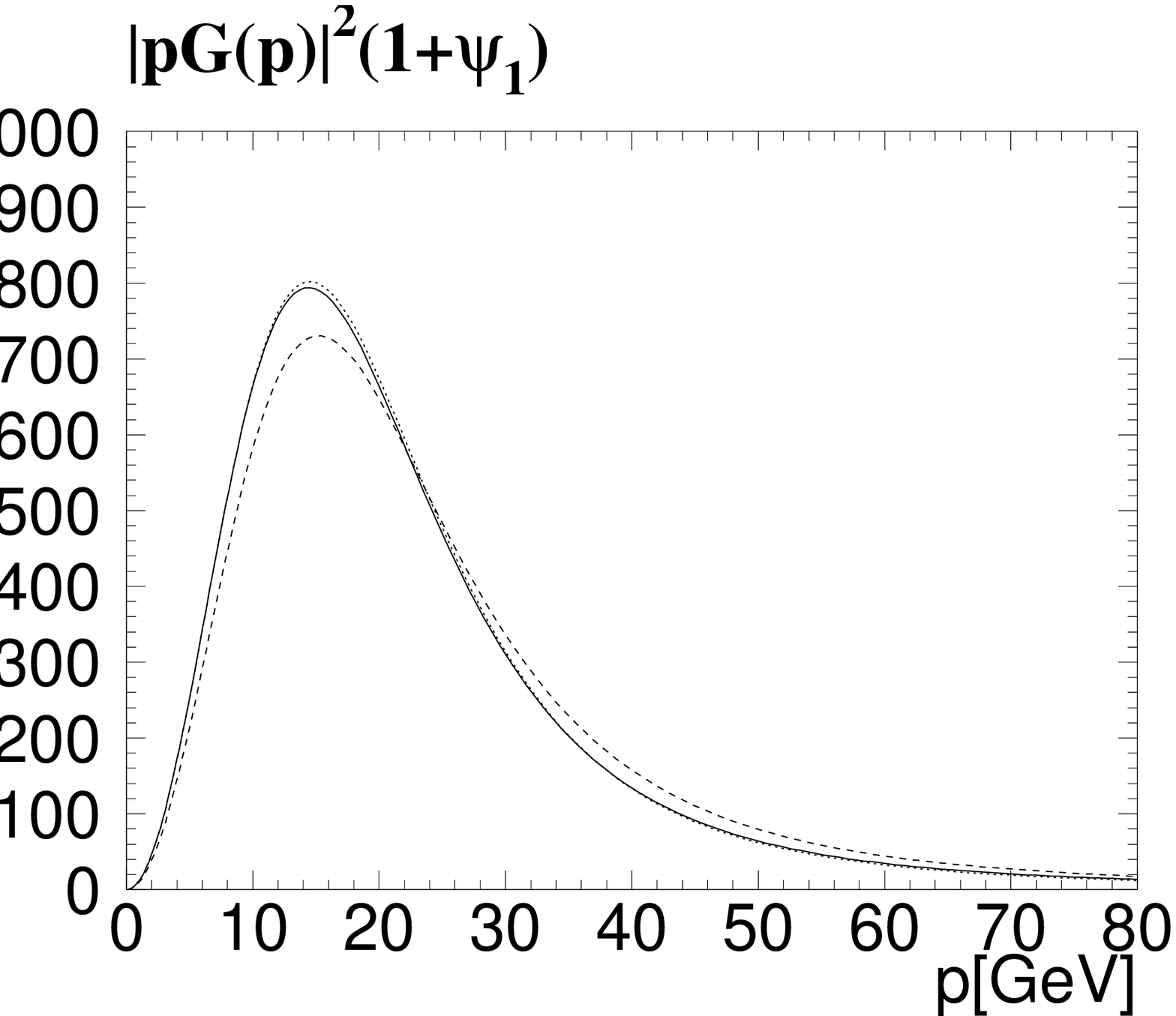}} & 
  \epsfxsize 70mm \mbox{\epsffile[0 0 567 454]{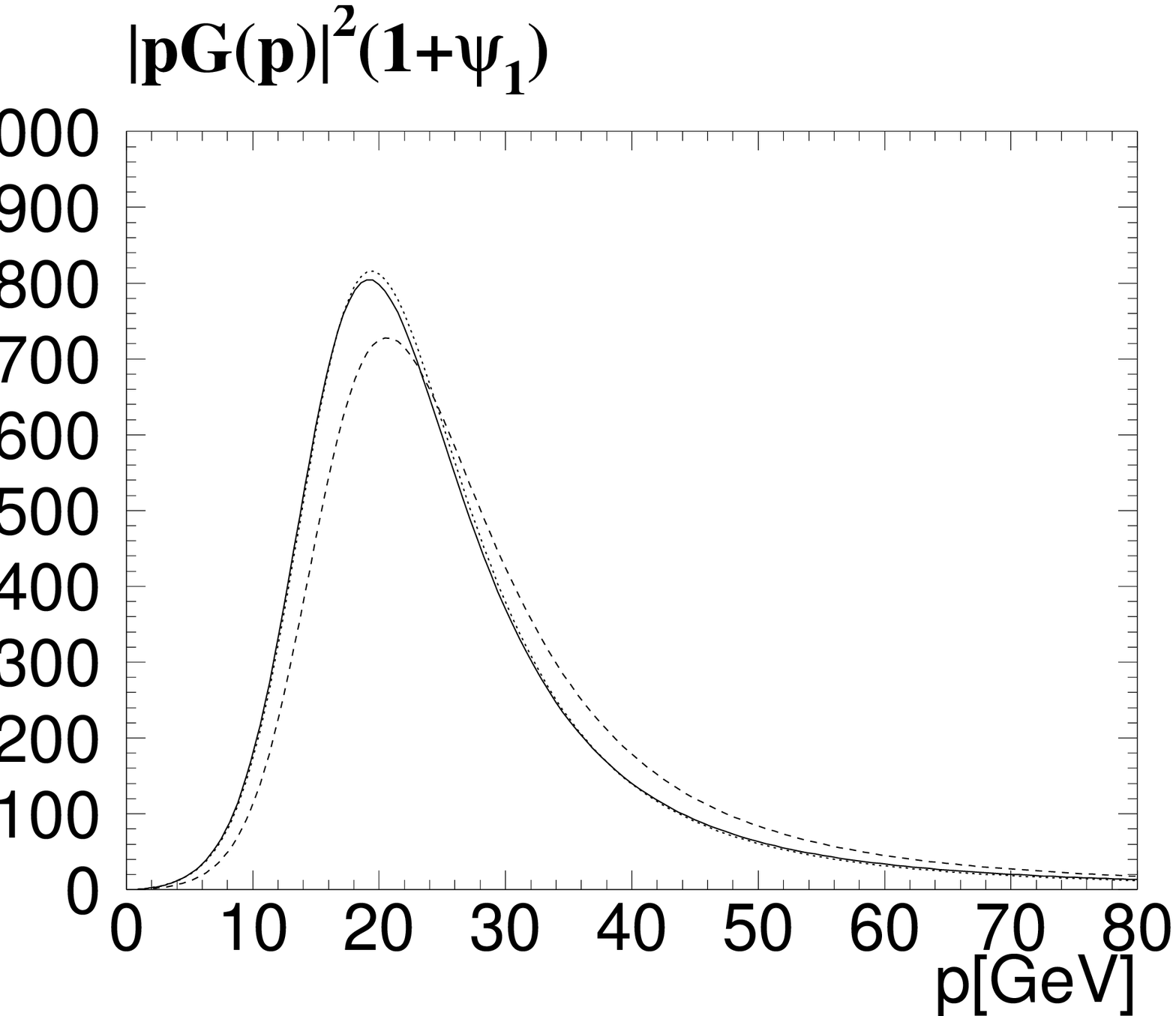}} \\  
    a) E=-3GeV & b) E=0GeV \\
  \epsfxsize 70mm \mbox{\epsffile[0 0 567 454]{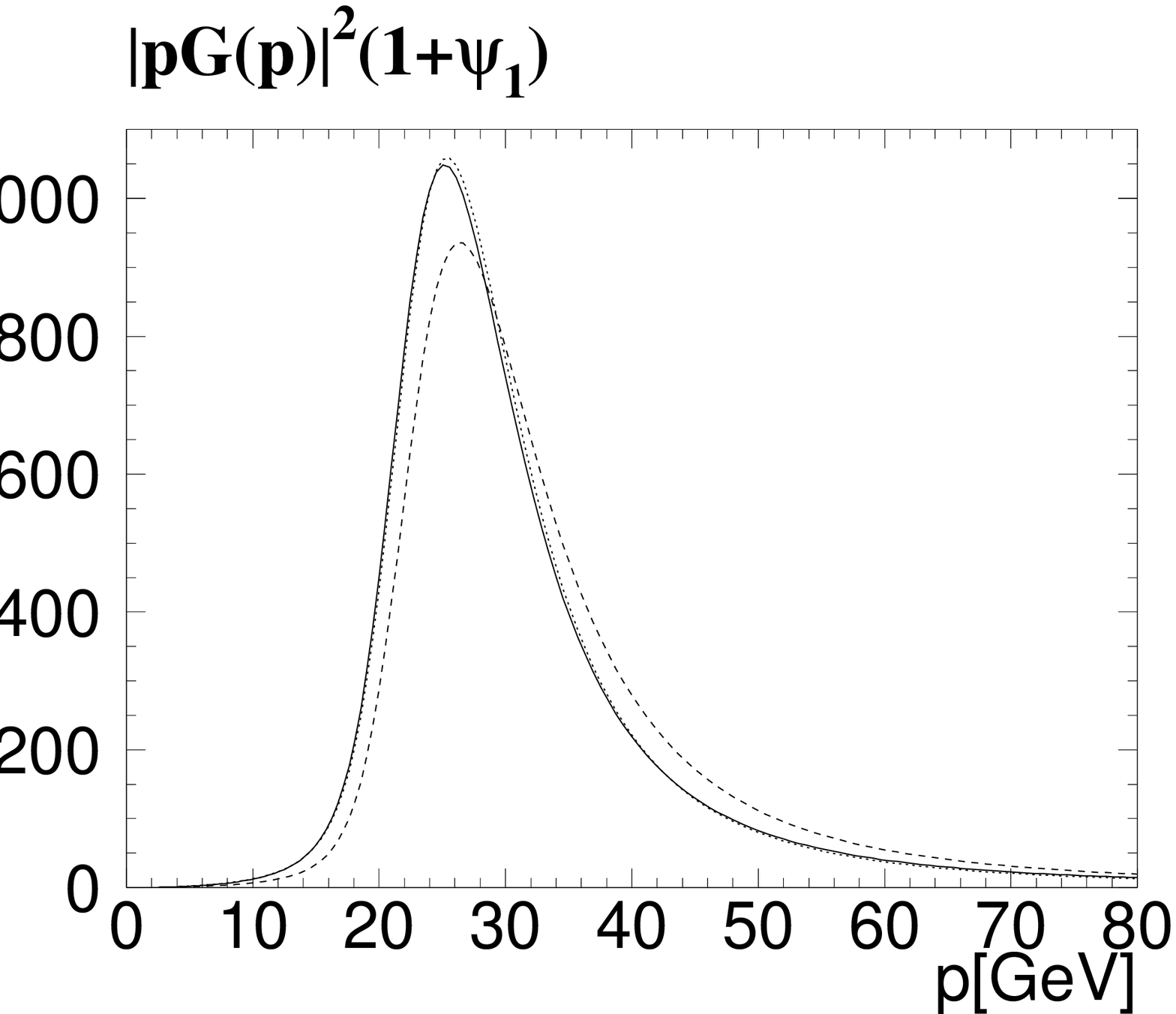}} & 
  \epsfxsize 70mm \mbox{\epsffile[0 0 567 454]{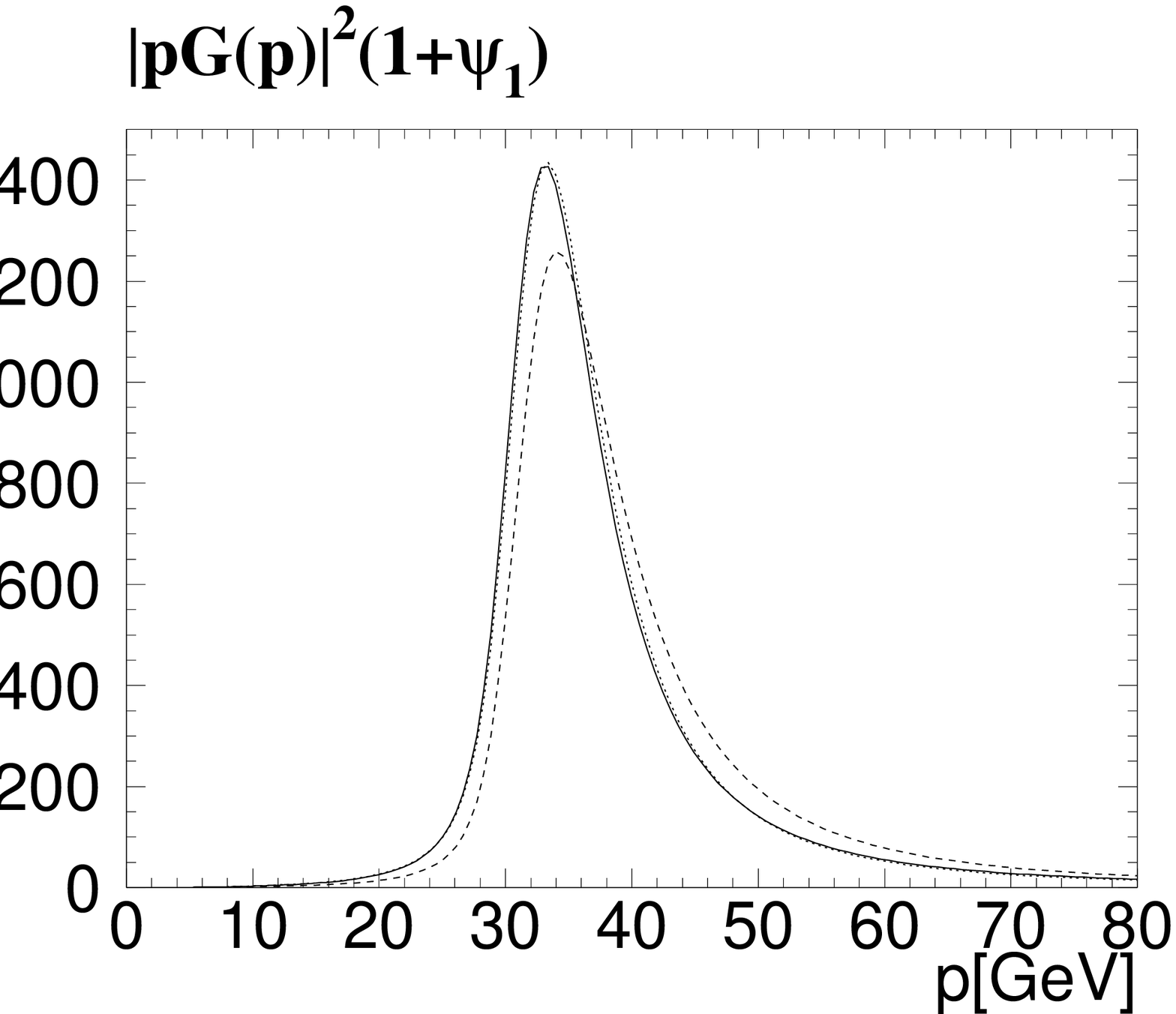}} \\
    c) E=2GeV  & d) E=5GeV
\end{tabular}
\caption{Modification of the momentum distribution through rescattering.
Dashed line: no rescattering corrections included; Solid line: rescattering
contribution with full potential included; dotted line: rescattering
contribution with pure Coulomb potential and $\alpha_{\rm s}=0.187$ included.
\label{rescdsig}}
\end{figure}

\begin{figure}\begin{center}
\begin{tabular}{cc}
  \epsfxsize 70mm \mbox{\epsffile[0 200 567 600]{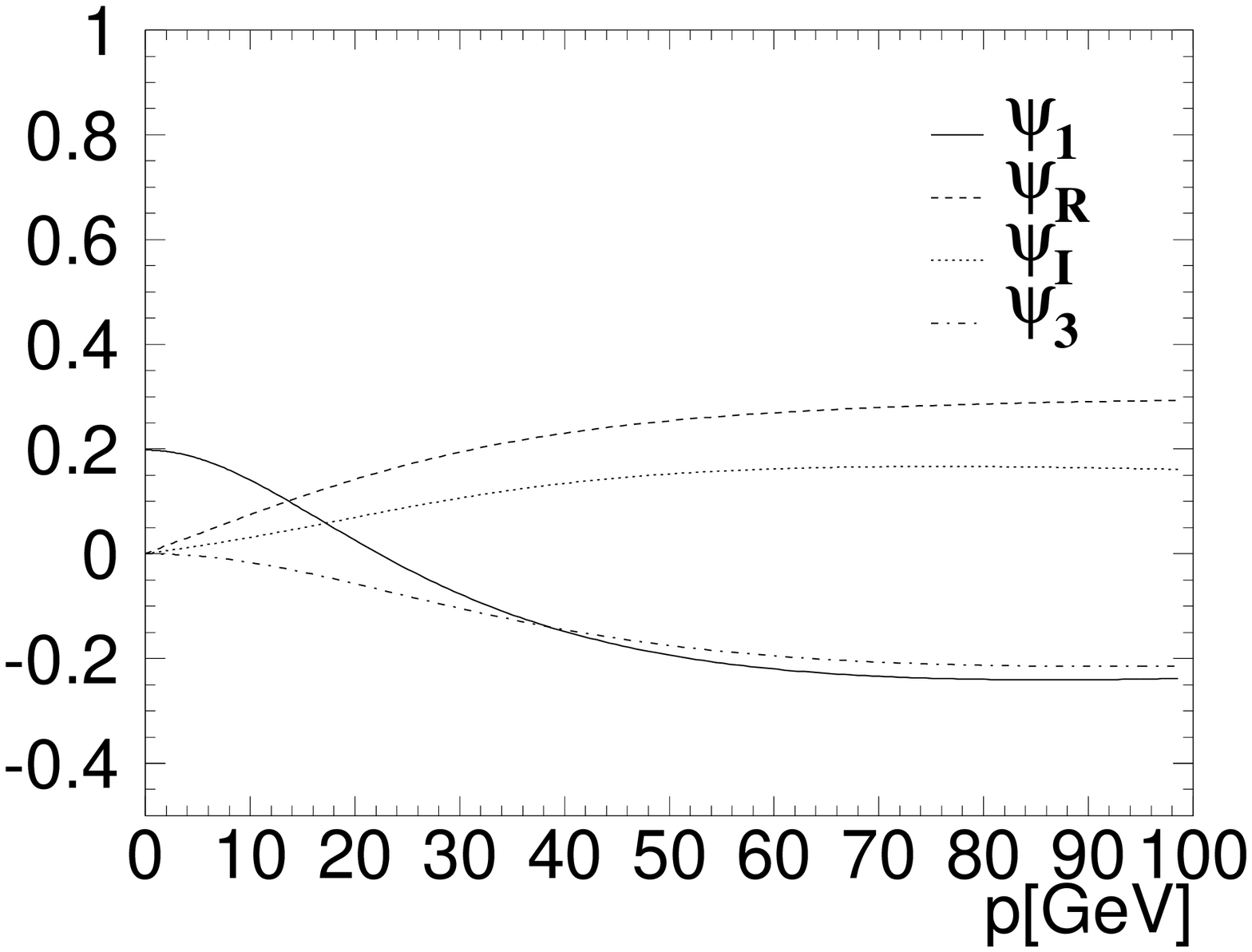}} &
  \epsfxsize 70mm \mbox{\epsffile[0 200 567 600]{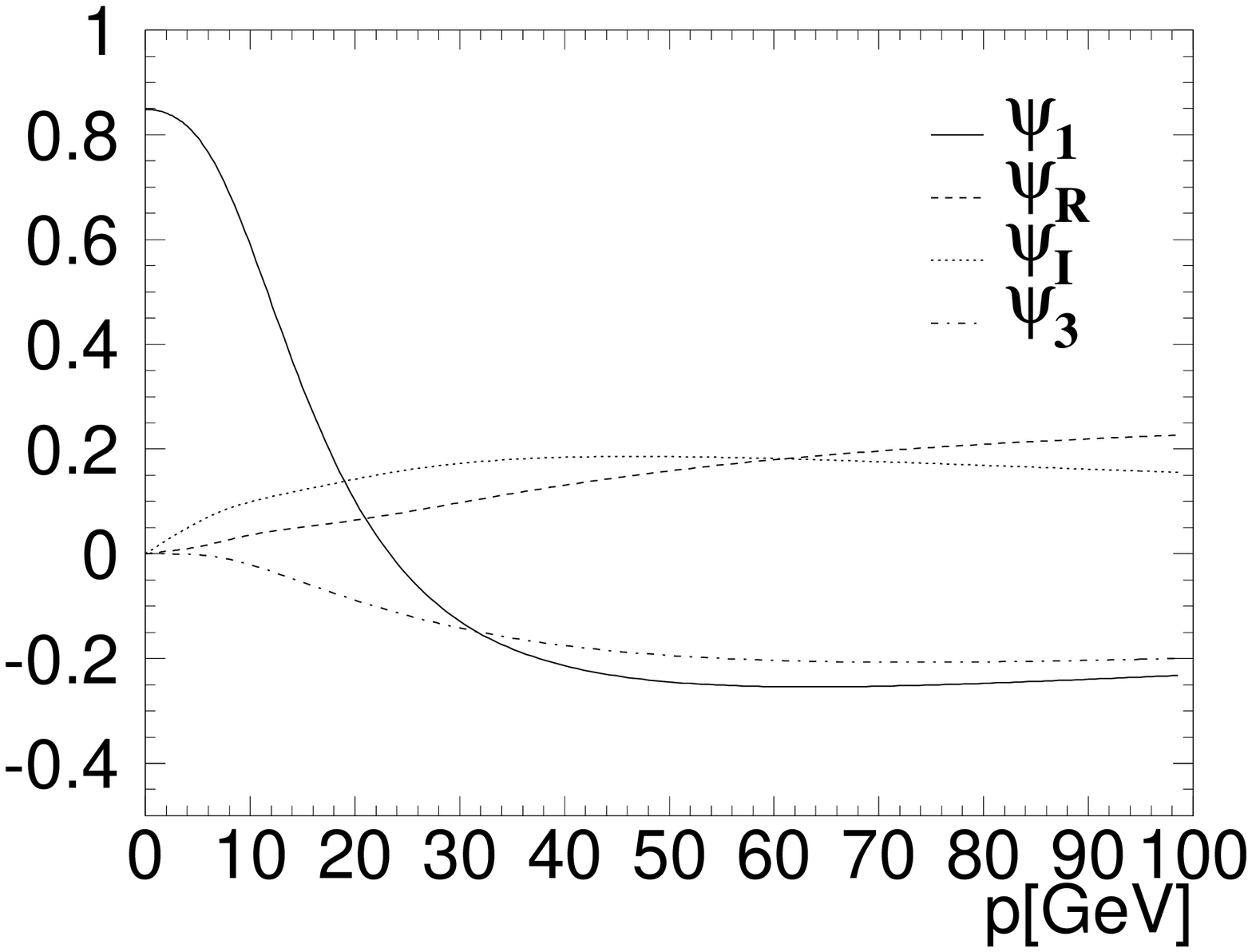}} \\
  a) E=-3 GeV & b) E=0 GeV \\
  \epsfxsize 70mm \mbox{\epsffile[0 200 567 600]{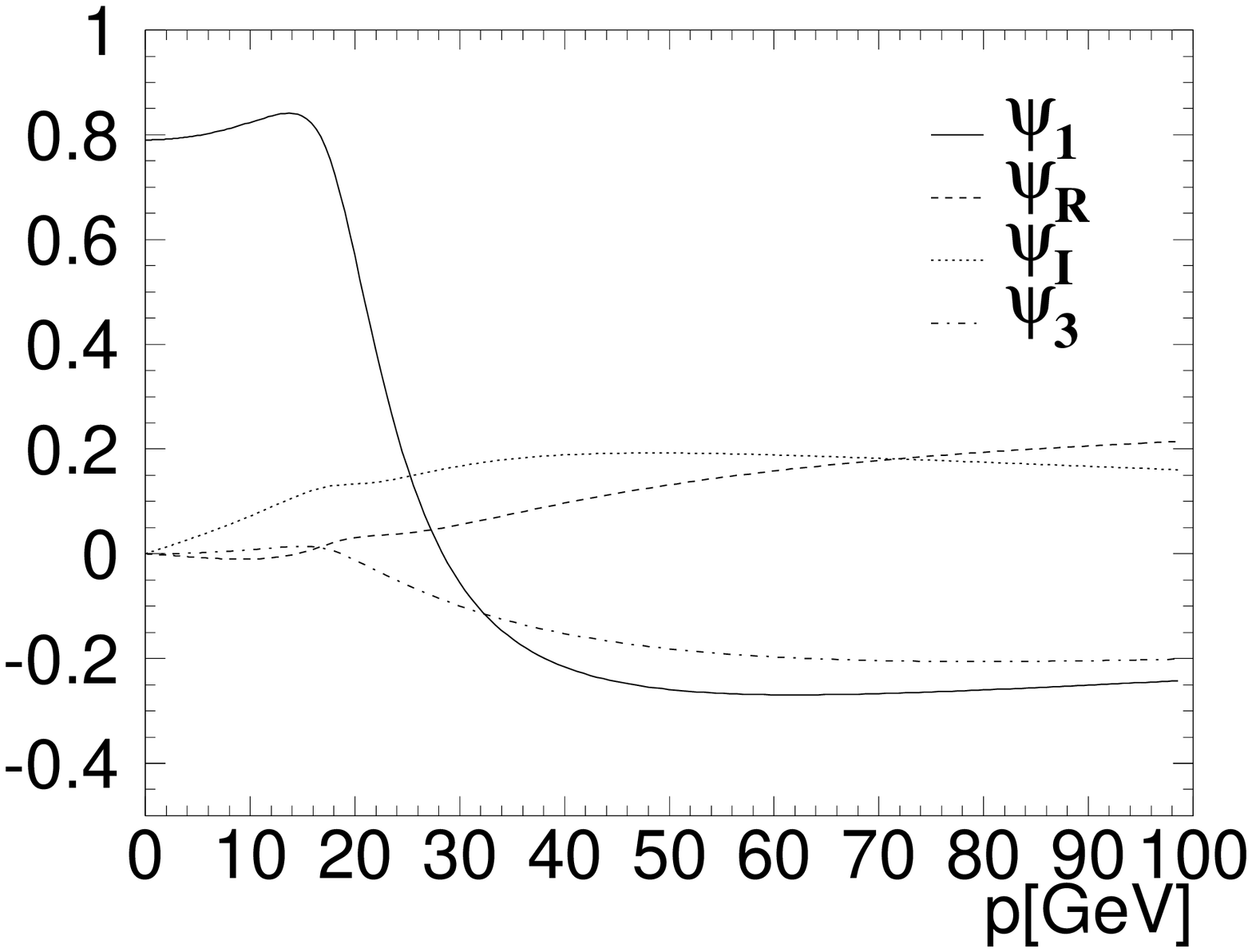}} &
  \epsfxsize 70mm \mbox{\epsffile[0 200 567 600]{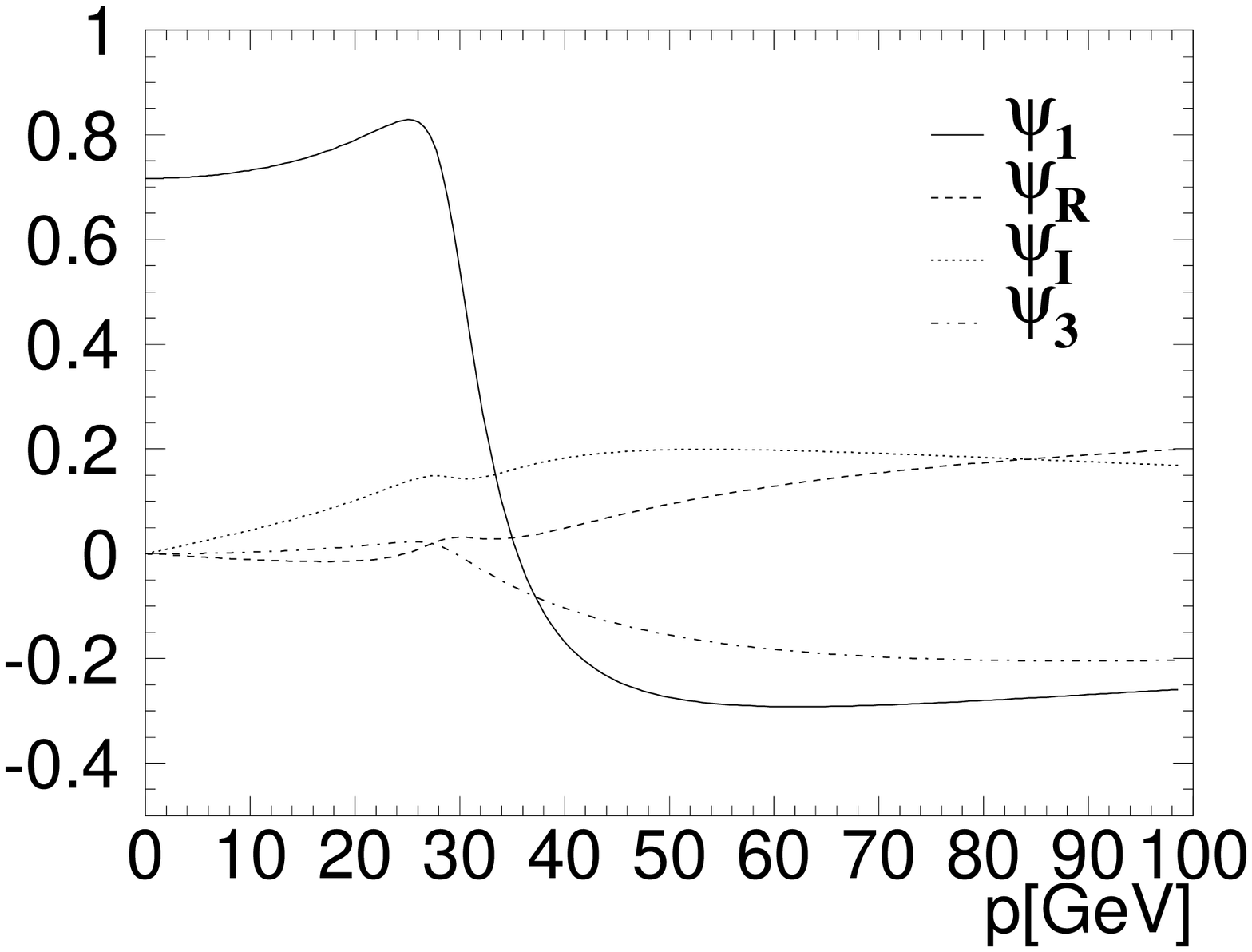}} \\
  c) E=2 GeV & d) E=5 GeV
\end{tabular}
\end{center}
\caption{\label{psisfig}The four functions $\psi_1$, 
$\psi_{\rm _R}=\Re\;\psi_2$, 
$\psi_{\rm _I}=\Im\;\psi_2$ and $\psi_3$ required for the inclusion of the
rescattering corrections.}
\end{figure}

Fig.~\ref{rescdsig} shows the influence of the rescattering
correction on the momentum distribution as given in
(\ref{dsig_lept}). The size of rescattering corrections is governed by
the three functions $\psi_1$, $\psi_2$ and $\psi_3$ which are displayed
in Fig.~\ref{psisfig}. From the definition and the way
they enter the different expressions the following interpretation is
possible: $\psi_1$ governs the size of the additional $S$-wave
amplitude, $\psi_2$ and $\psi_3$, respectively,
give the $P$- and $D$-wave amplitudes generated by
rescattering\footnote{ Note that their appearance has an origin
  different from the $S$-$P$--wave interference: the latter would
  vanish for $a_t=0$ whereas rescattering would remain.}.  

Fig.~\ref{rescdsig} clearly demonstrates that the attractive force of the
$t\bar t$ spectator on the decay products from the primary decay leads
to a shift towards smaller momenta (the peak positions are listed in
Table \ref{peakpos}). The dashed line corresponds to the Born result,
Fig.~\ref{ggfg.ps}(a), the other two include rescattering,
implemented however in different ways. The solid line was calculated
using the full potential, i.e.\ the same that appears in the
determination of the Green functions. For the dotted line, which is
hardly distinguishable from the solid curve, a pure Coulomb potential
with fixed coupling constant was used, similar to the previous
analyses of \cite{SThes,FMY94}. The calculation for the second
alternative is somewhat easier because $G({\bf p},E)$ depends on
$\mbox{p}=|{\bf p}|$ only and, for the simple choice of the potential,
the functions $\psi_i$ can be converted into one-dimensional integrals.
Only one integration will have to be performed numerically in this
case \cite{SThes}.

\begin{table}\begin{center}
\begin{tabular}{|r||c|c|c|}
  \hline
  E [GeV] & $p_{\rm Born}$ [GeV] & $p^{\rm full}_{\rm resc}$ [GeV] & 
          $p^{\rm fixed}_{\rm resc}$ [GeV] \\ \hline\hline
  -3      & 15.3 & 14.5 & 14.5 \\ \hline
   0      & 20.6 & 19.2 & 19.3 \\ \hline
   2      & 26.4 & 25.1 & 25.3 \\ \hline
   5      & 34.2 & 33.1 & 33.3 \\ \hline
\end{tabular}\end{center}
\caption{\label{peakpos}Approximate position of the peak in the momentum
  distribution.  $p^{\rm full}_{\rm resc}$ is the value obtained including
  rescattering with the full potential, correspondingly $p^{\rm fixed}_{\rm
    resc}$ with Coulomb potential, but $\alpha_{\rm s}=0.187$.}
\end{table}

\begin{figure}
\begin{center}\begin{tabular}{cc}
  \epsfxsize 65mm \mbox{\epsffile[0 0 567 454]{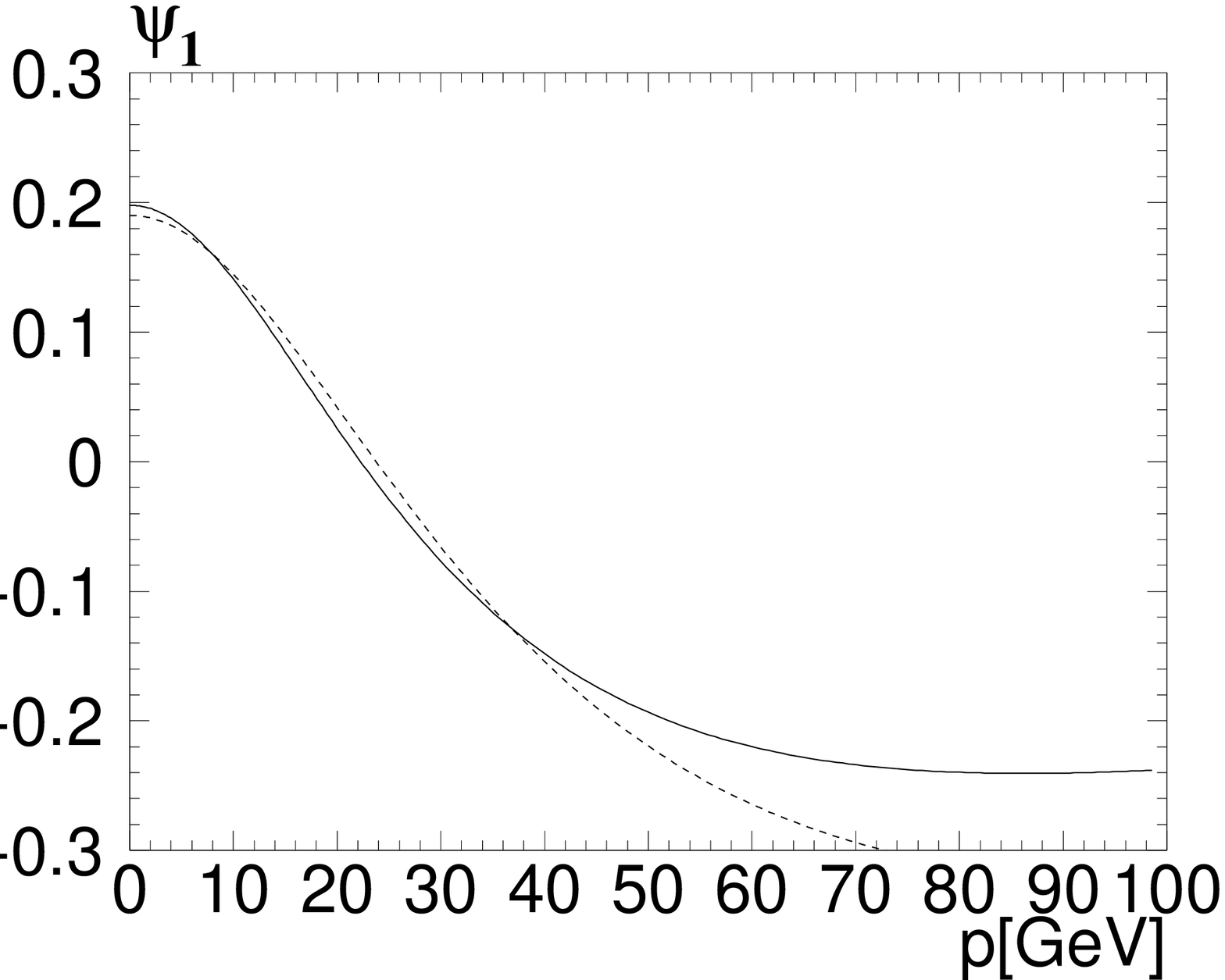}} &
  \epsfxsize 65mm \mbox{\epsffile[0 0 567 454]{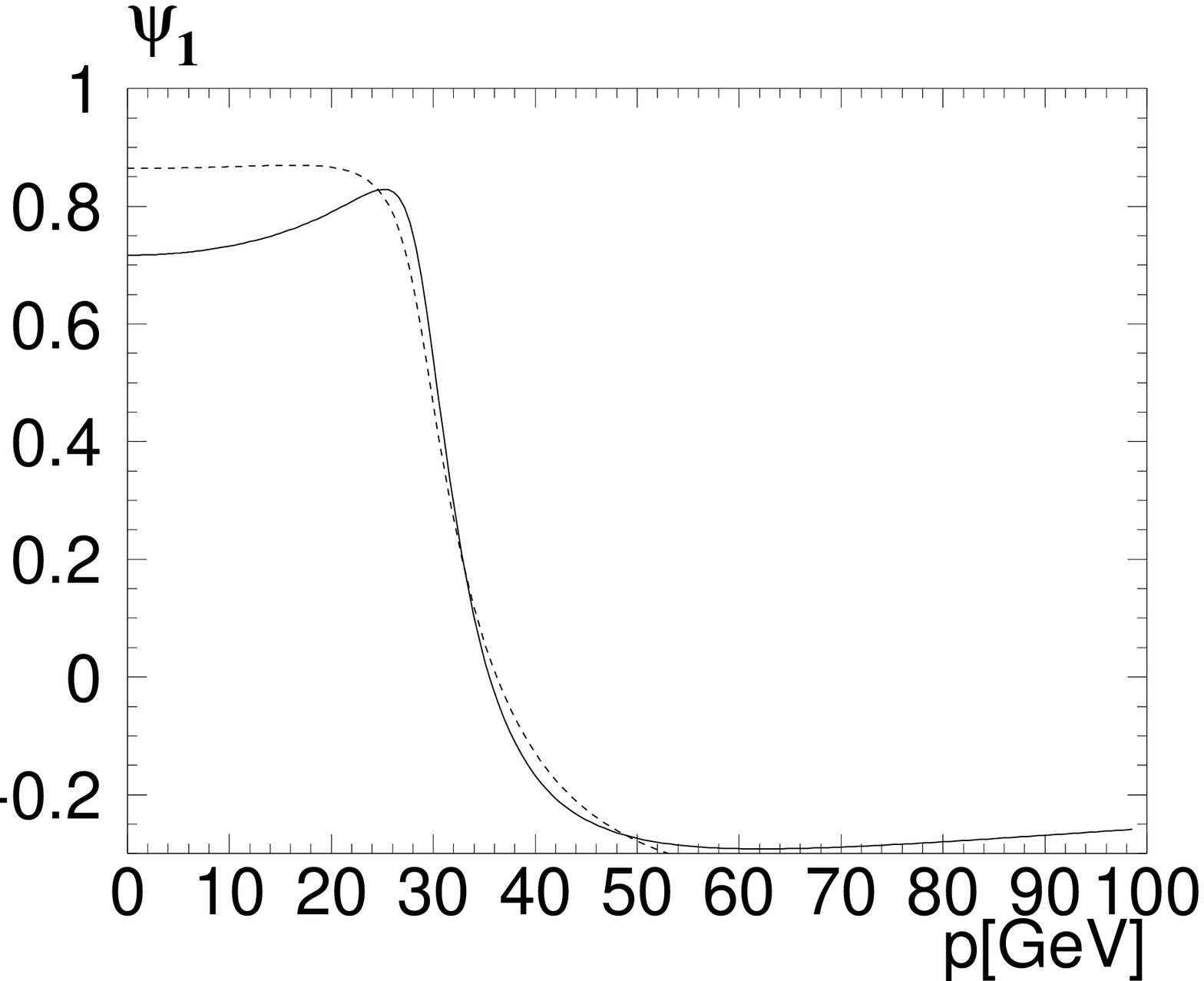}} \\
  $\psi_1(\rmp,\mbox{E=-3GeV})$ & $\psi_1(\rmp,\mbox{E=5GeV})$
\end{tabular}\end{center}
\caption{The function $\psi_1$ for two energies using different potentials.
Solid line: full potential used; dashed line : pure Coulomb potential
with fixed $\alpha_{\rm s}=0.187$ used.\label{psifig}}
\end{figure}

\begin{figure}
\begin{center}\begin{tabular}{cc}
  \epsfxsize 65mm \mbox{\epsffile[0 0 567 454]{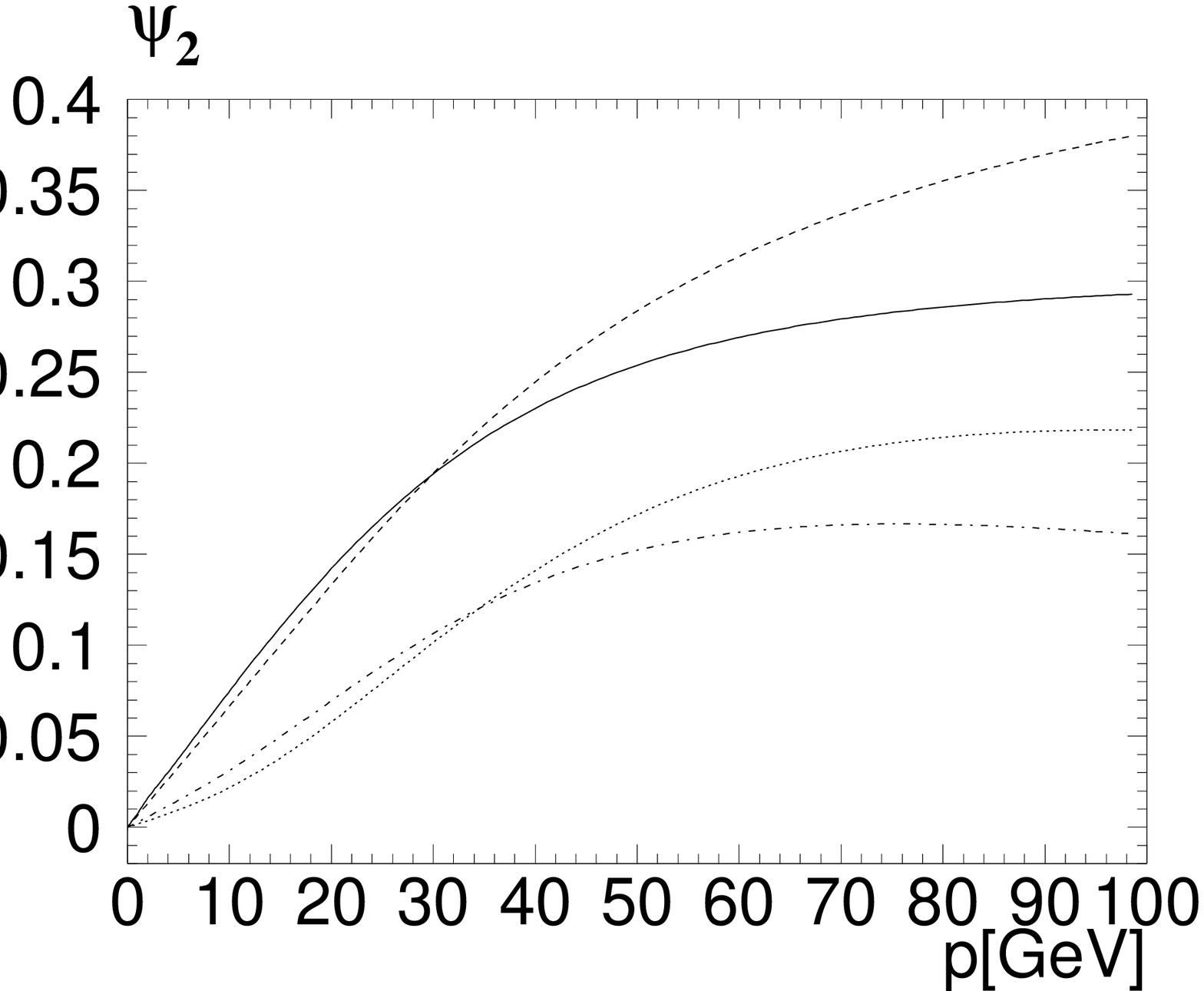}} &
  \epsfxsize 65mm \mbox{\epsffile[0 0 567 454]{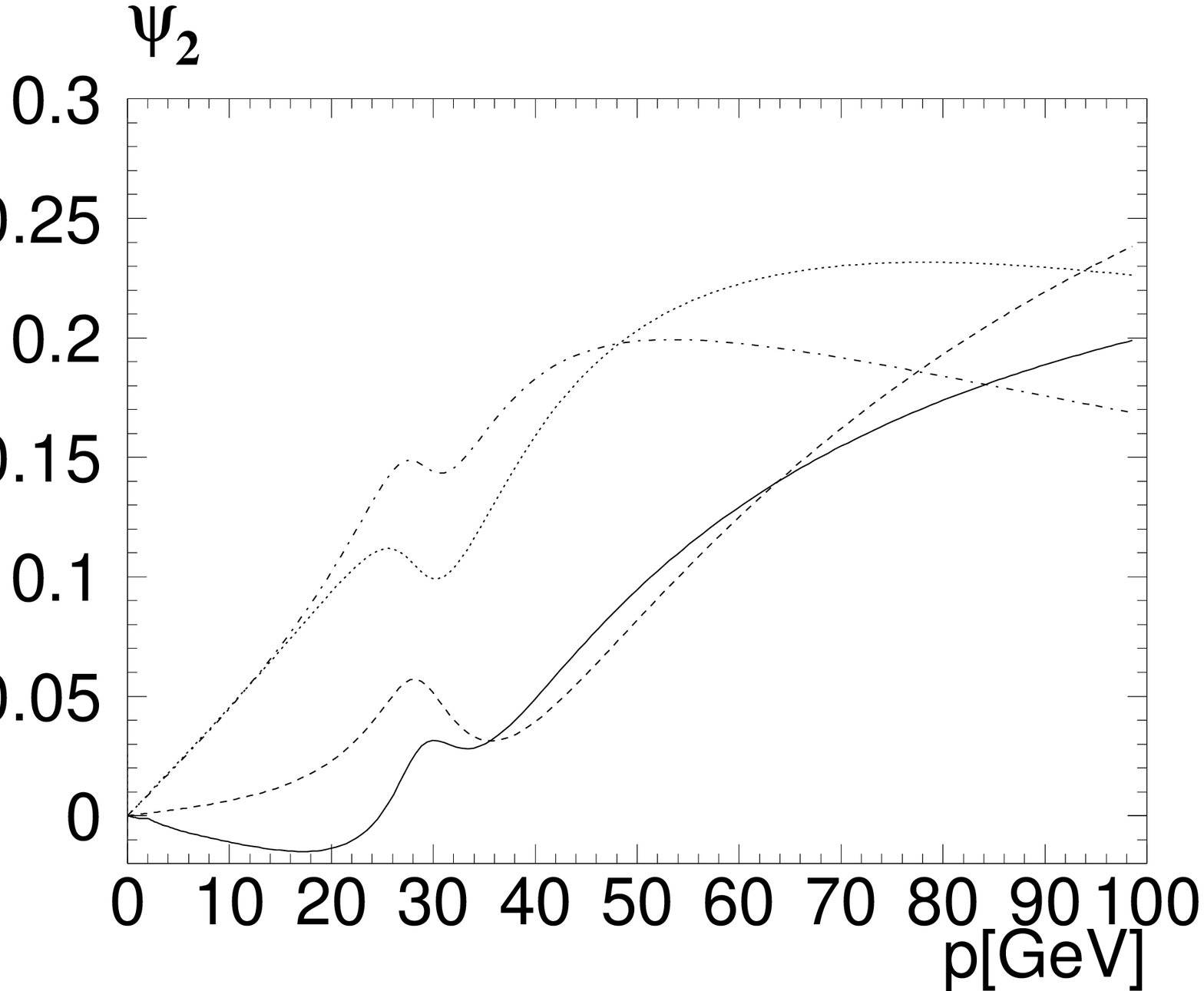}} \\
  $\psi_{\rm _R}(\rmp,\mbox{E=-3GeV})$ and 
    $\psi_{\rm _I}(\rmp,\mbox{E=-3GeV})$ &
  $\psi_{\rm _R}(\rmp,\mbox{E=5GeV})$ and 
    $\psi_{\rm _I}(\rmp,\mbox{E=5GeV})$
\end{tabular}\end{center}
\caption{The functions $\psi_{\rm _R}$ and $\psi_{\rm _I}$ 
    for two energies using different
    potentials. Solid line: $\psi_{\rm _R}$, 
    full potential used; dashed line: $\psi_{\rm _R}$,
    pure Coulomb potential with fixed $\alpha_{\rm s}=0.187$ used.  
    Dash-dotted line:
    $\psi_{\rm _I}$, full potential used; dotted line: $\psi_{\rm _I}$, 
    pure Coulomb potential with fixed $\alpha_{\rm s}=0.187$ used.
    \label{psirfig}}
\end{figure}

However, to obtain such good an agreement between the dotted and the
solid curve, a value $\alpha_{\rm s}=0.187$ has been adopted 
{\em a posteriori}.
This large value can easily be understood: the scale governing the
rescattering corrections should be of the order of top momentum
instead of top mass and thus a larger coupling has to be expected.
The satisfactory agreement between the two schemes is also evident from
Figs.~\ref{psifig} and \ref{psirfig} for the two functions $\psi_1$
and $\psi_2$. Note, however, that $\alpha_{\rm s}$ has been fixed for this
comparison a posteriori to the value $0.187$ which differs from
$\alpha_{\rm s}(\alpha_{\rm s}^2(M_Z^2)m_t^2)\approx0.15$ 
used in \cite{MurSum2}.

\subsubsection{The forward-backward asymmetry}
\begin{figure}
 \begin{center}
  \leavevmode
   \epsfxsize=12cm
   \epsffile[70 80 510 735]{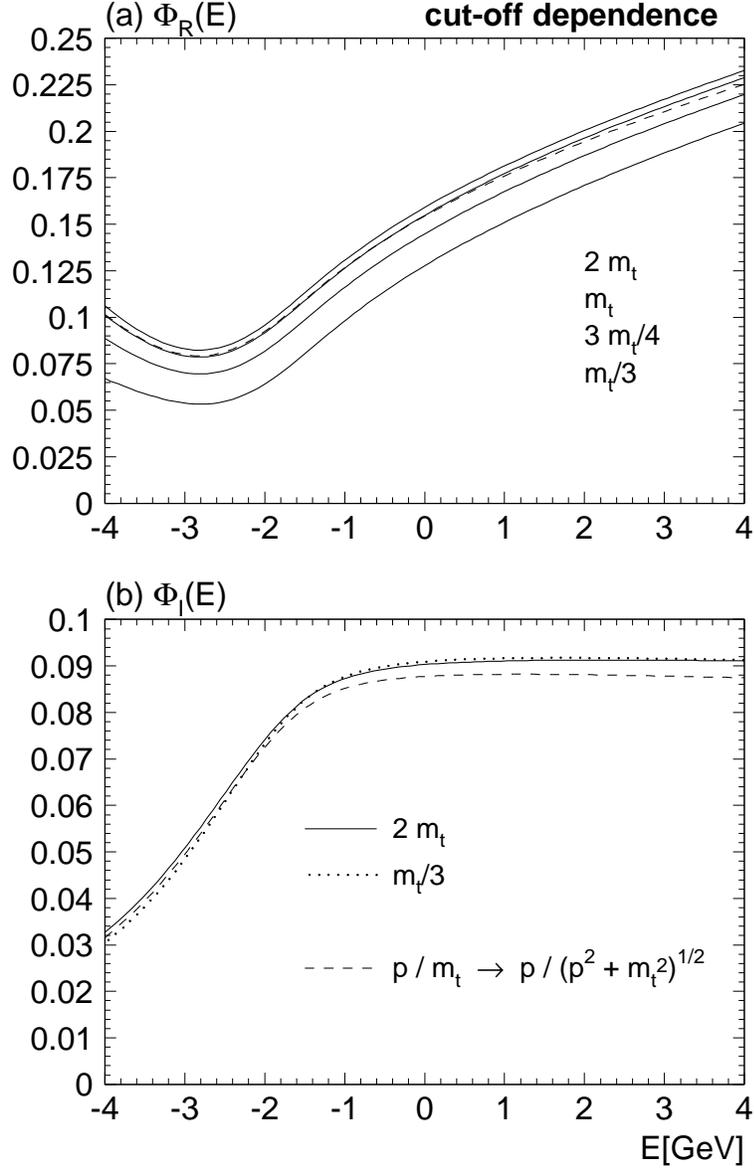}
  \caption[]{\label{cutdep.ps}\sloppy The integrated function $\Phi(E)$,
        (a) real part, (b) imaginary part. The different solid lines in 
        (a) correspond to cutoffs of $p_m=2m_t,m_t,3m_t/4$ and $m_t/3$, with
        the highest curve for $p_m=2m_t$. For the dashed lines in (a) and (b) 
        the relativistic prescription was adopted.}
 \end{center}
\end{figure}

Ignoring rescattering corrections for the moment the forward-backward
asymmetry is governed by the relative magnitude of $P$- versus $S$-wave
amplitudes.  The ratio between the corresponding two distributions is shown in
Fig.~\ref{phi.ps}.  For noninteracting stable quarks $\varphi(\bfp,E) =
$p$/m_t$, and this behaviour persists even in the presence of the QCD
potential. The energy dependence of $\varphi$ is not very strong even in the
presence of the potential, and most of the energy variation of $\Phi$ visible
in Fig.~\ref{cutdep.ps} is driven by the energy dependence of the Green
function which through the averaging procedure samples different 
momentum regions.

It should be stressed that the rate and thus the experimental sensitivity
varies strongly as a function of p (see Fig.~\ref{ggfg.ps}) and thus
only a limited momentum range will be explored for any given energy. In fact,
given limited statistics only integrated quantities will be measured in a first
round. These are characterized by the function $\Phi(E)$ whose real and
imaginary parts are shown in Fig.~\ref{cutdep.ps}.  The result depends slightly
on the cutoff procedure --- in a real measurement the cutoff prescription will
be dictated by experimental considerations. For definiteness all subsequent
results will be given for the ``relativistic'' cutoff prescription%
\footnote{We note that this treatment incorporates the bulk of the
``relativistic'' corrections discussed in \cite{moed}.} where p$/m_t$ is
replaced by p$/E$.

The energy dependence of both the real and the imaginary parts of $\Phi$ can be
understood as follows: Above threshold the interaction is of minor importance
and $\Phi_{\rm R}$ increases just like p$/m_t=\sqrt{E/m}$.  
It reaches its minimum
roughly in the region of the would-be-$1S$-resonance where the $S$ wave is
still fairly prominent and the $P$-wave is nearly absent. For lower energies
$\Phi$ starts to increase, again roughly $\propto \sqrt{|E|/m_t}$, a
consequence of the uncertainty principle (see also \cite{jkt_or}).  The
imaginary part of $\Phi$ which will induce the normal component of the top
quark polarization levels off at a constant value of around $0.09$, well
consistent with the perturbative prediction (\ref{cap_phi_gt0})
$\Phi_{\rm I}=2\alpha_{\rm s}($p$)/3$ with 
$\rmp \approx \sqrt {E^2 + \Gamma_t^2}$.  It
vanishes rapidly below $E=0$: real intermediate $t\bar t$-states are no longer
accessible, rescattering is absent and the imaginary part vanishes.  Throughout
most of the threshold region real and imaginary part of $\Phi$ are of
comparable magnitude, and so are normal and transverse polarization.

\begin{figure}
 \begin{center}
  \leavevmode
  \epsfxsize=135mm
  \epsffile[70 100 510 735]{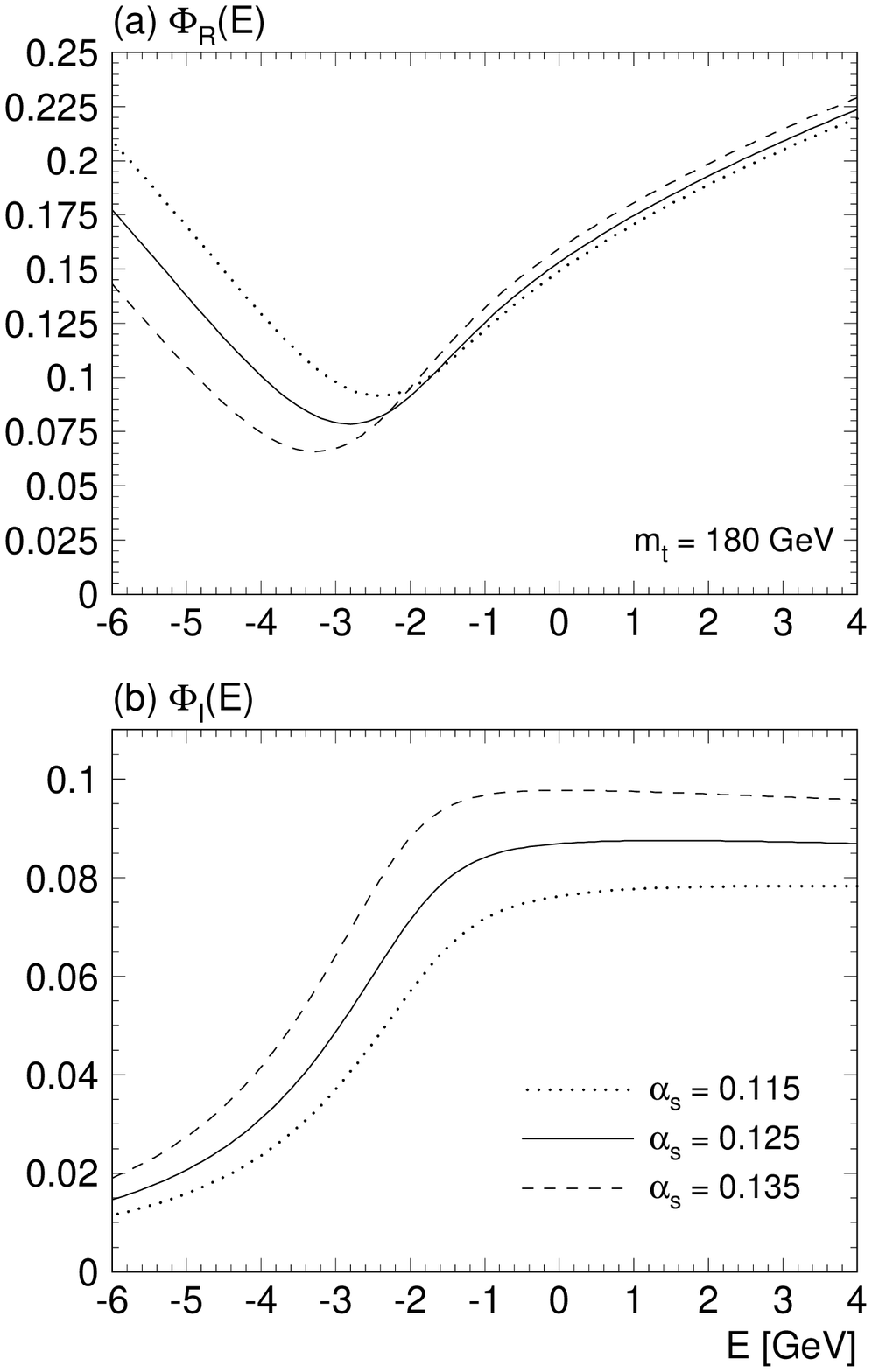}
  \caption[]{\label{alph_dep.ps}\sloppy Real and imaginary part of $\Phi(E)$
        for three different values of $\alpha_{\rm s}$.}
 \end{center}
\end{figure}

The $\alpha_{\rm s}$ dependence of $\Phi$ is displayed in
Fig.~\ref{alph_dep.ps}.  Above threshold the real part is practically
independent of $\alpha_{\rm s}$ and entirely determined by the kinematic
relation.  The minimum of the curve is lowered and shifted towards lower values
of $E$, reflecting the decrease of the $1S$ peak and the larger separation
between the ``would be'' $S$- and $P$-wave resonance. The imaginary part of
$\Phi$ is essentially proportional to $\alpha_{\rm s}$, a result obvious from
(\ref{cap_phi_gt0}).

\begin{figure}
 \begin{center}
  \epsfxsize=135mm
  \leavevmode
  \epsffile[70 110 510 735]{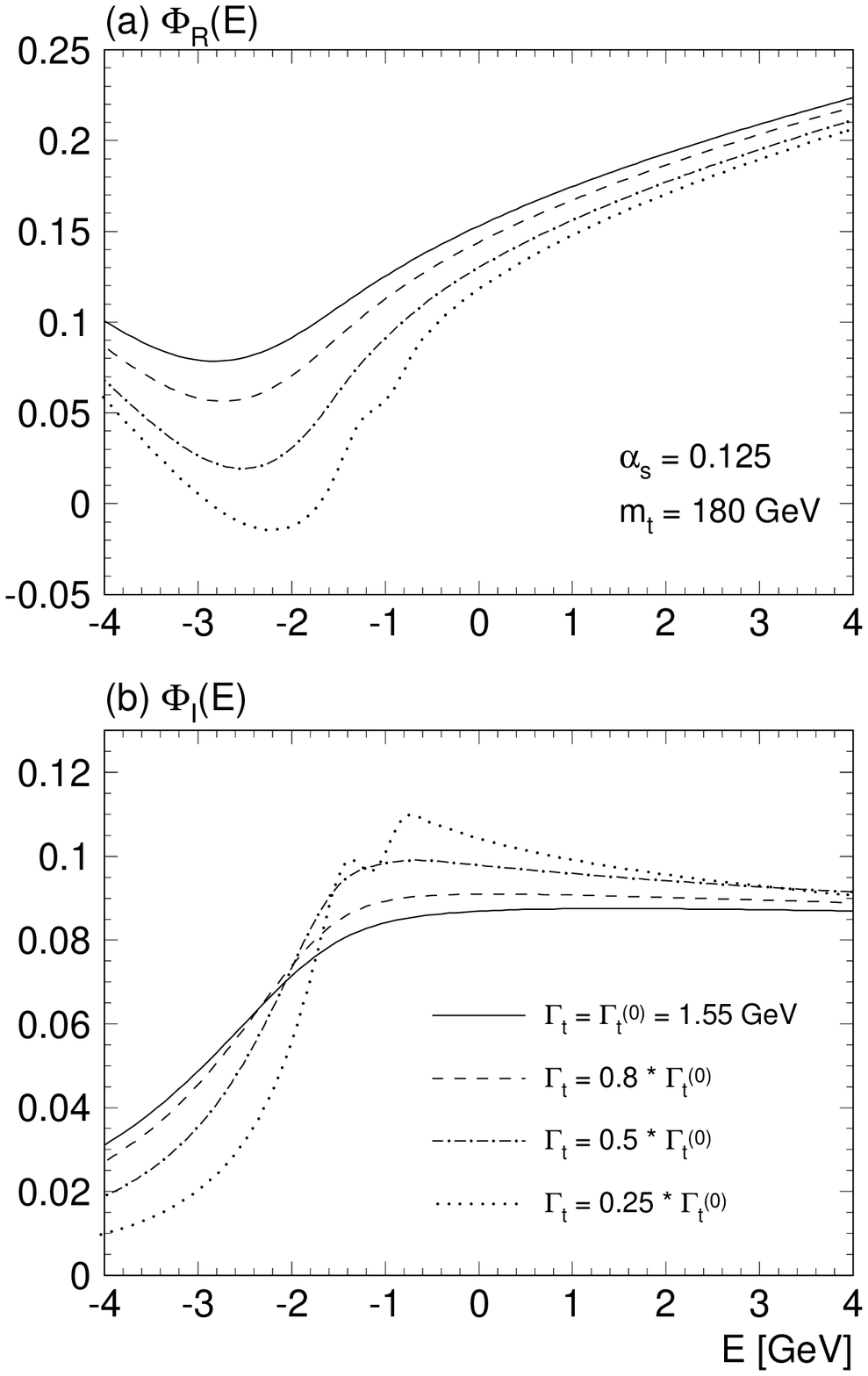}
  \caption[]{\label{width_Phi.ps}\sloppy Real and imaginary part of $\Phi(E)$
        for artificially lowered width. $\Gamma_t^{(0)}$ is the standard model
        width given in \cite{jezk}.}
 \end{center}
\end{figure}

The width of the top quark is fixed in the Standard Model: $\Gamma_t=1.55$ GeV
for a top mass of 180 GeV \cite{JK}.  New decay modes could increase the
width, large CKM angles for the mixing between the third and a fourth family
could lower the width. The sensitivity of $\Phi$ towards a reduction of
$\Gamma_t$ is demonstrated in Fig.~\ref{width_Phi.ps}. For an artificially
reduced width the real part of $\Phi$ is significantly reduced in the
resonance region. It is evident that both $E$ and $\Gamma_t$ contribute to the
effective momentum. The oscillatory pattern reflects the reappearance of
resonance structures. The changes for the imaginary part can also be easily
interpreted: The suppression of rescattering below threshold becomes more
pronounced for the reduced width, leading to a reduction of $\Phi_{\rm I}$ 
for $E$ below $-2$ GeV. 
Above this energy rescattering is accessible and even enhanced
for the case of reduced width: the effective momentum is smaller and the QCD
coupling correspondingly enhanced.
 
\begin{figure}
 \begin{center}
  \leavevmode
  \epsfxsize=135mm
  \epsffile[70 110 510 735]{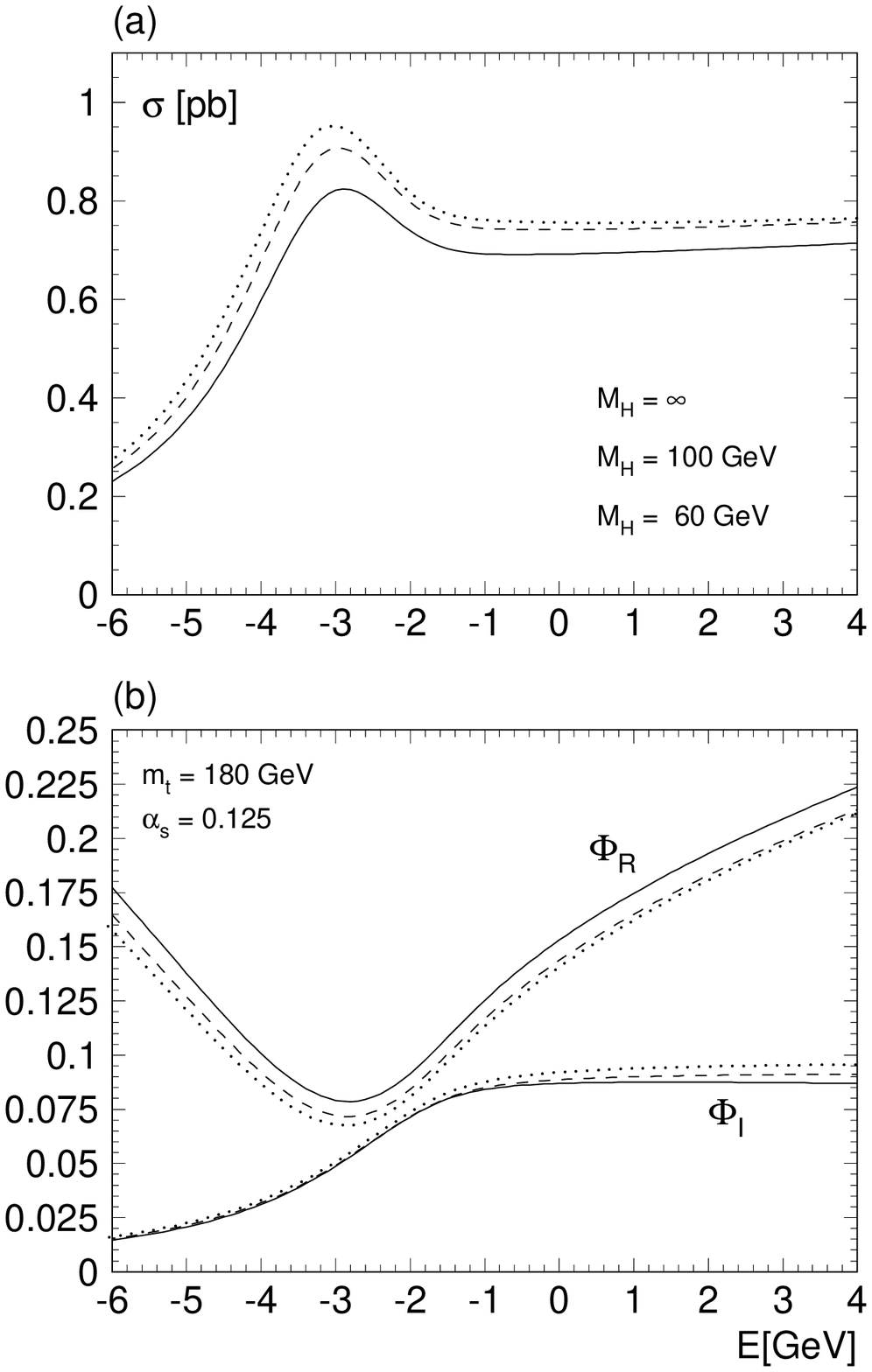}
  \caption[]{\label{singlet.ps}\sloppy (a) Total cross section and 
        (b) $\Phi_{\rm R} = \Re\Phi$ and $\Phi_{\rm I}= \Im \Phi$ for
        two different values of the Higgsmass $M_{\rm H}$ and without
        Higgs (solid line).}
 \end{center}
\end{figure}

Fig.~\ref{singlet.ps} displays the influence of Higgs exchange on these
predictions. The Yukawa potential induced by Higgs exchange plus the hard
vertex correction according to \cite{jezk} is included in the prediction for
the cross section --- the prediction for $\Phi$ contains the Yukawa potential
only.  The modification of the cross section is quite sizeable, the influence
on angular distribution and top polarization is negligible.

\begin{figure}
\begin{tabular}{cc}
  \epsfxsize 70mm \mbox{\epsffile[0 60 567 760]{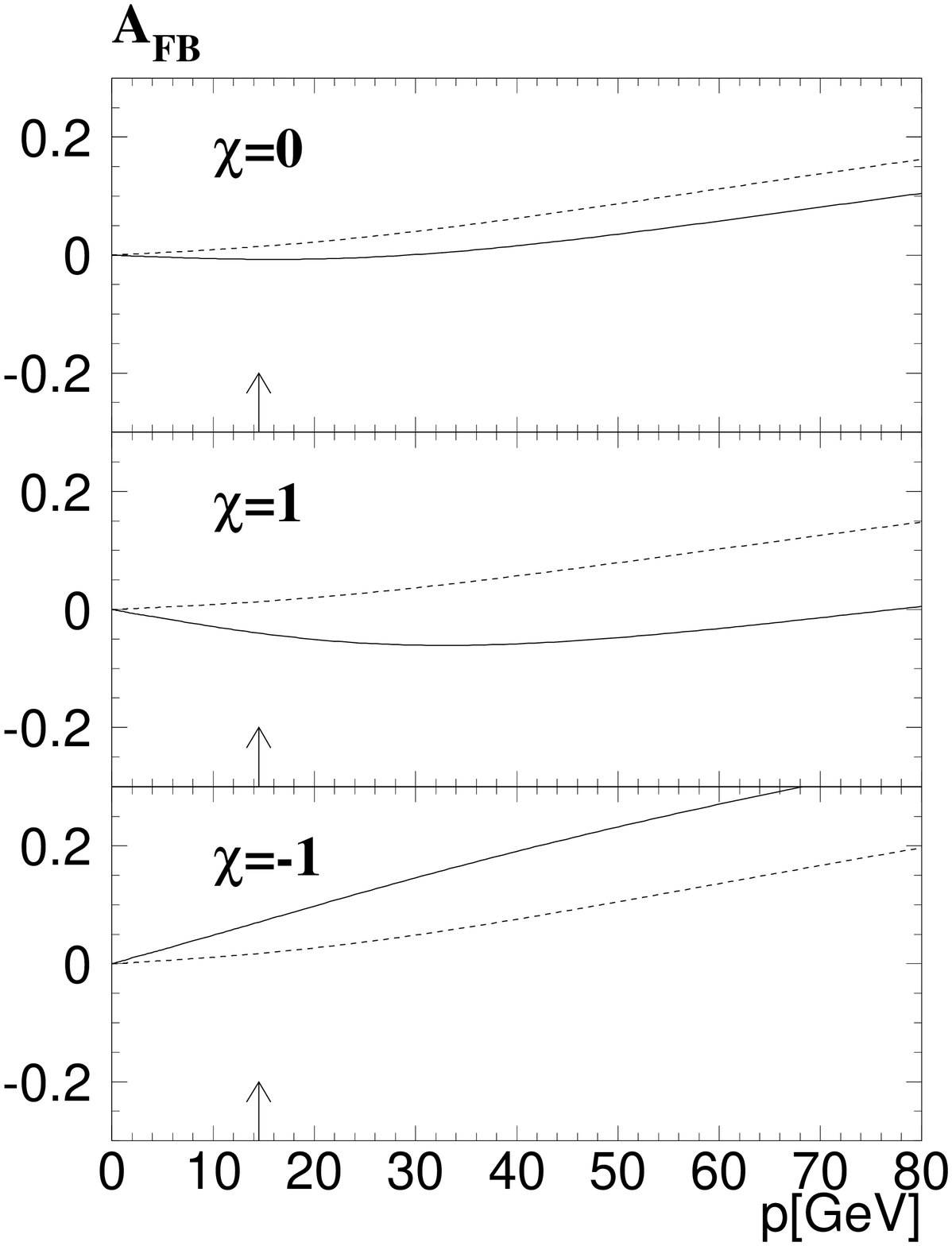}} &
  \epsfxsize 70mm \mbox{\epsffile[0 60 567 760]{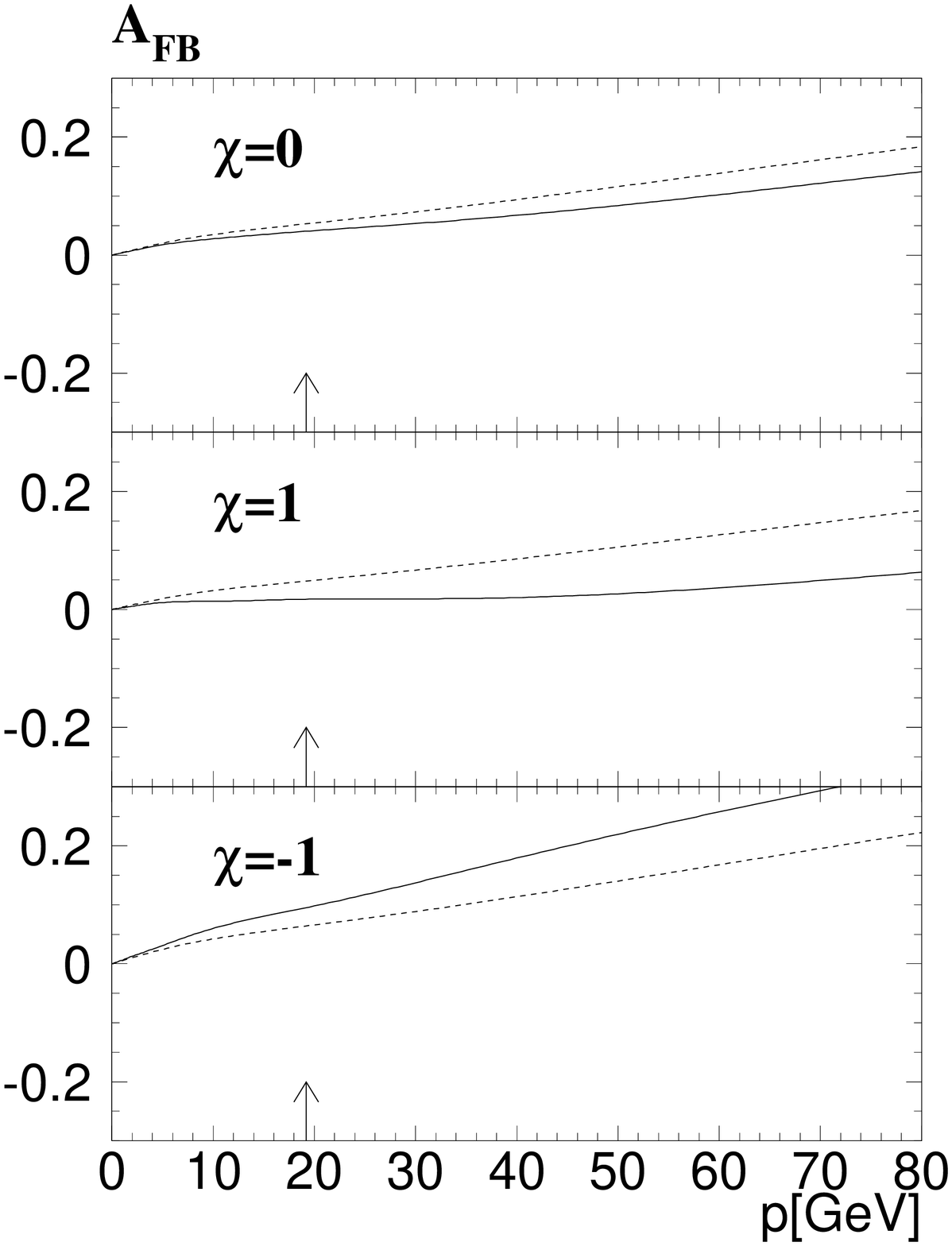}} \\
  a) E=-3GeV &  b) E=0GeV \\
  \epsfxsize 70mm \mbox{\epsffile[0 60 567 760]{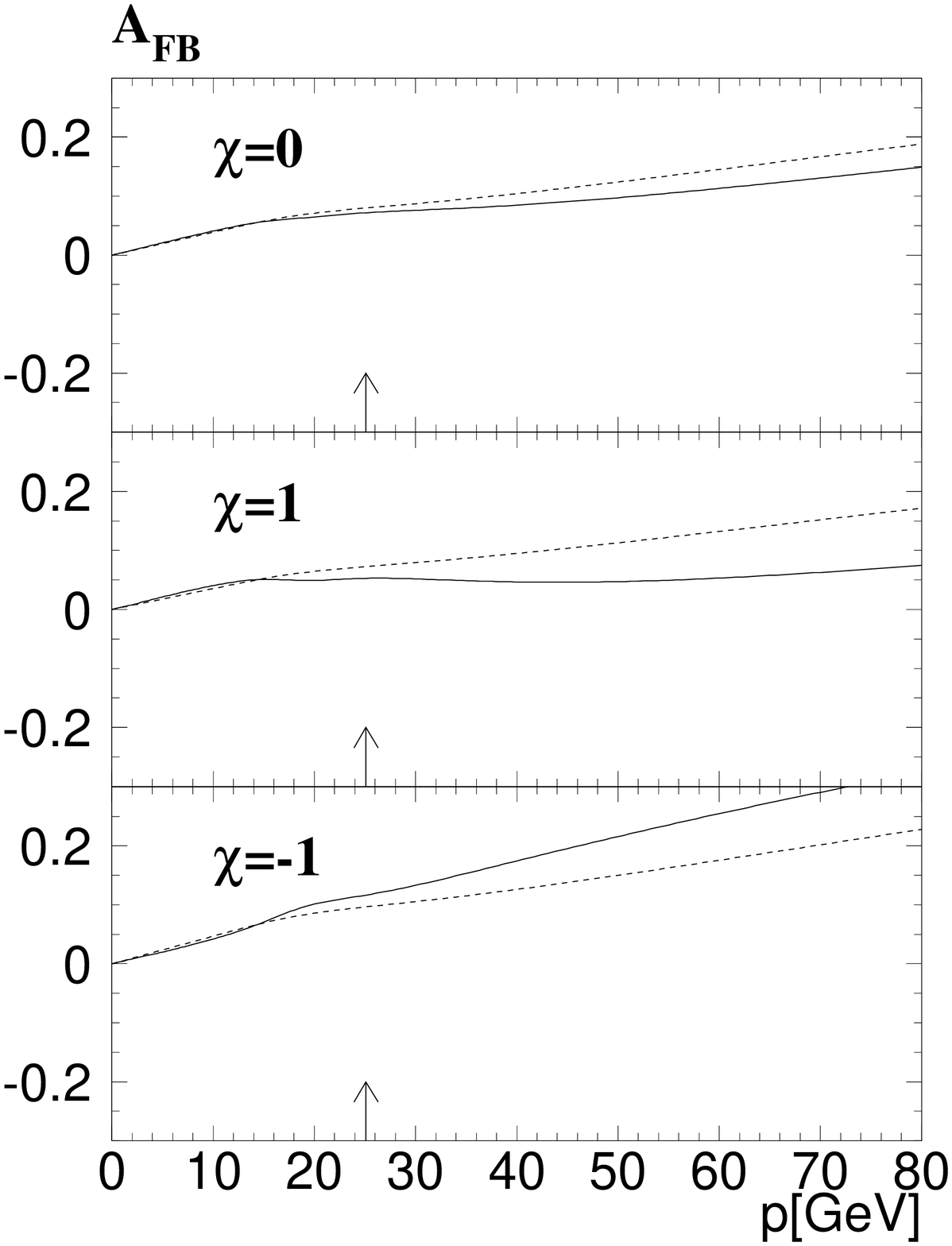}} &
  \epsfxsize 70mm \mbox{\epsffile[0 60 567 760]{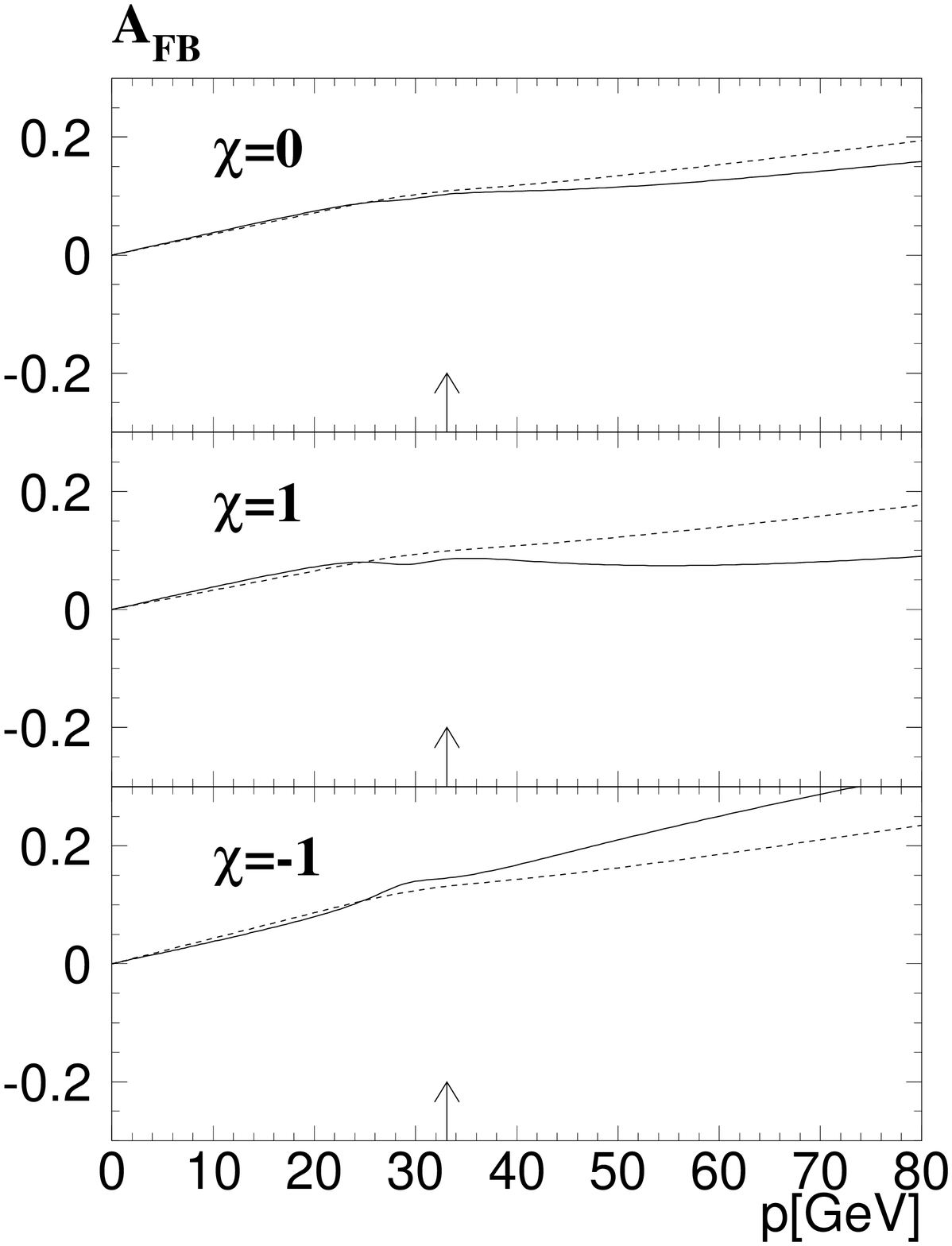}} \\
  c) E=2GeV & d) E=5GeV
\end{tabular}
\caption{\label{rescafb}Rescattering corrections to the forward-backward
asymmetry. Dashed line: $S$-$P$--wave interference contribution; 
solid line:
full result. The little arrows indicate the position of the peak in the
momentum distribution.}
\end{figure}

With these ingredients it is straightforward to predict the forward-backward 
asymmetry ${\cal A}_{\rm FB}$. It depends on the beam
polarization $\chi$, the cms energy E and the top quark momentum p.
The predictions without (dashed line) and with (solid line)
rescattering are displayed in Fig.~\ref{rescafb}. The relative size
and sign of the rescattering contribution also depend on momentum,
energy and beam polarization. Note that ${\cal A}_{\rm FB}$ is
affected by both $\psi_1$ and $\psi_2$ --- in contrast to the momentum
distribution {\em per se}, which was influenced solely by $\psi_1$. This is
consistent with the interpretation of $\psi_1$ and $\psi_2$ as $S$-
and $P$-wave decay amplitudes mentioned above.

In the region where most of the rate is concentrated (see Table \ref{peakpos})
rescattering corrections can practically be ignored and the qualitative
discussion based on $\varphi$ alone is applicable. For precision studies,
however, rescattering may become important.

\subsubsection{Top quark polarization}

Three remarks are important regarding the effect of rescattering on the
top quark polarization. First, what will be discussed in the following is the
quantity that can be measured by the moments of the lepton spectrum, which
should be understood as a practical way of defining top quark polarization.
Second, an extremely useful result is that the normal component 
${\cal P}_{\rm N}$
is not affected by rescattering and can thus be obtained immediately from
Figs.~\ref{pol_coefs.ps} and \ref{phi.ps}.
As a rule of thumb one predicts 
${\cal P}_{\rm N}($p$,E)\sim-0.07\sin\vartheta$. 

Third, the trivial angular dependence of the various spin components arising
from $S$-$P$--interference seems to be spoilt by rescattering due the $\psi_3$
terms. In the present situation, however, $\psi_3$ is always
multiplied by a numerically quite small factor, i.e.\ for $m_t=180$GeV
\[ y=\frac{M_W^2}{m_t^2}\sim0.2,\qquad\frac{1-4y+3y^2}{1+2y+3y^2}
    \sim0.22 \ .
\]
Since all $\psi_i$ are of comparable magnitude, the $\psi_3$ contribution
can be neglected. Hence the angular dependence of the longitudinal and
transverse polarization essentially remains as before. 

\begin{figure}
\begin{center}\begin{tabular}{cc}
  \epsfxsize 70mm \mbox{\epsffile[0 0 567 454]{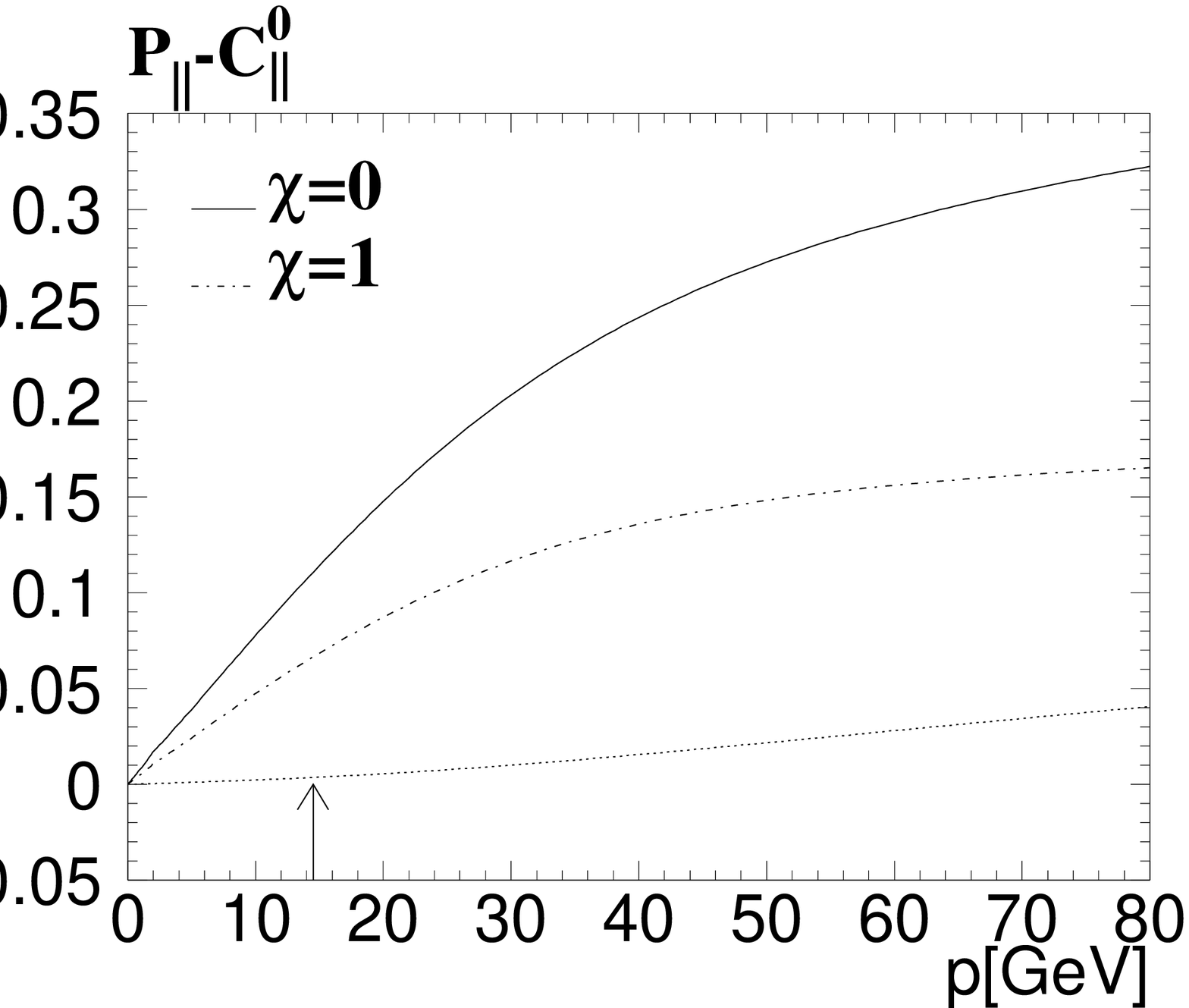}} &
  \epsfxsize 70mm \mbox{\epsffile[0 0 567 454]{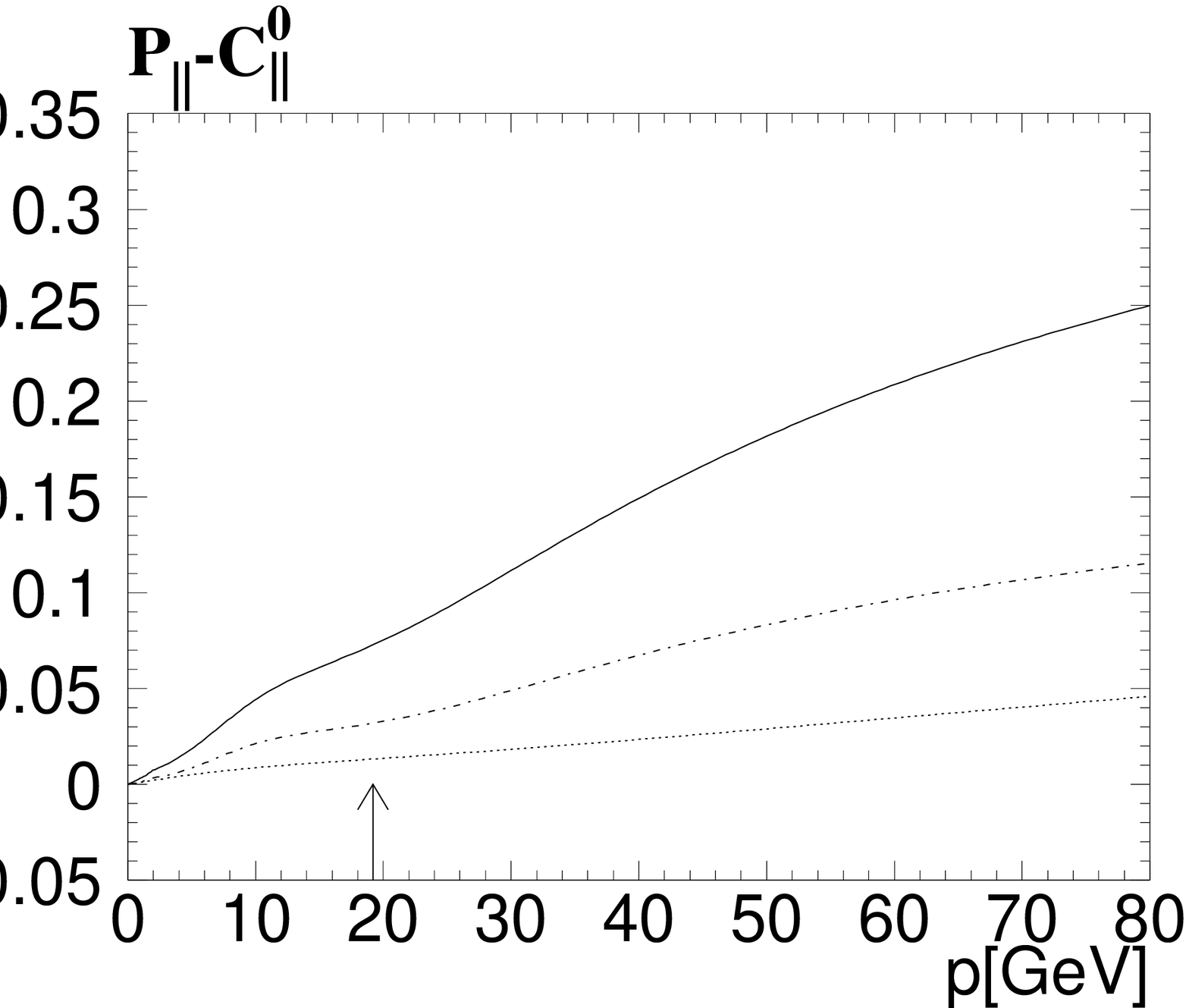}} \\
  a) E=-3.GeV & b) E=0GeV \\
  \epsfxsize 70mm \mbox{\epsffile[0 0 567 454]{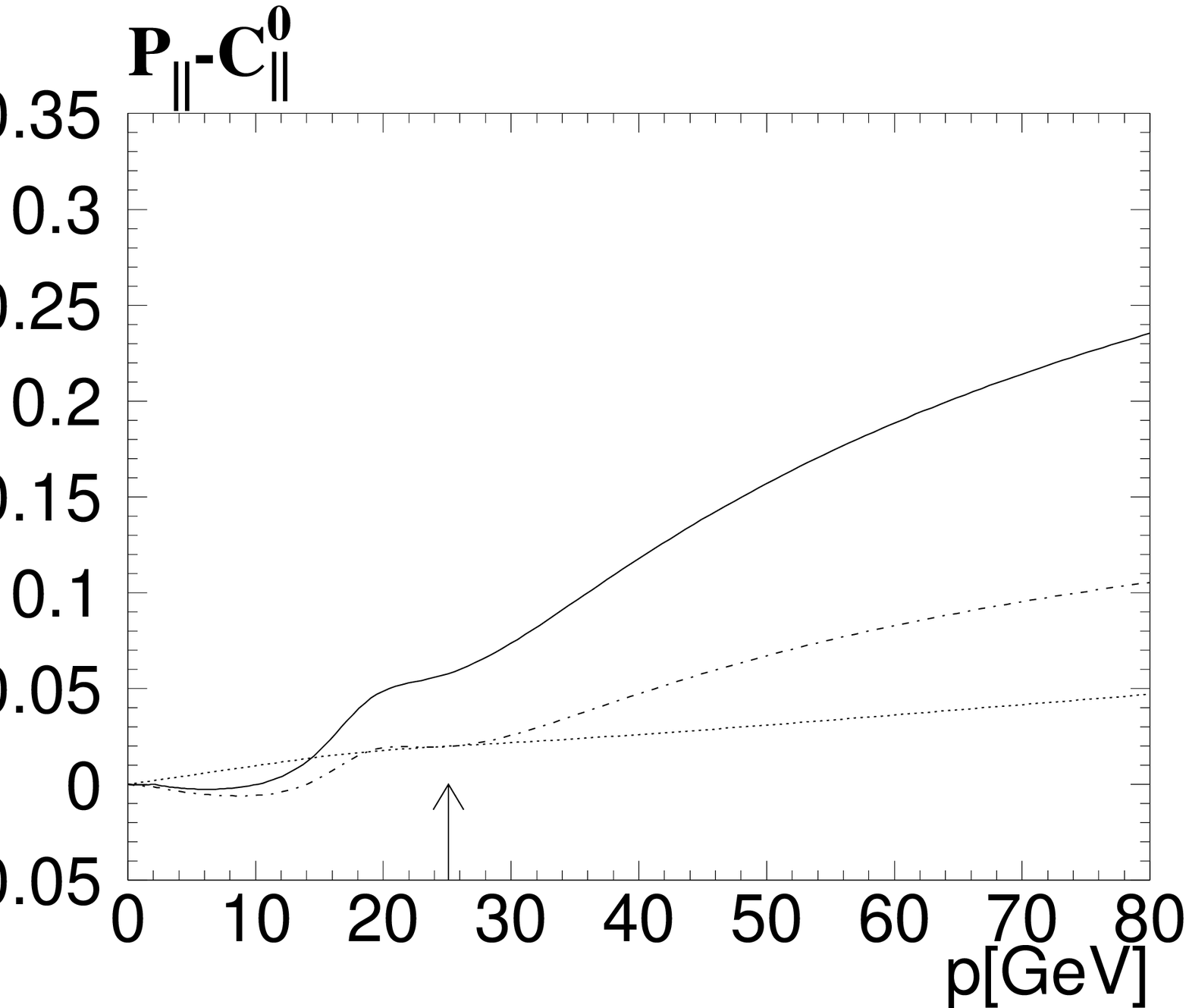}} &
  \epsfxsize 70mm \mbox{\epsffile[0 0 567 454]{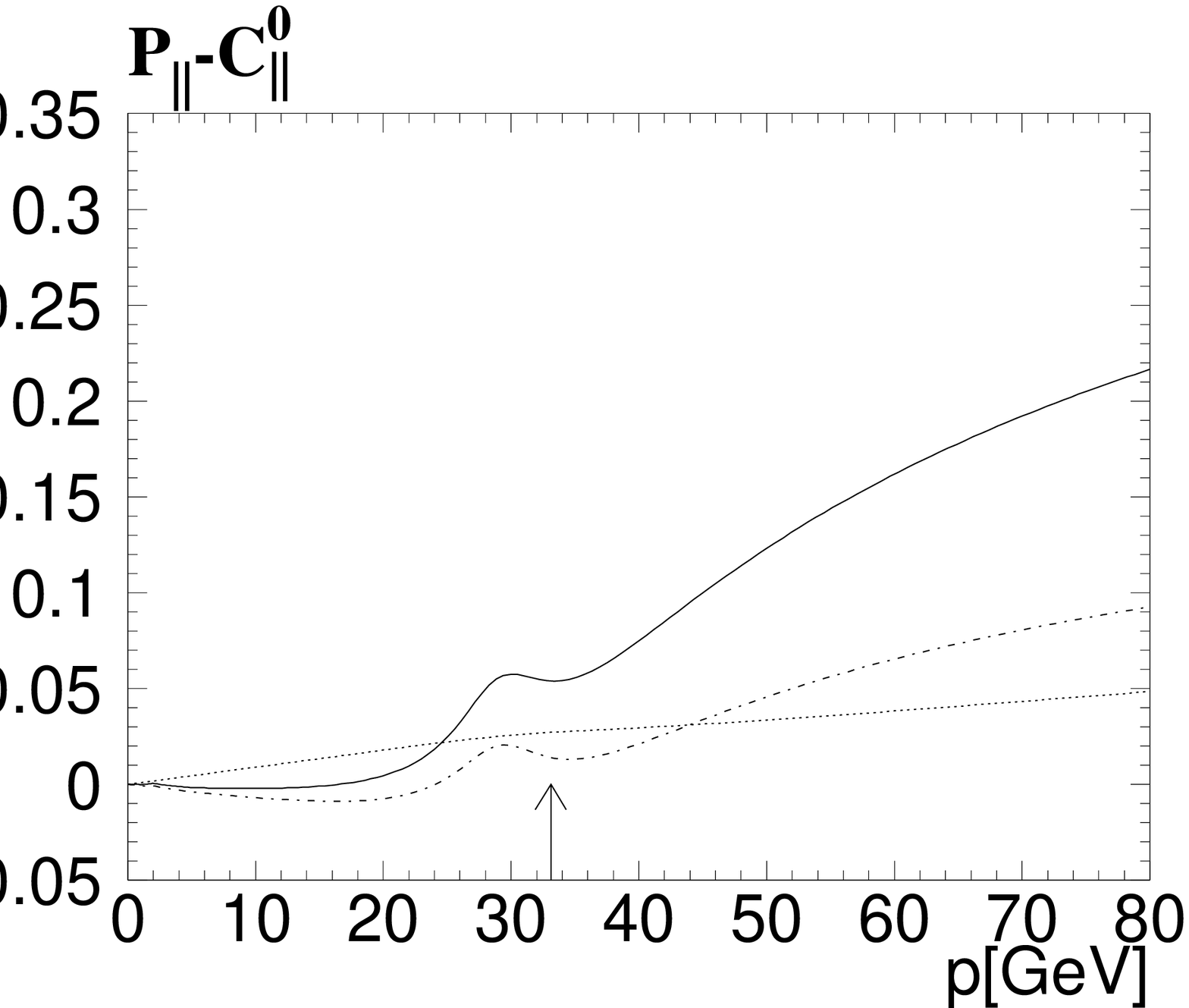}} \\
  c) E=2GeV & d) E=5GeV
\end{tabular}\end{center}
\caption{\label{pol_p.ps}Subleading part of the longitudinal component of the
  polarization vector ${\cal P}_\parallel$ for $\vartheta=0$.
  The solid line shows the complete result for unpolarized beams, the dotted
  line the $S$-$P$--interference contribution. The dash-dotted line shows
  the complete result for fully polarized beams, where the
  $S$-$P$--wave contribution vanishes. The arrows indicate the position of the
  peak in the momentum distribution.}
\end{figure}

\begin{figure}
\begin{center}\begin{tabular}{cc}
  \epsfxsize 70mm \mbox{\epsffile[0 0 567 454]{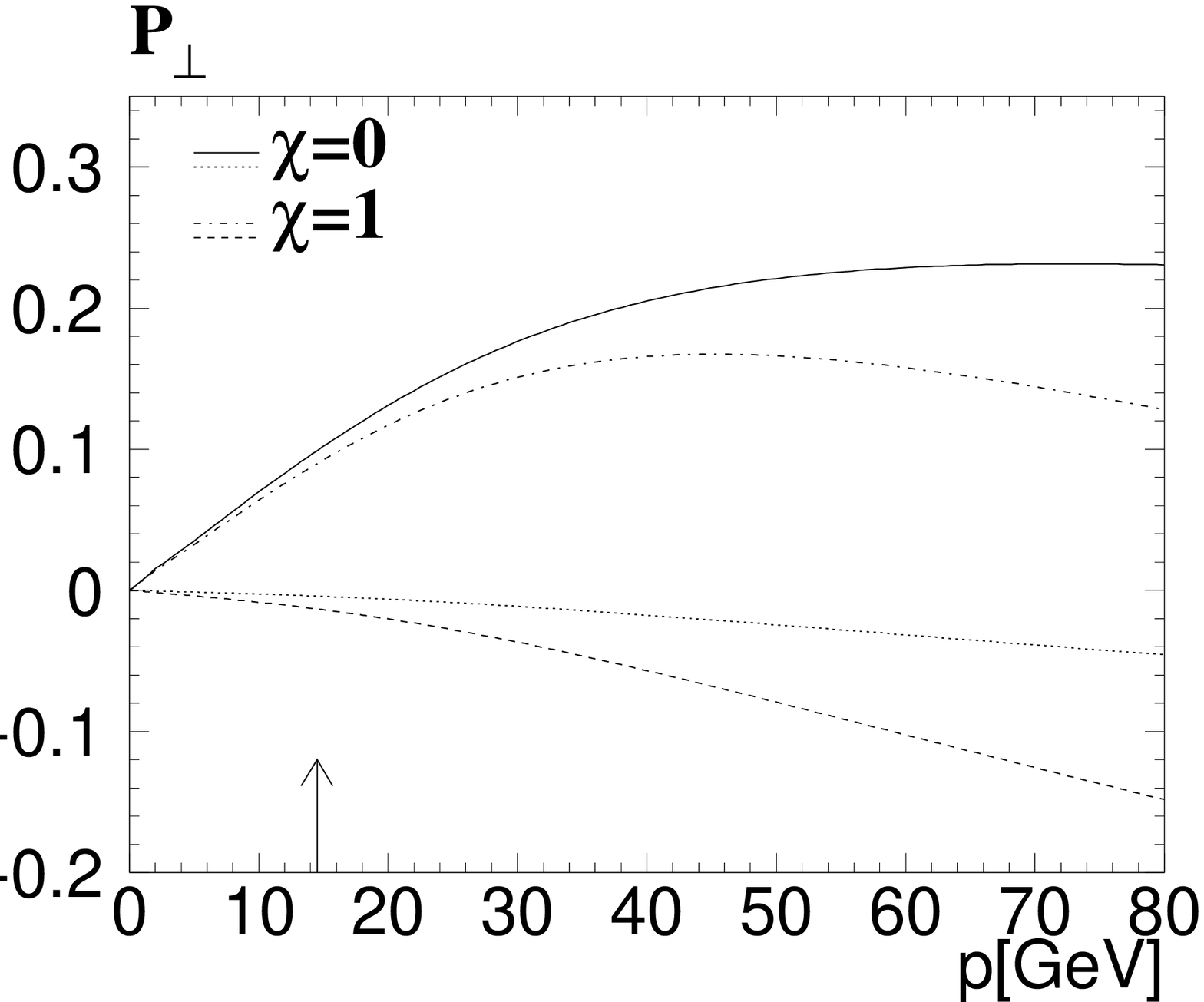}} &
  \epsfxsize 70mm \mbox{\epsffile[0 0 567 454]{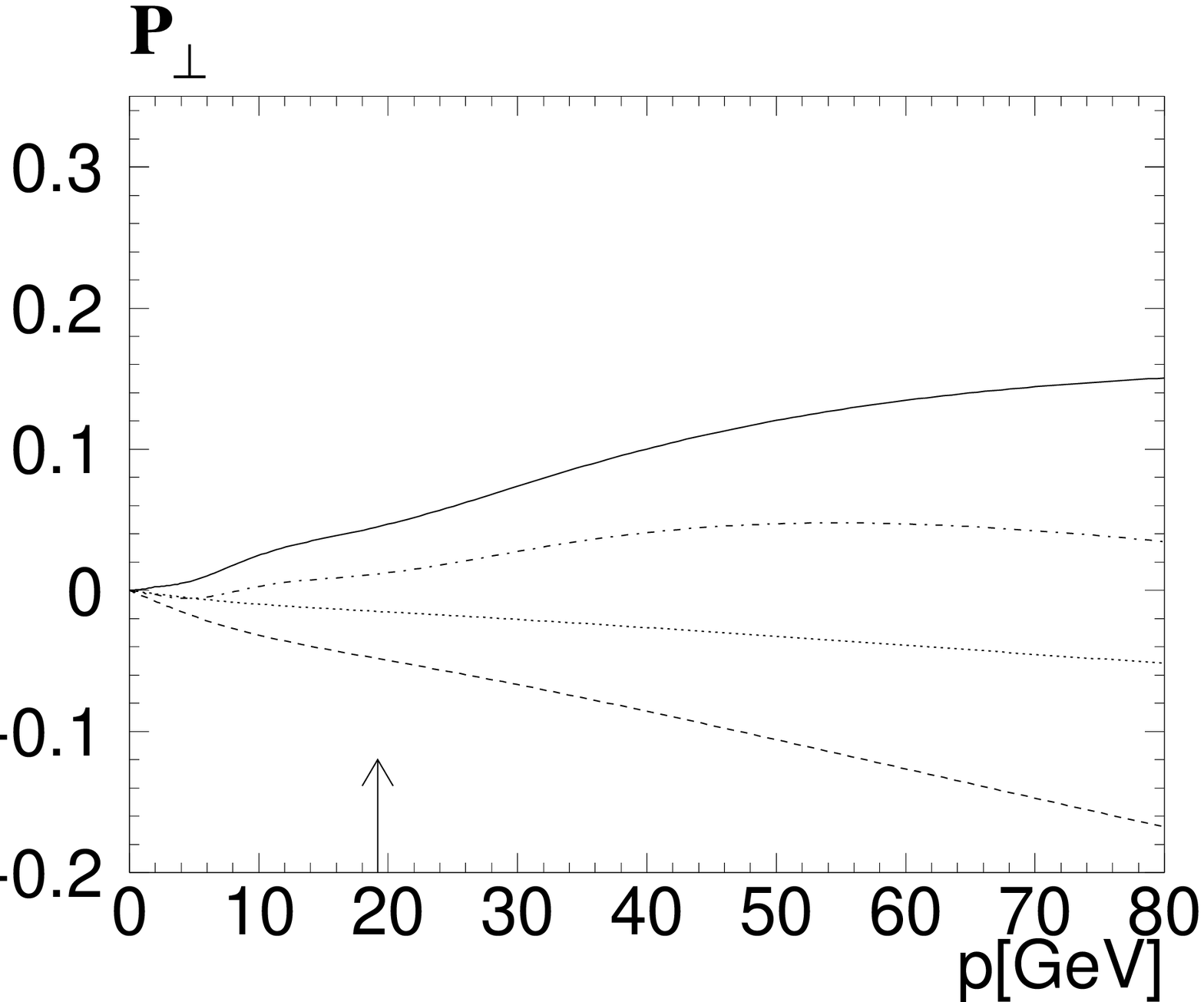}} \\
  a) E=-3GeV & b) E=0GeV \\
  \epsfxsize 70mm \mbox{\epsffile[0 0 567 454]{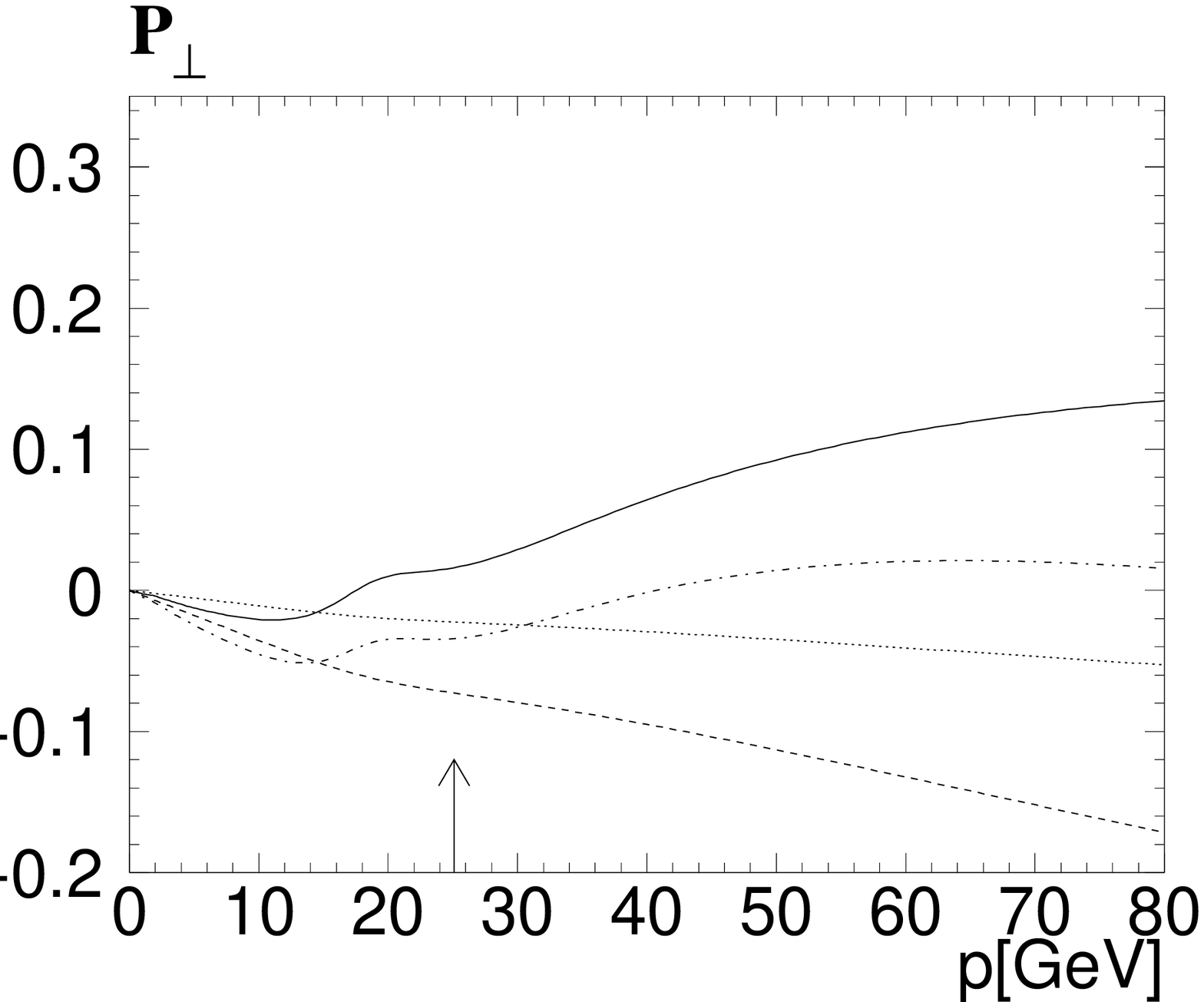}} &
  \epsfxsize 70mm \mbox{\epsffile[0 0 567 454]{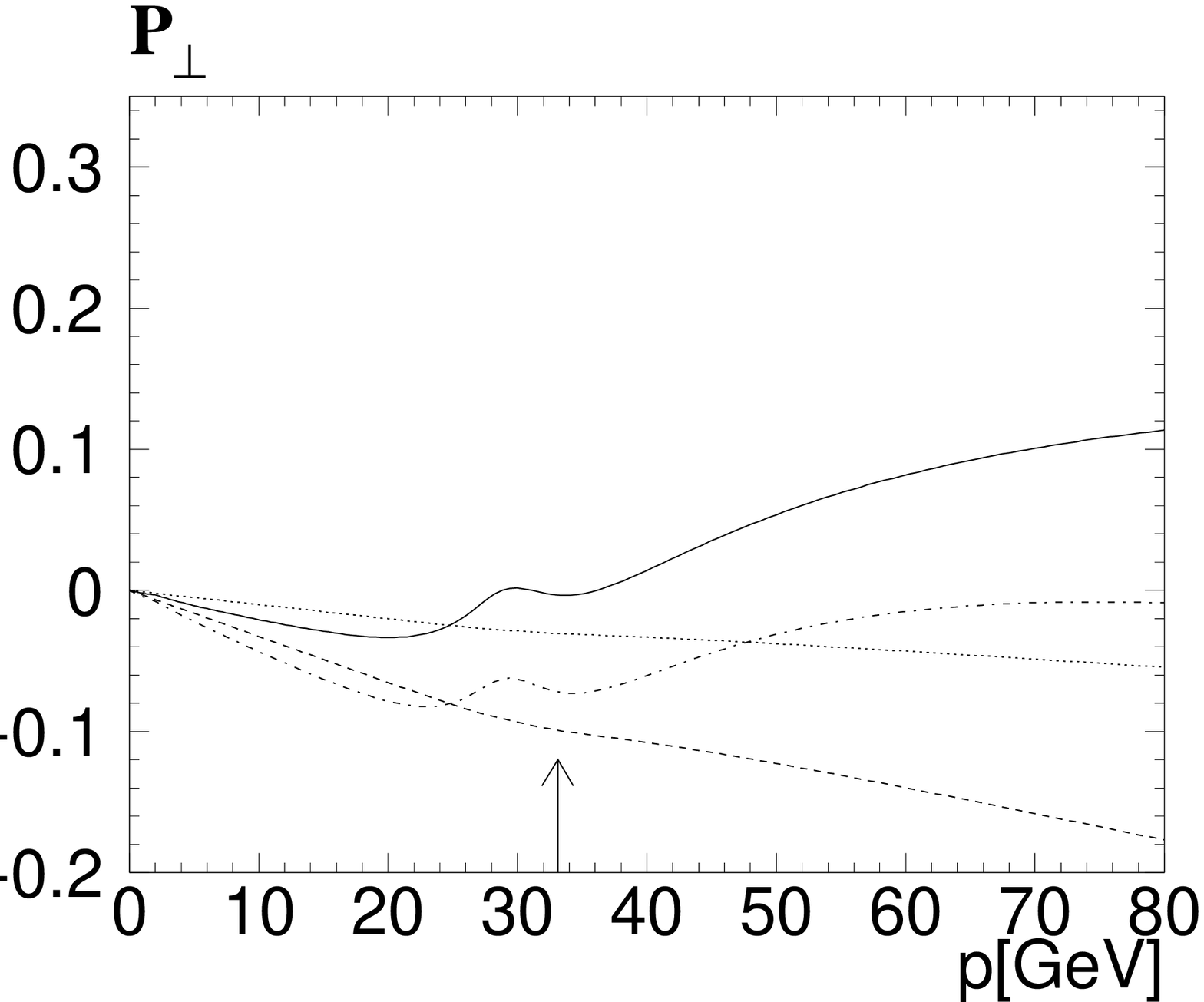}} \\
  c) E=2GeV & d) E=5GeV
\end{tabular}\end{center}
\caption{\label{pol_t.ps}Transverse component of the polarization vector
  ${\cal P}_\perp$, for $\vartheta=\pi/2$. The solid and dotted lines
  correspond to unpolarized beams, the dash-dotted and dotted lines to fully
  polarized beams. The upper two curves show the full result, the lower two the
  $S$-$P$--interference contribution. The arrows indicate the position of
  the peak in the momentum distribution.}
\end{figure}

In contrast, the influence of $\psi_2$ terms
on the longitudinal (Fig.~\ref{pol_p.ps}) and transverse 
(Fig.~\ref{pol_t.ps}) 
components of the ``polarization vector'' is dramatic.
The subleading piece of ${\cal P}_\parallel$, i.e.\ the coefficient of
$\cos\vartheta$, is dominated by rescattering, essentially because the
$S$-$P$--interference contribution is proportional to the small quantity
$C_\parallel^1$ and even vanishes for $|\chi|=1$, a consequence 
of the normalization condition for the polarization vector. 
This condition does no longer
hold once rescattering corrections are taken into account by calculating
moments of the lepton spectrum (it would of course remain true in the spin
projection formalism). 
The dominant piece of ${\cal P}_\parallel$ which is given by $C_\|^0$
remains however unaffected and
the subleading term which is shown in Fig.~\ref{pol_p.ps} vanishes after
integration over the top direction.

The transverse component is also dominated by the rescattering term as it is
evident from Figs.~\ref{pol_coefs.ps} and \ref{phi.ps} that
$S$-$P$--interference alone would lead to a negative result for non-negative
$\chi$.

To summarize: 
Our analysis has shown that the forward-backward asymmetry
and the normal component of the polarization vector are (relatively) 
stable against rescattering corrections. 
They determine the top couplings $a_3$ and $a_4$ via $C_{\rm FB}$ (or
possibly some hints on non-standard CP-violating interactions) and both the
real and imaginary part of the function $\Phi$ that contains important
information on $\Gamma_t$ (and thus on new physics or the existence of a
fourth generation) and the strong coupling constant $\alpha_{\rm s}$. 
The longitudinal component of the polarization vector is split into two parts.
The leading piece
$C_\parallel^0$ is independent of the production angle and measures the
couplings $a_1$ and $a_2$. The subleading angular dependent part and
the transverse component are small. They are sensitive to rescattering but
vanish after angular integration.

\appendix

\section{\label{RApp}Rescattering corrections}

In this appendix we will give a more detailed description of
rescattering using the formalism presented in Sect.\
\ref{rescatt}. For simplicity we will restrict ourselves to diagram
\ref{tbbar}a and mention only briefly where differences occur with
respect to diagram \ref{tbbar}b.

The quantity of interest is the interference contribution between the leading
order amplitude (with $\rmp=|\bf p|$, which should not be confused with
the four-vector $p$, and correspondingly for $k$)
\begin{eqnarray} 
  {\cal M}^{(0)}_j(s_+) &=&-\Big(\frac{ig_w}{\sqrt8}\Big)^2\bar u(b)\not\!\epsilon_+
        (1-\gamma^5)\frac{1+\gamma_5\!\not\!s_+}{2}\Lambda_+\gamma_j\Lambda_-
        \not\!\epsilon_-(1-\gamma^5)v(\bar b) \nonumber \\ &&
        \cdot G(\rmp,E)\Big(G_t(p,E)+G_{\bar t}(p,E)\Big)
\end{eqnarray}
and the rescattering amplitude
\begin{eqnarray}
  {\cal M}^{a}_i &=& \Big(\frac{ig_w}{\sqrt8}\Big)^2\int\!\frac{d^4k}{
     (2\pi)^4}\bar u(b)\not\!\epsilon_+(1\!-\gamma^5)i\Lambda_+\gamma^\mu
     \Lambda_+\gamma_i\Lambda_-\!\not\!\epsilon_-(1-\gamma^5)S_{\rm F}(k\!-\!p
     -\bar b) \nonumber \\ && \gamma^\nu v(\bar b)\cdot 
     G_t(p,E)G(\rmk,E)\Big(G_t(k,E)+G_{\bar t}(k,E)\Big)D_{\mu\nu}(k-p)
     g_s^2C_F
\end{eqnarray}
to the hadronic tensor
\begin{equation}
   H_{ij}=\int\!\frac{dp_0}{2\pi}\int\!d\mbox{PS}_2(t;bW^+)\int\!d\mbox{PS}_2
        (\bar t;\bar bW^-){\cal M}_i{\cal M}_j^\dagger~+~\mbox{c.c.}
\end{equation}
$G_t$ and $G_{\bar t}$ are the nonrelativistic propagators
$$
  G_{t/\bar t}(p,E) = \frac{1}{E/2\pm p_0+i\Gamma_t/2-{\bf p}^2/2m_t}
$$
and $S_{\rm F}(k-p-\bar b)$ is the $\bar b$ propagator. Neglecting
terms of the order $(k_0-p_0)^2$ and $({\bf k-p})^2$ it can be written as
$$
 S_{\rm F}(k-p-\bar b) \approx -i\frac{\not\!k-\not\!p-\not\!\bar b}
    {2E_{\bar b}(
    k_0-p_0-{\bf n}_{\bar b}\cdot({\bf k-p})-i\varepsilon/2E_{\bar b})}.
$$
The $k_0$ integration may then be performed immediately: closing the contour
in the lower half-plane and using Coulomb gauge to avoid spurious poles
in the gluon propagator, one only picks up the pole of $G_t(k,E)$ (putting
antitop on its mass-shell):
$$
  k_0 = \frac{1}{2}(-i\Gamma_t-E+{\bf k}^2/2m_t).
$$
The appearance of $\Lambda_+\gamma^\mu\Lambda_+
D_{\mu\nu}(k-p)=\Lambda_+D_{00}(k-p)g_{\nu0}$ implies that only
Coulomb gluon exchange contributes and the replacement
$g_s^2C_FD_{00}(k-p)\to-iV({\bf k-p})$ is at hand. We thus find
\begin{eqnarray*}
  {\cal M}^{a}_i &=& \Big(\frac{ig_w}{\sqrt8}\Big)^2\int\!\frac{d^3k}{
     (2\pi)^3}\bar u(b)\frac{\not\!\epsilon_+(1\!-\gamma^5)\Lambda_+\gamma_i
     \Lambda_-\!\not\!\epsilon_-(1-\gamma^5)(\not\!\bar b+\not\!p-\not\!k)
     \gamma^0}{2E_{\bar b}(-p_0-{\bf n}_{\bar b}\cdot({\bf k-p})-i\Gamma_t/2)}
     \\
  && v(\bar b)G_t(p,E)G(\rmk,E)V({\bf k-p})
\end{eqnarray*}
and insert this into the expression for the hadronic tensor. The $(bW^+)$
phase space integration that arises is trivial due to (\ref{psp}), and
we arrive at (only displaying one of the two interference terms)
\begin{eqnarray*}
 H_{ij} & = & \int\!\frac{dp_0}{2\pi}\int\!d\mbox{PS}_2(\bar t;\bar bW^-)
    \int\frac{d^3k}{(2\pi)^3}\frac{g_w^2}{4} \\ &&
    G^*(\rmp,E)G(\rmk,E)V({\bf k\!-\!p})G_t(p,E)\Big(G^*_t(p,E)\!+\!
    G^*_{\bar t}(p,E)\Big) \\
 && \frac{\Gamma_t}{2m_t E_{\bar b}}\sum_{\epsilon_-}
   \frac{\mbox{Tr}\Big[\not\!t(1-\gamma^5)
   \Lambda_+\gamma_i\Lambda_-\!\not\!\epsilon_-(\not\!\bar b+\not\!p-\not\!k)
   \gamma^0\not\!\bar b\epsilon_-(1-\gamma^5)\Lambda_-\gamma_j\Lambda_+
   \frac{1+\gamma_5\!\not\!s_+}{2}\Big]}
   {p_0+{\bf n}_{\bar b}\cdot({\bf k-p})+i\Gamma_t/2}.
\end{eqnarray*}
Now the $p_0$ integration has become simple: closing the contour in the upper
half-plane, only the pole of $G^*_t(p,E)$ is enclosed (putting top on its
mass shell):
$$
   p_0 = \frac{1}{2}(i\Gamma_t-E+{\bf p}^2/2m)\longrightarrow
   G_t(p,E)=\frac{1}{i\Gamma_t} \, .
$$
The terms proportional to $k_0$ or $p_0$ in the numerator are thus obviously
subleading, but the same is also true for the space components. Hence only
$$
  \frac{g_w^2}{4}\int\!d\mbox{PS}_2(\bar t;\bar bW^-)
  \frac{\gamma^\rho\!\not\bar b\gamma^0\!
  \not\bar b\gamma^\sigma(1-\gamma^5)}{2E_{\bar b}({\bf n}_{\bar b}\cdot
  ({\bf k-p})
  +i\Gamma_t)}\Big(-g_{\rho\sigma}+\frac{W_\rho W_\sigma}{m_W^2}\Big)
$$ $$
  = \Big( A({\bf k,p})\gamma^0-\frac{\mbox{\boldmath$\gamma$}\cdot\bf(k-p)}
    {|{\bf k-p}|}B({\bf k,p})\Big)(1-\gamma^5)
$$
remains to be determined, which is relatively straightforward and yields
\begin{eqnarray}
  A({\bf k},{\bf p}) & = & -i\frac{\Gamma_t}{|{\bf k}-{\bf p}|}\arctan\frac{|
        {\bf k}-{\bf p}|}{\Gamma_t} \\
  B({\bf k},{\bf p}) & = & -\kappa\frac{\Gamma_t}{|{\bf k}-{\bf p}|}
        \Big(1-\frac{\Gamma_t}{|{\bf k}-{\bf p}|}\arctan\frac{|{\bf k}-{\bf p}|
        }{\Gamma_t}\Big)
\end{eqnarray}
with $y=m_W^2/m_t^2$ and $\kappa=(1-2y)/(1+2y)$.
So we finally find the hadronic tensor to be
\begin{eqnarray*}
  H_{ij} &=& \frac{1}{m_t}\int\frac{d^3k}{(2\pi)^3}G^*(\rmp,E)
     G(\rmk,E)V({\bf k\!-\!p})\mbox{Tr}\Big[\not\!t(1-\gamma^5)\Lambda_+
     \gamma_i\Lambda_- \\ &&
     \Big(A({\bf k},{\bf p})\gamma^0-\frac{\mbox{\boldmath$\gamma$}
     \cdot\bf(k-p)}{|{\bf k-p}|}
     B({\bf k,p})\Big)(1-\gamma^5)\Lambda_-\gamma_j\Lambda_+
   \frac{1+\gamma_5\!\not\!s_+}{2}\Big]
\end{eqnarray*}
plus the corresponding complex conjugate term. The trace can be further
simplified using $\not t=m_t\gamma^0$ which is valid to the order
required here, and thus
$$ \Lambda_+\frac{1+\gamma^5\!\not\!s_+}{2}\frac{\not t}{m}(1-\gamma^5)
   \Lambda_+ = \Lambda_+\frac{1+{\bf s_+\cdot\mbox{\boldmath$\gamma$}}
   \gamma^5}{2} \, ,
$$
leading to
\begin{eqnarray} \lefteqn{
  H_{ij} = -\int\frac{d^3k}{(2\pi)^3}G^*(\rmp,E)G(\rmk,E)
    V({\bf k\!-\!p})} && \nonumber \\
  && \times\mbox{Tr}\Big[\Lambda_+\gamma_i
     \Big(A({\bf k},{\bf p})-\frac{{\bf (k-p)}\cdot\mbox{\boldmath$\gamma$}
     \gamma^5}{|{\bf k-p}|}
     B({\bf k},{\bf p})\Big)\gamma_j\frac{1+{\bf s}_+\cdot\mbox{\boldmath$\gamma$
     }\gamma^5}{2}\Big].
\end{eqnarray}
Performing the trace, contracting with the leptonic tensors and taking
twice the real part to include the complex conjugate term, one finally
obtains the formula given in Sect.~\ref{rescatt}. The relation between the
functions $A$, $B$ on one hand and $\psi_1$, $\psi_2$ on the other is
quite obvious:
\begin{eqnarray}
   \psi_1(\rmp,E)\Gamma_t &=& 
        2\Re\int\!\frac{d^3k}{(2\pi)^3}V(|{\bf k}-{\bf p}|)
        \frac{G(\rmk,E)}{G(\rmp,E)}A({\bf k},{\bf p}) \\
   -\kappa\psi_2(\rmp,E)\Gamma_t &=& 2\int\!\frac{d^3k}{(2\pi)^3}V(|{\bf
        k}-{\bf p}|)\frac{{\bf p}\cdot({\bf k-p})}{|{\bf p}||{\bf k- 
        p}|}\frac{G(\rmk,E)}{G(\rmp,E)}B({\bf k},{\bf p}) .
\end{eqnarray}

The calculation of the second diagram is essentially the same, the
main difference arising from the change in the ordering of the
vertices which produces another trace:
$$ \mbox{Tr}\Big[\Lambda_+\frac{1+\gamma_5\!\not\!s_+}{2}\Big(A({\bf k},
  {\bf p})\gamma^0-\frac{\mbox{\boldmath$\gamma$}\cdot\bf(k-p)}
  {|{\bf k-p}|}B({\bf k,p})\Big)
  (1-\gamma^5)\Lambda_+\gamma_i\gamma_j\Big]
$$
and  consequently leads to a different contribution to the differential
cross section. 

\section{\label{app:moments}Calculation of the moments of the lepton spectrum} 

It has already been mentioned in Sect.~\ref{moments} that for the
calculation of the moments of the lepton spectrum, the hadronic tensor
is replaced by either
\begin{equation}
   H_{ij} = \int\!\frac{dp_0}{2\pi}\!\int\!\frac{dW^2}{2\pi}\!
            \int\!\!d\mbox{PS}_2(t;bW^+) \int\!\!d\mbox{PS}_2(W^+;l\nu)
            \int\!\!d\mbox{PS}_2(\bar t;\bar bW^-)\sum_{\epsilon_-}\,
            {\cal M}_i{\cal M}_j^\dagger
\end{equation}
or
\begin{equation}
   H_{ij;n} = \int\!\frac{dp_0}{2\pi}\!\int\!\frac{dW^2}{2\pi}\!
            \int\!\!d\mbox{PS}_2(t;bW^+) \int\!\!d\mbox{PS}_2(W^+;l\nu)\,
            (nl)\,
            \int\!\!d\mbox{PS}_2(\bar t;\bar bW^-)\sum_{\epsilon_-} 
            {\cal M}_i{\cal M}_j^\dagger
\end{equation}
where the first expression leads to the differential cross section, the
second one to the moments. The amplitudes ${\cal M}$ describe production
of the $t\bar t$ pair and its subsequent decays:
the three-body transition $t\to bl\nu$ and the
two-body transition $\bar t\to \bar bW^-$.

Let us consider for illustration the simplest case first -- the
leading vector-vector contribution to the hadronic tensors, neglecting
terms of order $\beta$. For this contribution
\begin{eqnarray} 
{\cal M}_i{\cal M}_j^\dagger &=&
  g_w^2\,D_W\,\Big(\frac{g_w^2}{4}\Big)^2|G(\rmp,E)|^2|G_t(p,E)+
   G_{\bar t}(p,E)|^2\, T^{\alpha\beta}\, 
  \nonumber\\
  &\times&\mbox{Tr}\Big[\gamma_\beta\not\!b\gamma_\alpha\Lambda_+\gamma_i
  \Lambda_-\not\!\epsilon_-\not\!\bar b\not\!\epsilon_-(1-\gamma^5)
  \Lambda_-\gamma_j\Lambda_+(1+\gamma^5)\Big]
\end{eqnarray}
where 
\begin{equation}
  T^{\alpha\beta} = \frac{1}{4}\mbox{Tr}[\not\!\nu\gamma^\alpha\not\! l
        \gamma^\beta(1-\gamma^5)].
\end{equation}
and the narrow width approximation is employed for the propagator of $W^+$
\begin{equation}
  D_W=\Big|\frac{1}{W^2-m_W^2+im_W\Gamma_W}\Big|^2\approx\frac{\pi}{m_W
      \Gamma_W}\delta(W^2-m_W^2).
\end{equation}
Integration over the two-body phase space $\mbox{PS}(\bar t;\bar
bW^-)$ has already been described in Sect.~\ref{rescatt}, see
(\ref{psp}), and over $W^2$ is trivial.  The $p_0$ integral can
be solved by closing the integration contour in the upper half-plane,
thus enclosing the two poles
$$ p_0 = \frac{1}{2}(E+i\Gamma_t-{\bf p}^2/m_t) \qquad\mbox{of }G_{\bar t} $$
$$ p_0 = \frac{1}{2}(-E+i\Gamma_t+{\bf p}^2/m_t)\qquad\mbox{of }G_t^* $$
with the result
\begin{equation}
  \int\frac{dp^0}{2\pi}|G_t(p,E)+G_{\bar t}(p,E)|^2 = \frac{2}{\Gamma_t}.
\end{equation}

Using Lorentz symmetry one obtains 
\begin{equation}
  \int\!d\mbox{PS}_2(W;l\nu)\,T^{\alpha\beta} =
        \frac{1}{24\pi}(W^\alpha W^\beta-W^2g^{\alpha\beta})
\end{equation}
and
\begin{equation}
  \int\!d\mbox{PS}_2(W;l\nu)\,T^{\alpha\beta}\,(nl) = \frac{1}{48\pi}\Big(
        (Wn)(W^\alpha W^\beta-W^2g^{\alpha\beta})-\frac{i}{2}W^2\epsilon^{
        \alpha\beta\rho\delta}n_\rho W_\delta\Big)
\end{equation}
and integration over the two-body phase space $\mbox{PS}_2(t;bW^+)$ also
presents no problem if one exploits Lorentz symmetry.  We have to
calculate
$$
  \int d\mbox{PS}_2(t;bW^+)\Big(\not\!W^+\not\!b\not\!W^+ + 
       2m_t^2 y\not\!b\Big) =
  \int d\mbox{PS}_2(t;bW^+)\Big(\not\!t\not\!b\not\!t+2m_t^2 y\not\!b\Big)
$$
where $b^2=0$ is assumed, and
$$
  \int d\mbox{PS}_2(t;bW^+)
  \Bigg[(Wn)\Big(\not\!t\not\!b\not\!t+2m^2 y\not\!b\Big)
  -i\frac{m^2y}{2}\epsilon^{\alpha\beta\rho\delta}n_\rho W_\delta
   \gamma_\beta\not\!b\gamma_\alpha\Bigg]~. 
$$
The calculation of these expressions can be reduced to the calculation of
tensor integrals of the form
\begin{equation}\label{bwpspa}
  \int d\mbox{PS}_2(t;bW^+)b^\alpha
  =\frac{1-y}{2}t^\alpha\cdot\mbox{PS}(t;bW^+)
\end{equation}
and
\begin{equation}\label{bwpspb}
  \int d\mbox{PS}_2(t;bW^+)\, b^\alpha W^\beta = \frac{1-y}{12}\Big[
   2(1+2y)t^\alpha t^\beta+(1-y) m_t^2 g^{\alpha\beta}\Big]
   \cdot\mbox{PS}_2(t;bW^+)
\end{equation}
Putting everything together and identifying BR$(W\to l\nu)$ with
BR$(t\to bl\nu)$, we find
\begin{eqnarray}
  H_{ij} & = & 2\mbox{BR}(t\to bl\nu)\frac{|G(\rmp,E)|^2}{m_t^2}\mbox{Tr}
     \Big[\not\!t\Lambda_+\gamma_i\Lambda_-\not\!\bar t(1-\gamma^5)\Lambda_-
     \gamma_j\Lambda_+(1+\gamma^5)\Big] \\
  H_{ij;n} & = & 2\mbox{BR}(t\to bl\nu)\frac{|G(\rmp,E)|^2}{m_t^2}
     \frac{1+2y+3y^2}{4(1+2y)}\frac{1}{3} \nonumber \\
  && \mbox{Tr}\Big[\Big(4(nt)\not\!t-m_t^2\not\!n\Big)\Lambda_+\gamma_i\Lambda_-
     \not\!\bar t(1-\gamma^5)\Lambda_-\gamma_j\Lambda_+(1+\gamma^5)\Big].
\end{eqnarray}
The traces can now be performed. Contractions of the resulting expressions 
for the tensors $ H_{ij}$ and $ H_{ij;n}$
with the leptonic tensors $L^{ij}_{s,a}$, c.f.\ (\ref{eq:Lij}),
yield the results given in Sect.~\ref{moments}.

Contributions to the moments from rescattering in the $t\bar b$
system, c.f.\ Fig.~\ref{tbbar}a, can be calculated in a way analogous
to that described in App.~\ref{RApp} for the integrals over $k_0$,
$p_0$ and $\mbox{PS}_2(\bar t;\bar bW^-)$ whereas integration over
$\mbox{PS}_3(t;bl\nu)$ is performed as described in this appendix. It is
not surprising therefore that one arrives at final expressions which
are analogous to those obtained using the formalism of
App.~\ref{RApp}. The same is true for all factorisable contributions,
in particular for $S$-$P$--wave interference ones.
  
A new feature arises for $b\bar t$ rescattering, see
Fig.~\ref{tbbar}b, which is non-factorisable.  Instead of
(\ref{bwpspa},\ref{bwpspb}) two new tensor integrals appear in the
calculation
\begin{eqnarray}
  I_1^\alpha\gamma_\alpha &\!\!=\!\!& \int\!\!d\mbox{PS}_2(t;bW^+)
   \frac{\not\!b\gamma^0 \not\!b}{(b+k-p)^2} \approx 
    m_t\frac{1\!-\!y}{2}\!\int\!\!d\mbox{PS}_2(t;bW^+)\frac{\not\!b}{2b(k-p)}
   \\
  I_2^{\alpha\beta}\gamma_\alpha &\!\!=\!\!& \int\!\!d\mbox{PS}_2(t;bW^+)
   \frac{\not\!b\gamma^0\!\not\!b\;b^\beta}{(b+k-p)^2} \approx
    m_t\frac{1\!-\!y}{2}\! \int\!\!d\mbox{PS}_2(t;bW^+)\frac{\not\!b\;b^\beta}
    {2b(k-p)}
\end{eqnarray}
The $k_0$ and $p_0$ integrations can again be performed by closing the 
integration contours in the complex planes. 
However, in the present case the contours are closed in the upper 
$k^0$ half-plane and the lower $p^0$ half-plane, 
and this choice implies  
$k_0-p_0\approx i\Gamma_t$.

The integrals $I_1^\alpha$ are proportional to $A({\bf k,p})$ or $B({\bf k,p})$
again. One finds
\begin{eqnarray}
  I_1^0 &=& A({\bf k},{\bf p}) \frac{m_t}{\Gamma_t}\frac{1-y}{16\pi}\\
  I_1^j &=& \frac{1+2y}{1-2y}\frac{(k-p)^j}{|{\bf k}-{\bf p}|}
       B({\bf k},{\bf p})\frac{m_t}{\Gamma_t}\frac{1-y}{16\pi}. 
\end{eqnarray}
The integrals $I_2^{\alpha\beta}$ are given by the following formul\ae:
\begin{eqnarray}
  I_2^{00} &=& 2A({\bf k},{\bf p}) \frac{m_t}{\Gamma_t}\frac{(1-y)^2}{64\pi} \\
  I_2^{0i} &=& 2\frac{1+2y}{1-2y}\frac{(k-p)^i}{|{\bf k}-{\bf p}|}
                 B({\bf k},{\bf p}) \frac{m_t}{\Gamma_t}\frac{(1-y)^2}{64\pi}\\
  I_2^{ij} &=& \Big(\delta_{ij}-\frac{(k-p)^i(k-p)^j}{|{\bf k}-{\bf p}|^2}
                  \Big)A({\bf k},{\bf p}) 
                  \frac{m_t}{\Gamma_t}\frac{(1-y)^2}{64\pi}\nonumber \\
    && -\frac{1+2y}{1-2y}\frac{i\Gamma_t}{|{\bf k}-{\bf p}|}\Big(\delta_{ij}
       -3\frac{(k-p)^i(k-p)^j}{|{\bf k}-{\bf p}|^2}\Big)B({\bf k},{\bf p})
       \frac{m_t}{\Gamma_t}\frac{(1-y)^2}{64\pi}.
\end{eqnarray}
The second term in $I_2^{ij}$ is subleading, because the
relevant momentum transfer should satisfy 
$|{\bf k}-{\bfp}|\gg\Gamma_t$, 
i.e.\ the top width only acts as infrared regulator
and can be put equal to zero where possible (the overall factor
$1/\Gamma_t$ common to all integrals drops out in the final result).
The first term must be kept however, and the appearance of this second
rank tensor is the reason why the function $\psi_3$ has to be
introduced.


\def\app#1#2#3{{\it Acta~Phys.~Polonica~}{\bf B #1} (#2) #3}
\def\apa#1#2#3{{\it Acta Physica Austriaca~}{\bf#1} (#2) #3}
\def\fortp#1#2#3{{\it Fortschr.~Phys.~}{\bf#1} (#2) #3}
\def\npb#1#2#3{{\it Nucl.~Phys.~}{\bf B #1} (#2) #3}
\def\plb#1#2#3{{\it Phys.~Lett.~}{\bf B #1} (#2) #3}
\def\prd#1#2#3{{\it Phys.~Rev.~}{\bf D #1} (#2) #3}
\def\pR#1#2#3{{\it Phys.~Rev.~}{\bf #1} (#2) #3}
\def\prl#1#2#3{{\it Phys.~Rev.~Lett.~}{\bf #1} (#2) #3}
\def\prc#1#2#3{{\it Phys.~Reports }{\bf #1} (#2) #3}
\def\cpc#1#2#3{{\it Comp.~Phys.~Commun.~}{\bf #1} (#2) #3}
\def\nim#1#2#3{{\it Nucl.~Inst.~Meth.~}{\bf #1} (#2) #3}
\def\pr#1#2#3{{\it Phys.~Reports }{\bf #1} (#2) #3}
\def\sovnp#1#2#3{{\it Sov.~J.~Nucl.~Phys.~}{\bf #1} (#2) #3}
\def\yadfiz#1#2#3{{\it Yad.~Fiz.~}{\bf #1} (#2) #3}
\def\jetp#1#2#3{{\it JETP~Lett.~}{\bf #1} (#2) #3}
\def\zpc#1#2#3{{\it Z.~Phys.~}{\bf C #1} (#2) #3}
\def\ptp#1#2#3{{\it Prog.~Theor.~Phys.~}{\bf #1} (#2) #3}
\def\nca#1#2#3{{\it Nouvo~Cim.~}{\bf #1A} (#2) #3}

\end{document}